\pdfoutput=1
\newcommand*{\ATLASLATEXPATH}{latex/}
\documentclass[UKenglish,texlive=2016,txfonts=true,cernpreprint]{\ATLASLATEXPATH atlasdoc}

\usepackage[lineno=false,csquotes=false,subcaption=false,subfigure=true]{\ATLASLATEXPATH atlaspackage}

\usepackage{\ATLASLATEXPATH atlasbiblatex}

\usepackage{\ATLASLATEXPATH atlascontribute}

\usepackage{\ATLASLATEXPATH atlasphysics}

\graphicspath{{logos/}{figures/}}

\AtlasTitle{Electron efficiency measurements with the ATLAS detector using 2012 LHC proton--proton collision data}

\author{The ATLAS Collaboration}

\date{\today}

\AtlasRefCode{PERF-2016-01}

\PreprintIdNumber{CERN-EP-2016-262}

\AtlasJournalRef{Eur. Phys. J. C (2017) 77}
\AtlasDOI{10.1140/epjc/s10052-017-4756-2}

\AtlasAbstract{%
This paper describes the algorithms for the reconstruction and identification of electrons in the central region of the ATLAS detector at the Large Hadron Collider (LHC). These algorithms were used for all ATLAS results with electrons in the final state that are based on the 2012 $pp$ collision data produced by the LHC at $\sqrt{\mathrm{s}}$ = 8~\TeV. 
The efficiency of these algorithms, together with the charge misidentification rate, is measured in data and evaluated in simulated samples using electrons from $Z\rightarrow ee$, $Z\rightarrow ee\gamma$ and $J/\psi \rightarrow ee$ decays. For these efficiency measurements, the full recorded data set, corresponding to an integrated luminosity of 20.3 fb$^{-1}$, is used. 
Based on a new reconstruction algorithm used in 2012, the electron reconstruction efficiency is 97\% for electrons with $\et=15$~\GeV\ and 99\% at $\et = 50$~\GeV.
Combining this with the efficiency of additional selection criteria to reject electrons from background processes or misidentified hadrons, the efficiency to reconstruct and identify electrons at the ATLAS experiment varies from 65\% to 95\%, depending on the transverse momentum of the electron and background rejection.
}

\hypersetup{pdftitle={ATLAS document},pdfauthor={The ATLAS Collaboration}}

\usepackage{amssymb} 
\usepackage{graphicx}
\usepackage{caption}
\usepackage{multirow}
\usepackage{epstopdf}
\usepackage{epsfig}
\usepackage{mathrsfs}
\usepackage{rotating}

\begin{document}

\maketitle

\tableofcontents

\section{Introduction}
\label{sec:Intro}

In the ATLAS detector~\cite{ATLASdetector}, electrons in the central detector region are triggered by and reconstructed from energy deposits in the electromagnetic (EM) calorimeter
that are matched to a track in the inner detector (ID). Electrons are distinguished from other particles 
using identification criteria with different levels of background rejection and signal efficiency. The identification criteria rely on the shapes of EM showers in the calorimeter as well as on tracking quantities and the quality of the matching of the tracks to the clustered energy deposits in the calorimeter. They are based either on independent requirements or on a single requirement, the output of a likelihood function built from these quantities.  

In this document, measurements of the efficiency to 
reconstruct and identify prompt electrons and their EM charge in the central region of the ATLAS detector\footnote{
ATLAS uses a right-handed coordinate system
with its origin at 
the nominal $pp$ interaction point at the centre of the detector.
The positive $x$-axis is defined by the direction
from the interaction point to the centre of the LHC ring, with the
positive $y$-axis pointing upwards, while the
beam direction defines the $z$-axis. The azimuthal angle $\phi$ is measured
around the beam axis and the polar angle $\theta$ is the angle from
the $z$-axis. The pseudorapidity is defined as $\eta = -\ln
\tan(\theta/2)$. The radial distance between two objects is defined as
$\Delta R = \sqrt{{(\Delta\eta)}^2+{(\Delta\phi)}^2}$.
Transverse energy is computed as $\et = E \cdot \sin\theta$.}
with pseudorapidity $|\eta|<2.47$ 
 are presented for \textit{pp} collision data produced by the Large Hadron Collider (LHC) in 2012 at a centre-of-mass energy of $\sqrt{\mathrm{s}}$ = 8 \TeV,  and compared to the prediction from Monte Carlo (MC) simulation. The goal is to extract correction factors and their uncertainties for measurements of final states with prompt electrons in order to adjust the MC efficiencies to those measured in data. Electrons from semileptonic heavy-flavour decays are treated as background.
 
The efficiency measurements follow the methods introduced in Ref.~\cite{Aad:2011Paper} for the 2011 ATLAS electron performance studies but are improved in several
respects 
 and adjusted for the 2012 data-taking conditions. 
The measurements are based on the \tandp\ method using the \Zboson\ and the \JPsi\ resonances, requiring  the
presence of an isolated identified electron as the \textit{tag}. Additional selection criteria are applied to obtain a
high purity sample of electron candidates that can be used as \textit{probes} to measure the reconstruction or identification
efficiency. The measurements span different but overlapping kinematic regions and are studied as a function of the electron's transverse momentum and pseudorapidity. The results are combined taking into account
bin-to-bin correlations.

After briefly describing the ATLAS detector in Section~\ref{sec:Detector},
the algorithms to reconstruct and identify electrons are summarized in Sections~\ref{sec:Reconstruction} and \ref{sec:EleID}.
The general methodology of \tandp\ efficiency measurements and the decomposition of the efficiency into its different components are
reviewed in Section~\ref{sec:Method}. The data and MC samples used in this work are summarized in Section~\ref{sec:Samples}.
Sections~\ref{sec:MeasureID} and \ref{sec:Rejection} describe the identification efficiency measurements for signal electrons as well as backgrounds. In Section~\ref{sec:MisID}, the measurement of the electron charge misidentification rate is presented. Section~\ref{sec:MeasureReco} details the reconstruction efficiency measurement, which extends the identification measurement methodology, and Section~\ref{sec:CombiRecoID} describes the final results of the combined identification and reconstruction efficiency measurements. Section~\ref{sec:Conclusions} concludes with a summary of the results.

\section{The ATLAS detector}
\label{sec:Detector}

A complete description of the ATLAS detector is provided in Ref.~\cite{ATLASdetector}. A brief description of the subdetectors that are relevant for the detection of electrons is given in this section.

The ID provides precise 
reconstruction of tracks within $|\eta| < 2.5$. It consists of three layers of pixel
detectors close to the beam-pipe, four layers of 
silicon microstrip detector modules with pairs of single-sided sensors glued back-to-back
(SCT) providing eight hits per track at intermediate radii, and a transition
radiation tracker (TRT) at the outer radii, providing on average 35 hits per track
 in the range $|\eta| < 2.0$. The TRT offers substantial discriminating
power between electrons and charged hadrons between energies of 0.5 \GeV\ and 100
\GeV, via the detection of X-rays produced by transition radiation.
The innermost pixel layer in 2012 and earlier, also called the b-layer, is located outside the beam-pipe at a radius of 50 mm. Together with the other layers, it provides precise vertexing and significant rejection of
photon conversions through the requirement that a track has a hit in this 
layer.

The ID is surrounded by a thin superconducting solenoid with a length of 5.3 m and diameter of 2.5 m. The solenoid provides a 2 T magnetic field for the measurement of the curvature of charged particles to determine their charge and momentum. The solenoid design attempts to minimize the amount of material by integrating it into a vacuum vessel shared with the LAr calorimeter. The magnet thus only contributes a total of 0.66 radiation lengths of material at normal incidence.

 The main EM calorimeter is a lead/liquid-argon sampling
 calorimeter with accordion-shaped electrodes and lead absorber plates.
 It is divided into a barrel section (EMB) covering $|\eta|<1.475$
 and two endcap sections (EMEC) covering $1.375<|\eta|<3.2$. 
 For $|\eta|<2.5$, it is divided into three layers longitudinal in shower depth (called strip, middle and back layers)
 and offers a fine segmentation in the lateral
 direction of the showers.
 At high energy, most of the EM shower energy is collected in the middle layer which has a lateral
 granularity of 0.025 $\times$ 0.025 in $\eta$--$\phi$ space. The first (strip) layer consists
 of strips with a finer granularity in the $\eta$-direction and with a coarser granularity in $\phi$. It provides
 excellent $\gamma$--$\pi^0$ discrimination and a precise estimation of the pseudorapidity of the
 impact point.
 The back layer collects the energy deposited in the tail of high-energy EM showers.
 A thin presampler detector,
 covering $|\eta|<1.8$, is used to correct for fluctuations in upstream energy losses.
 The transition region between the EMB and EMEC calorimeters, $1.37<|\eta|<1.52$, suffers from a large  amount of material. 

Hadronic calorimeters 
with at least three segments longitudinal in shower depth surround the EM calorimeter and are used in this context to reject hadronic jets.
The forward calorimeters 
cover the range $3.1 < |\eta| < 4.9$ and also 
have EM shower identification capabilities given
their fine lateral granularity and longitudinal segmentation into three layers.

\section{Electron reconstruction}\label{sec:Reconstruction}

Electron reconstruction in the central region of the ATLAS detector ($|\eta|<2.47$) starts from energy deposits
(clusters) in the EM calorimeter which are then matched to reconstructed
tracks of charged particles in the ID. 

\subsection{Electron seed-cluster reconstruction}

The $\eta$--$\phi$ space of the EM calorimeter system is divided into a
grid of $N_\eta \times N_\phi = 200 \times 256$ towers of size $\Delta\eta^\mathrm{tower}
\times \Delta\phi^\mathrm{tower} = 0.025 \times 0.025$, 
corresponding to the granularity of the EM accordion calorimeter middle layer.
The energy of the calorimeter cells in all shower-depth layers
(the strip, middle and back EM accordion calorimeter layers and for $|\eta|<1.8$ also the
presampler detector) is summed
to get the tower energy. The energy of a cell which spans several towers is distributed evenly among the towers without taking into account any geometrical weighting.

To reconstruct the EM clusters, seed clusters of 
towers with total cluster transverse energy above 2.5~\GeV\ are
searched for by a sliding-window algorithm~\cite{topoclusters}. 
The window size is $3 \times 5$ towers 
in $\eta$--$\phi$ space.
A duplicate-removal algorithm is applied to nearby seed clusters.

Cluster reconstruction is expected to be very efficient
for true electrons. 
In MC samples passing the full ATLAS simulation chain, the efficiency is about 95\% for electrons with a transverse energy of
$\ET=7$~\GeV\ and reaches 99\% at $\ET=15$~\GeV\ and 99.9\% at $\ET=45$~\GeV,
placing a requirement only on the angular distance between the generated electron and the reconstructed 
electron cluster.
The efficiency decreases with increasing pseudorapidity in the endcap region $|\eta|>1.37$.

\subsection{Electron-track candidate reconstruction}
\label{sec:electron-track}

Track reconstruction for electrons was improved for the 2012 data-taking period with respect to the one used for 2011 data-taking, especially
for electrons which undergo significant energy loss due to bremsstrahlung in
the detector, to achieve a high and uniform efficiency.

Table~\ref{tab:IDcuts} shows the definition of shower-shape and track-quality variables, including $R_\eta$ and $R_\mathrm{Had}$. For each seed EM cluster\footnote{As in the 2011 electron reconstruction algorithm, 
clusters must satisfy loose requirements 
on the maximum fraction of energy deposited in the different layers of the EM calorimeter system: 
0.9, 0.8, 0.98, 0.8 for the presampler detector, 
the strip, the middle and the back EM accordion calorimeter layers, respectively.} passing loose shower-shape requirements of 
$R_\eta > 0.65$ and $R_\mathrm{Had} < 0.1$ a region-of-interest (ROI) is defined as a cone of size $\Delta R$ = 0.3 around the
seed cluster barycentre. The collection of these EM cluster ROIs is retained for use in the track reconstruction.

Track reconstruction proceeds in two steps: pattern recognition and track fit. In 2012, in addition to the standard track-pattern recognition and track fit, an electron-specific pattern recognition and track fit were introduced in order to recover losses from bremsstrahlung and therefore improve the reconstruction of electrons. Either of these algorithms, the pattern recognition and the track fit, use a particle-specific hypothesis for the particle mass and respective probability for the particle to undergo bremsstrahlung, referred to in the following either as pion or electron hypothesis.

The standard pattern recognition~\cite{ATLASPatternReco} uses the pion hypothesis for energy loss in the material of the detector. If a track\footnote{The transverse momentum threshold for tracks
reconstructed with the pion hypothesis is 400 \MeV\ based on the pattern recognition.} seed (consisting of
three hits in different layers of the silicon detectors) with a transverse
momentum larger than 1~\GeV\ cannot be successfully extended to a full
track with at least seven hits using the pion hypothesis and it falls within
one of the EM cluster ROIs,
it is retried with the new pattern recognition using an electron
hypothesis that allows for
energy loss. This modified pattern recognition algorithm (based on a Kalman filter--smoother formalism~\cite{KalmanFilter}) 
allows up to 30\% energy loss at each material surface to account for bremsstrahlung. Below 1~\GeV, no refitting is performed. Thus, an electron-specific algorithm has been integrated into the standard track reconstruction; it improves the performance for electrons and has minimal interference with the main track reconstruction.

Track candidates are fitted using either the pion or the electron hypothesis
(according to the hypothesis used in the pattern recognition)
with the ATLAS Global $\chi^2$ Track Fitter~\cite{ATLASTrackFitter}. The electron hypothesis employs the same track fit as for the pion hypothesis except that it adds an extra term to compensate for the increase in $\chi^2$ due to bremsstrahlung losses, in order to be able to fit the track with an acceptable $\chi^2$ such that it can be further used in the electron reconstruction.
 If a track candidate fails the pion hypothesis track fit due to a large $\chi^2$ (for example caused by large energy losses), it is refitted
using the electron
hypothesis.

Tracks are then considered as loosely matched to an EM cluster, if they pass either of the
following two requirements:

\begin{itemize}
\item[(i)] 
 Tracks with at least four silicon hits are extrapolated from their  
 measured perigee 
 to the middle layer of the EM accordion calorimeter. 
In the middle layer of the calorimeter, the extrapolated tracks have to be  either within 0.2 in $\phi$ of the EM cluster on the side the track is bending towards 
 or within 0.05 on the opposite side. They also have to be within 0.05 in $\eta$ of the EM cluster. 
 TRT-only tracks, i.e. tracks with less than four silicon hits, are extrapolated from the last
 measurement point. They are retained at this early stage as they are used later in the reconstruction chain to reconstruct photon conversions. Clusters without any associated tracks with silicon hits are eventually considered as photons and are not used to reconstruct prompt-electron candidates. TRT-only tracks have to pass the same requirement for the difference in $\phi$ between track and cluster as tracks with silicon hits 
 but no requirement is placed on the difference in $\eta$ between track and cluster at this stage 
 as their $\eta$ coordinate is not measured precisely.
\item[(ii)]  
 The track extrapolated to the middle layer of
 the EM accordion calorimeter, after rescaling its momentum to the measured cluster energy, 
 has to be either within 0.1 in $\phi$ of the EM cluster on the side the track is bending towards 
 or within 0.05 on the opposite side. Furthermore, non-TRT-only 
 tracks must be within 0.05 in $\eta$ of the calorimeter cluster. 
 As in (i), the track extrapolation is made from the last measurement point for TRT-only tracks 
 and from the 
 point of closest approach with respect to the primary collision vertex
 for tracks with silicon hits.
\end{itemize}

Criterion (ii) aims to recover tracks of typically large curvature that have 
potentially suffered significant energy loss before reaching the calorimeter. Rescaling the momentum of the track to that of the reconstructed cluster allows retention of tracks whose measured momentum in the ID does not match the energy reconstructed in the calorimeter because they have undergone bremsstrahlung. The bremsstrahlung is assumed to have occurred in the ID or the cryostat and solenoid before the calorimeter (for tracks with silicon hits) or in the cryostat and solenoid before the calorimeter (for TRT-only tracks). 

At this point, all electron-track candidates are defined. 
The track parameters of these candidates, for all but the TRT-only tracks, are precisely re-estimated using an optimized 
electron track fitter, the Gaussian Sum Filter
(GSF)~\cite{GSF} algorithm, which is a non-linear generalization of the Kalman 
filter~\cite{KalmanFilter} algorithm. It yields a
better estimate of the electron track parameters, especially those in the
transverse plane, by accounting for non-linear bremsstrahlung effects. 
TRT-only tracks and the very rare tracks (about 0.01\%) that fail the GSF fit keep 
the parameters from the Global $\chi^2$ Track Fit. 
These tracks are then used to perform the final track--cluster matching
to build electron candidates and also to provide information for particle 
identification.

\subsection{Electron-candidate reconstruction}
\label{sec:elCandReco}
An electron is reconstructed if at least one track is matched to the seed
cluster. The efficiency of this matching and subsequent track quality requirements is measured as the reconstruction efficiency in Section~\ref{sec:MeasureReco}. 
The track--cluster matching proceeds as described for the previous step in Section~\ref{sec:electron-track}, 
but with the GSF refitted tracks and tighter requirements: the separation in $\phi$ must be less than 0.1 (and not 0.2). Additionally, TRT-only tracks must satisfy loose 
track--cluster matching criteria in $\eta$ and tighter ones in $\phi$: in the TRT barrel $|\Delta \eta| < 0.35$ and in the TRT endcap $|\Delta\eta| < 0.02$. In both the barrel and the endcaps the requirements are $|\Delta\phi| < 0.03$ on the
 side the track is bending towards and  $|\Delta\phi| < 0.02$ on the other side.
In this procedure, more than one track can be associated with a cluster. 

Although all tracks assigned to a cluster are kept for further analysis, the
best-matched one is chosen as the primary track which is used to determine the kinematics and charge of the electron
and to calculate the electron identification decision. 
Thus choosing the primary track is a crucial step in the electron reconstruction chain.
To favour the primary electron track and to avoid
random matches between nearby tracks in the case of cascades due to bremsstrahlung,
tracks with at least one hit in the pixel detector are preferred.
If more than one associated track has pixel hits, the following sorting criteria are considered. First, the distance between the
track and the cluster is considered for any pair of tracks, which are referred to as $i$ and $j$ in the following. 
Then two angular distance variables are defined
in the $\eta$--$\phi$ plane. $\Delta R$ is the distance between the cluster
barycentre and the extrapolated track in the middle layer of the EM accordion calorimeter,
while $\Delta R_\mathrm{rescaled}$ is the distance between the cluster
barycentre and the extrapolated track when the track momentum is rescaled to the
measured cluster energy before the extrapolation to the middle layer. If 
$|\Delta R_{\mathrm{rescaled},i} - \Delta R_{\mathrm{rescaled},j}| > 0.01$,  
the track with the smaller $\Delta R_\mathrm{rescaled}$ is 
chosen. If 
$| \Delta R_{\mathrm{rescaled},i} - \Delta R_{\mathrm{rescaled},j}| \leq 0.01$
and $|\Delta R_i - \Delta R_j| > 0.01$, the track with smaller $\Delta R$ is
taken.
For the rest of the cases, the two tracks have both similar $\Delta R_\mathrm{rescaled}$ and 
similar $\Delta R$, and the track with more pixel hits\footnote{Throughout the paper, when counting hits in the pixel and SCT detectors, non-operational modules that are traversed by the track are counted as hits.} 
is chosen as the primary track.
A hit in the first layer of the pixel detector counts twice to prefer tracks with
early hits. If there are two best tracks with exactly the same numbers of hits, the track with smaller $\Delta R$ is taken.

All seed clusters together with their matching tracks, if there is at least one of them, are treated as electron
candidates. 
Each of these electron clusters is then rebuilt in all four layers sequentially, starting from the middle layer, using $3 \times 7$ ($5 \times 5$) cells in $\eta \times \phi$ in the barrel (endcaps) of the EM accordion calorimeter. The cluster position is adjusted in each layer to take into account the distribution of the deposited energy. The fixed sizes of $3 \times 7$ ($5 \times 5$) cells for electron clusters 
were optimized to take into account the different
overall energy distributions in the barrel (endcap) accordion calorimeters specifically for electrons.\footnote{Unconverted (converted) photon clusters, which are used in the reconstruction efficiency measurement in Section~\ref{sec:MeasureReco}, are built using  $3 \times 5$ ($3 \times 7$) cells in the barrel and $5 \times 5$ ($5 \times 5$) cells in the endcap.}

Up to this point, neither the electron clusters nor the cells inside the clusters are calibrated. The energy calibration \cite{many:1637529} is applied as the next step and was improved for 2012 data using multivariate analysis (MVA) techniques~\cite{TMVA} and an improved description of the detector~\cite{ATLASGeant4} by the \texttt{GEANT4}~\cite{Geant4} simulation. The calibration procedure is outlined briefly below.

After applying the electronic readout calibration to the calorimeter cells
with a global energy scale factor corresponding to the electron response,
a number of data pre-corrections are applied for
measured effects of the bunch train structure and imperfectly corrected response in specific regions.
The presampler energy scales and the EM accordion calorimeter strip-to-middle-layer energy-scale ratios are also corrected~\cite{many:1637529}.

The cluster energy is then determined from the energy in the three layers of the EM accordion calorimeter
by applying a correction factor determined by linear regression using an MVA trained on large samples of single-electron MC events produced with the full ATLAS simulation chain. 
The input quantities used for electrons and photons are the total energy measured in the accordion calorimeter, the
ratio of the energy measured in the presampler to the energy measured in the accordion, the shower depth,\footnote{The shower depth is defined as $X = \Sigma_i X_iE_i/ \Sigma_i E_i$
where $E_i$ is the cluster energy in layer $i$ and $X_i$ is the approximate calorimeter thickness (in radiation lengths) from the interaction point to the middle of layer $i$,
including the presampler detector layer where present.}
the pseudorapidity of the cluster barycentre in the ATLAS coordinate system, 
and the $\eta$ and $\phi$ positions of the cluster barycentre
in the local coordinate system of the calorimeter. 
Including the cluster barycentre position allows a correction to be made for the larger lateral energy leakage
for particles that hit a cell close to the edge and for the  variation of the response as a function of the
particle impact point with respect to the calorimeter absorbers.
 
In the last step, correction factors are derived in situ using a large sample of collected \Zee\ events. 
They are applied to the reconstructed electrons as a final energy calibration in data events. Electron energies are smeared in simulated events, as the simulated electrons have a better energy resolution than electrons in data.
 
The four-momentum of central electrons ($|\eta|<2.47$)
is computed using information from both the final
cluster and the track best matched to the original seed cluster. The energy is
given by the cluster energy. The $\phi$ and $\eta$ directions are taken from the
corresponding track parameters, except for TRT-only tracks for which the
cluster $\phi$ and $\eta$ values are used.

\section{Electron identification}
\label{sec:EleID}

Not all objects built by the electron reconstruction algorithms are prompt electrons which are considered signal objects in this publication. Background objects include hadronic jets as well as electrons from photon conversions, Dalitz decays and from semileptonic heavy-flavour hadron decays. In order to reject as much of
these backgrounds as possible while keeping the efficiency for prompt electrons high, electron identification algorithms are based on
discriminating variables, which are combined into a menu of selections with various background rejection powers. Sequential requirements and MVA techniques are employed.
 
Variables describing the longitudinal and lateral shapes of the
EM showers in the calorimeters, the properties of the tracks in the ID, as well as the matching between tracks
and energy clusters are used to discriminate against the different background sources. 
These variables \cite{Atlas:ElectronPubNote,Aad:2010Paper,Aad:2011Paper} are detailed in Table~\ref{tab:IDcuts}. 
Table~\ref{tab:IDcutsUsage} summarizes which variables are used for the different selections of the so-called cut-based and likelihood (LH) \cite{neymanpearson} identification menus.

\begin{table*}
\caption{Definition of electron discriminating variables. }
\label{tab:IDcuts}
\footnotesize
\renewcommand{\arraystretch}{1.3}
\begin{center}
\begin{tabular}{|l|l|l|}
\hline
Type & Description & Name \\
\hline
 Hadronic leakage & Ratio of \et\ in the first layer of the hadronic  calorimeter to \et\ of the EM cluster  & $R_\mathrm{Had1}$ \\ 
& (used over the range $|\eta| < 0.8$ or $|\eta| > 1.37$)  & \\
\cline{2-3}
  & Ratio of \et\ in the hadronic calorimeter to \et\ of the EM cluster& \rhad \\
 & (used over the range $0.8 <|\eta| < 1.37$) & \\
\hline
Back layer of & Ratio of the energy in the back layer to the total energy in the EM accordion & \fIII  \\
EM calorimeter &  calorimeter &  \\
\hline
Middle layer of  & Lateral shower width, $\sqrt{(\Sigma E_i \eta_i^2)/(\Sigma E_i) -((\Sigma E_i\eta_i)/(\Sigma E_i))^2}$, where $E_i$ is the & \weta \\
EM calorimeter  & energy and $\eta_i$ is the pseudorapidity of cell $i$  and the sum is calculated within &  \\
 & a window of $3 \times 5$ cells &  \\
\cline{2-3}
& Ratio of the energy in 3 $\times$ 3 cells to the energy in 3 $\times$ 7 cells centred at the  & \rphi \\
& electron cluster position & \\
\cline{2-3}
& Ratio of the energy in 3 $\times$ 7 cells to the energy in 7 $\times$ 7 cells centred at the & \reta \\
& electron cluster position &  \\
\hline
Strip layer of       & Shower width, $\sqrt{(\Sigma E_i (i-i_\mathrm{max})^2)/(\Sigma E_i)}$, where $i$ runs over all strips in a window  &  \wstot  \\  
EM calorimeter       & of $\Delta\eta \times \Delta\phi \approx 0.0625 \times 0.2$, corresponding  typically to 20 strips in $\eta$, and   &                   \\
		     & $i_\mathrm{max}$ is the index of the highest-energy strip        &                   \\
\cline{2-3}
                     & Ratio of the energy difference between the maximum energy deposit and the energy deposit          &   \deltaEmax   \\
                     &  in a secondary maximum in the cluster to the sum of these energies         &    \\
\cline{2-3}     
& Ratio of the energy in the strip layer to the total energy in the EM accordion & \fI \\
& calorimeter &  \\
\hline
Track quality        & Number of hits in the b-layer (discriminates against photon conversions)     &   $n_\mathrm{Blayer}$ \\
\cline{2-3}
                     & Number of hits in the pixel detector        &    $n_\mathrm{Pixel}$ \\
\cline{2-3}
                     & Total number of hits in the pixel and SCT   detectors  &   $n_{\mathrm{Si}}$    \\
\cline{2-3}
                     & Transverse impact parameter
		                                                  &       \trackdO  \\
\cline{2-3}
                     & Significance of transverse impact parameter defined as the ratio  of the magnitude of \trackdO\   &       \dOSig   \\
                     & to its uncertainty                     &                \\
\cline{2-3}
                     &  Momentum lost by the track between the perigee and the last &   \deltapoverp  \\
                     & measurement point divided by the original momentum & \\
\hline
TRT                 & Total number of hits in the TRT      & $n_\mathrm{TRT}$          \\
\cline{2-3}
                    & Ratio of the number of high-threshold hits to the    total number of  hits in the TRT &    \TRTHighTHitsRatio  \\
\hline
Track--cluster     & $\Delta\eta$ between the cluster position in the strip layer  and the extrapolated track &   \deltaeta \\
\cline{2-3}
  matching    & $\Delta\phi$ between the cluster position in the middle layer and the extrapolated track & \deltaphi\\
\cline{2-3}
&   Defined as  \deltaphi, but the track momentum is rescaled to the cluster energy &  \deltaphires  \\
&   before extrapolating the track to the middle layer of the calorimeter  &  \\
\cline{2-3}
                    & Ratio of the cluster energy to the track momentum            &       $E/p$    \\  
\hline
Conversions         & Veto electron candidates matched to reconstructed photon  conversions            &  isConv \\
\hline
\end{tabular}
\end{center}
\end{table*}

\begin{table*}
  \caption{The variables used in the different selections of the electron identification menu.  
  }
  
\label{tab:IDcutsUsage}
\footnotesize
\begin{center}
\begin{tabular}{|l|l|l|l|l|l|l|l|}
\hline
  & \multicolumn{4}{c|}{Cut-based} &  \multicolumn{3}{c|}{Likelihood} \\
 \hline
Name & \loose & \medium & \tight & \multilepton & \looseLLH & \mediumLLH & \veryTightLLH \\
\hline
$R_\mathrm{Had(1)}$ &  \checkmark &  \checkmark &  \checkmark &  \checkmark &  \checkmark &  \checkmark &  \checkmark \\
\hline
\fIII &  &  \checkmark &  \checkmark & \checkmark & \checkmark &  \checkmark & \checkmark\\
\hline
\weta & \checkmark &  \checkmark &  \checkmark &  \checkmark &  \checkmark &  \checkmark &  \checkmark \\
\reta & \checkmark &  \checkmark &  \checkmark &  \checkmark &  \checkmark &  \checkmark &  \checkmark \\
\rphi & & & & & \checkmark &  \checkmark &  \checkmark \\
\hline
\wstot & \checkmark &  \checkmark &  \checkmark &  \checkmark &   &   &   \\
\deltaEmax & \checkmark &  \checkmark &  \checkmark &  \checkmark &  \checkmark &  \checkmark &  \checkmark \\
\fI  & & & & & \checkmark &  \checkmark &  \checkmark \\
\hline
$n_\mathrm{Blayer}$ &  &  \checkmark &  \checkmark & \checkmark &  \checkmark &  \checkmark &  \checkmark \\
$n_\mathrm{Pixel}$  & \checkmark &  \checkmark &  \checkmark &  \checkmark &  \checkmark &  \checkmark &  \checkmark \\
$n_\mathrm{Si}$  & \checkmark &  \checkmark &  \checkmark &  \checkmark &  \checkmark &  \checkmark &  \checkmark \\
\trackdO  &  &  \checkmark &  \checkmark &  &  &  \checkmark & \checkmark\\
\dOSig & & & & & & \checkmark &  \checkmark \\
\deltapoverp & & &  &  \checkmark &  \checkmark &  \checkmark &  \checkmark \\
\hline
$n_\mathrm{TRT}$ &  &  \checkmark &  \checkmark & \checkmark &   &   &   \\
\TRTHighTHitsRatio &  &  \checkmark &  \checkmark & \checkmark &  \checkmark &  \checkmark &  \checkmark \\
\hline
\deltaeta  & \checkmark &  \checkmark &  \checkmark &  \checkmark &  \checkmark &  \checkmark &  \checkmark \\
\deltaphi & & &  \checkmark &  &  &  & \\
\deltaphires & & &  &  \checkmark &  \checkmark &  \checkmark&  \checkmark  \\
$E/p$ &  &  &  \checkmark &  &  &  & \\
\hline
isConv & & & \checkmark &  &  &  & \checkmark\\
\hline
\end{tabular}
\end{center}
\end{table*}

\subsection{Cut-based identification} 
\label{sec:cutbasedID} 
The cut-based selections, \loose, \medium, \tight\ and \multilepton, are optimized in 10 bins in $|\eta|$ and 11 bins in \et. This binning allows the identification to take into account the variation of the electrons' characteristics due to e.g.\ the dependence of the
shower shapes on the amount of passive material traversed before entering the EM calorimeter. Shower shapes and track properties also change with the energy of the particle.
The electrons selected with \tight\ are a subset of the electrons selected with \medium, which in turn are a subset of \loose\ electrons.
With increasing tightness, more variables are added and requirements are tightened on the variables already used in the looser selections.

Due to its simplicity, the cut-based electron identification \cite{Atlas:ElectronPubNote,Aad:2010Paper,Aad:2011Paper}, which is based on
sequential requirements on selected variables, has been used by the ATLAS Collaboration for identifying electrons since the
beginning of data-taking.  In 2011, for $\sqrt{s} = 7$ \tev\ collisions, its performance (defined in terms of efficiency and background rejection)
 was improved by loosening requirements and introducing
additional variables, especially in the looser selections \cite{Aad:2011Paper}.  In 2012, for $\sqrt{s} = 8$ \tev\ collisions,  due to higher
instantaneous luminosities provided by the LHC, the number of overlapping collisions (pile-up) and therefore the number of particles in
an event\footnote{Here an ``event'' refers to a triggered bunch crossing with all its hard and soft $pp$ interactions, as recorded by the detector.} increased. Due to the higher energy density per event, the shower shapes, even of isolated electrons, tend to look more background-like. In order to cope with this, requirements
were loosened on the variables  most sensitive to pile-up ($R_\mathrm{Had(1)}$ and \reta) and tightened on others to keep the performance
(efficiency/background rejection) roughly constant as a function of the number of reconstructed primary vertices. A requirement on \fIII\ was added in 2012, as well. Furthermore, a new selection was added, called \multilepton, which is optimized for the low-energy electrons in the $H \rightarrow ZZ^* \rightarrow 4\ell$ ($\ell = e, \mu$) analysis. For these electrons, \multilepton\ has a similar efficiency to the \loose\
selection, but provides a better background rejection. In comparison to \loose, requirements on the shower shapes are loosened and
more variables are added, including those sensitive to bremsstrahlung effects.

\subsection{Likelihood identification}

MVA techniques are powerful, since they allow the combined evaluation of several 
properties when making a selection decision. Out of the different MVA techniques, the LH
was chosen for electron identification because of its simple construction.

The electron LH makes use of signal and background probability density functions (pdfs) of the discriminating variables. 
Based on these pdfs, which are treated as uncorrelated, an overall probability is calculated for the object to be signal or background. 
The signal and background probabilities for a given electron candidate are combined into a discriminant $d_{\mathcal L}$:

\begin{equation}
d_{\mathcal L} = {\mathcal L_S \over \mathcal L_S + \mathcal L_B },\phantom{xxxxx}
\mathcal L_{S(B)}(\vec{x}) = \prod\limits_{i=1}^{n} P_{S(B),i}(x_i)
\end{equation}

where $\vec{x}$ is the vector of variable values and $P_{S,i}(x_i)$ is the value of the signal probability density function 
of the $i^\text{th}$ variable evaluated at $x_i$. In the same way, $P_{B,i}(x_i)$ refers to the background probability density function. 

Signal and background pdfs used for the electron LH identification are obtained from data.  As in the
\multilepton\ cut-based selection, variables sensitive to bremsstrahlung effects are included. 

Furthermore, additional variables with significant discriminating power but also a large overlap between signal and background that prevents explicit
requirements (like \rphi\ and \fI) are included.
The variables counting the hits on the track are not used as pdfs in the
LH, but are left as simple requirements, as every electron should have a high-quality track to allow a robust momentum 
measurement.

The \looseLLH, \mediumLLH, 
and \veryTightLLH\ selections are designed to
roughly match the electron efficiencies of the \multilepton , \medium\ and \tight\ cut-based selections, 
but to have better rejection of light-flavour jets and conversions.\footnote{Another selection, \TightLLH, was originally also developed with the background rejection matching the background rejection of the \tight\ cut-based selection, but it was never used. Therefore, for the LHC Run 2, \veryTightLLH\ was renamed to \TightLLH.
}

Each LH selection places a requirement on a LH discriminant, made with a different set of variables. The \looseLLH\ features variables most useful for discrimination against light-flavour jets (in addition, a requirement on $n_\mathrm{Blayer}$ is applied to reject conversions). 
In the \mediumLLH\ and \veryTightLLH\ regimes, additional variables (\trackdO, isConv) are added
for further rejection of heavy-flavour jets and conversions. 
Although different variables are used for the different selections, a sample of electrons selected using a tighter LH is 
a subset of the electron samples selected using the looser LH to a very good approximation.

The LH for each selection consists of 9 $\times$ 6 sets of pdfs, divided into 9 $|\eta|$ bins and 6 \et\ bins. This binning is similar to, but coarser than, the binning used for the cut-based selections. It is chosen to balance the available number of events with the variation of the pdf shapes in \et\ and $|\eta|$. 

\subsection{Electron isolation}
\label{sec:identification_iso}
In order to further reject hadronic jets misidentified as electrons, most analyses require electrons to pass some isolation requirement in addition to the identification requirements described above. 
The two main isolation variables are:
\begin{itemize}
\item Calorimeter-based isolation:\\  
The calorimetric isolation variable $E_\mathrm{T}^{\mathrm{cone} \Delta R}$ is defined as the sum of the transverse energy deposited in the calorimeter cells in a cone of size $\Delta R $ around the electron,
excluding the contribution within $\Delta\eta \times \Delta\phi = 0.125 \times 0.175$ around the electron cluster barycentre.
It is corrected for energy leakage from the electron shower into the isolation cone and for the effect of pile-up using a 
correction parameterized as a function of the number of reconstructed primary vertices.
\item Track-based isolation:\\
The track isolation variable $p_\mathrm{T}^{\mathrm{cone} \Delta R}$ is the scalar sum of the transverse momentum of the tracks with \pt~$>$~0.4~\GeV\ in a cone of $\Delta R $ around the electron,
excluding the track of the electron itself. The tracks considered in the sum must originate from the primary vertex associated with the electron track
and be of good quality; i.e. they
must have at least nine silicon hits, one of which must be in the innermost pixel layer.
\end{itemize}

Both types of isolation are used in the \tnp\ measurements, mainly in order to tighten the selection criteria of the tag. 
Whenever isolation is applied to the probe electron candidate in this work (this only happens in the \jpsi\ analysis described in Section~\ref{sec:JpsiMain}), 
the criteria are chosen such that the effect on the measured identification efficiency is estimated to be small. 

\section{Efficiency measurement methodology}
\label{sec:Method}

\subsection{The \tnp\ method}

Measuring the identification and reconstruction efficiency requires a clean and unbiased sample of electrons. The
method of choice is the \tandp\ method, which makes use of the characteristic signatures of \Zee\ and
\Jpsiee\ decays. In both cases, strict selection criteria are applied on one of the two decay electrons, called tag, and the second electron, the probe, is used for the efficiency measurements. Additional
event selection criteria are applied to further reject background. Only events satisfying data-quality criteria, in particular concerning the ID and the calorimeters, are considered.
Furthermore, at least one reconstructed primary vertex with at least three tracks must be present in the event. 
The \tandp\ pairs must also pass requirements on their reconstructed invariant mass.  
In order to not bias the selected probe sample, each valid combination of electron pairs in the event is considered;  
an electron can be the tag in one pair and the probe in another. 

The probe samples are contaminated by background objects (for example, hadrons misidentified as electrons, electrons from semileptonic heavy
flavour decays or from photon conversions). This contamination is estimated using either background template shapes or
combined fits of background and signal analytical models to the data. The number of electrons is independently estimated at the
probe level and at the level where the probe electron candidate satisfies the tested criteria. The efficiency $\epsilon$ is defined as the fraction of
probe electrons satisfying the tested criteria.

The efficiency to detect an electron 
is divided into different components, namely trigger,
reconstruction and identification efficiencies, as well as the efficiency to satisfy additional analysis criteria,
like isolation. The full
efficiency $\epsilon_{\mathrm{total}}$ for a single electron can be written as:

\begin{align}
\begin{split}
\epsilon_{\mathrm{total}} &=
\epsilon_{\mathrm{reconstruction}}\times\epsilon_{\mathrm{identification}}\times\epsilon_{\mathrm{trigger}}\times\epsilon_{\mathrm{additional}} \\
&= \frac{N_\mathrm{reconstruction}}{N_\mathrm{clusters}}\times\frac{N_\mathrm{identification}}{N_\mathrm{reconstruction}}\times\frac{N_\mathrm{trigger}}{N_\mathrm{identification}}\times\frac{N_\mathrm{additional}}{N_\mathrm{trigger}}.
\end{split}
\end{align}

The efficiency components are defined and measured in a specific order to preserve consistency: the reconstruction efficiency, $\epsilon_{\mathrm{reconstruction}}$,
is measured with respect to electron clusters reconstructed in the EM calorimeter $N_\mathrm{clusters}$; the identification efficiency 
$\epsilon_{\mathrm{identification}}$ is determined with respect to reconstructed electrons $N_\mathrm{reconstruction}$. Trigger efficiencies are
calculated for reconstructed electrons satisfying a given identification criterion $N_\mathrm{identification}$. Therefore, for each identification selection a
dedicated set of trigger efficiency measurements is performed.
Additional selection
criteria are often imposed in analyses of collision data, for example on the isolation of electrons (introduced in Section~\ref{sec:identification_iso}). 
Neither trigger nor isolation efficiency measurements are covered here.

The determination of $\epsilon_{\mathrm{identification}}$ and $\epsilon_{\mathrm{reconstruction}}$ is 
described in Sections~\ref{sec:MeasureID} and \ref{sec:MeasureReco}. The efficiencies are measured in data and in simulated \Zee\ and \Jpsiee\ samples. To 
compare the data values with the estimates of the MC simulation, the same requirements are used to select the probe electrons. However, no background needs to be subtracted from the simulated samples; instead, the reconstructed electron track must be matched to an electron trajectory provided by the MC simulation within $\Delta R < 0.2$. 
Matched electrons from converted photons that are radiated 
off an electron originating from a $Z$ or $\Jpsi$ decay are also accepted by the analyses. 
The denominator of the reconstruction efficiency includes electrons that were not properly reconstructed. If electrons in the simulated \Zee\ samples are reconstructed as clusters without a matching track, the $Z$ decay electrons provided by the MC simulation are matched to the reconstructed cluster within $\Delta R < 0.2$. 

\subsubsection{Data-to-MC correction factors}\label{sec:sfs}

The accuracy with which the MC detector simulation models the electron efficiency plays an important
role in cross-section measurements and various searches for new physics. In order to achieve reliable results, the
simulated MC samples need to be corrected to reproduce the measured data efficiencies as closely as possible. 
This is achieved by a multiplicative correction factor defined as the ratio of the efficiency measured in data to that in the simulation.
These data-to-MC correction factors are usually close to unity. 
Deviations come from the mismodelling of tracking properties or shower shapes in the
calorimeters. 

Since the electron efficiencies depend on the transverse energy and pseudorapidity, 
the measurements are performed in two-dimensional bins in (\et, $\eta$). These bins follow the detector geometry and the binning used for optimization and are as narrow as the size of the respective data set allows. Residual effects, due to the finite bin widths and kinematic differences of the
physics processes used in the measurements, are expected to cancel in the data-to-MC efficiency ratio. 
Therefore, the combination of the different efficiency measurements is
carried out using the data-to-MC ratios instead of the efficiencies themselves. The procedure for the combination is described in Section \ref{sec:combi}.

\subsection{Determination of central values and uncertainties}\label{sec:SysUncert}

For the evaluation of the results of the measurements and their uncertainties using a given final state (\Zee, \Zeegamma\ or \Jpsiee), the following approach was chosen. The details of the efficiency measurement methods are varied in order to estimate the impact of the analysis choices and potential imperfections in the background modelling. Examples of these variations are the selection of the tag electron or the background estimation method. For the measurement of the data-to-MC correction factors, the same variations of the selection are applied consistently in data and MC simulation. Uncertainties due to charge misidentification of the \tandp\ pairs are neglected. 

The final result (the central value) of a given efficiency measurement using one of the \Zee, \Zeegamma\ or \Jpsiee\ processes is taken 
to be the average value of the results from all variations (including the use of different background subtraction methods, e.g. \Ziso\ and \Zmass\ for the \Zee\ final state as described in Sections~\ref{sec:Zmass_bkgsub} and \ref{sec:Ziso_bkgsub}).

The systematic uncertainty is estimated to be equal to the root mean square (RMS) of the measurements  with the intention of modelling a 68\% confidence interval. However, in many bins the RMS does not cover at least 68\% of all the
variations, so an empirical factor of 1.2 is applied to the determined uncertainty in all bins.

The statistical uncertainty is taken to be the average of the statistical uncertainties 
over all investigated variations of the measurement. 
The statistical uncertainty in a single variation of the measurement is calculated following the approach in
Ref.~\cite{cdf7168_eff_uncertainties}. 

\section{Data and Monte Carlo samples}
\label{sec:Samples}

The results in this paper are based on 8~\TeV\ LHC $pp$ collision data collected
 with the ATLAS detector in 2012. 
After requiring good data quality, in particular concerning the
ID and the EM and hadronic calorimeters,
the integrated luminosity used for the measurements 
is 20.3~fb$^{-1}$. 

The measurements are compared to predictions  
from MC simulation.
The \Zee\ and \Zeegamma\ MC samples are generated with the \texttt{POWHEG-BOX}~\cite{Powheg1,Powheg2,Powheg3} generator interfaced to \texttt{PYTHIA8}~\cite{Pythia8}, using the \texttt{CT10} NLO PDF set \cite{Lai:2010vv} for the hard process, the \texttt{CTEQ6L1} PDF set \cite{Pumplin:2002vw}  and a set of tuned parameters called the \texttt{AU2CT10} showering tune~\cite{ATLAS:2012uec} for the parton shower.
The \Jpsiee\ events are simulated using \texttt{PYTHIA8} both 
for prompt ($pp \rightarrow J/\psi+X$) and for non-prompt ($b\bar{b} \rightarrow J/\psi+X$) production. The \texttt{CTEQ6L1} LO PDF set is used, as well as the \texttt{AU2CTEQ6L1} parameter set for the showering~\cite{ATLAS:2012uec}.
All MC samples are processed through the full ATLAS 
detector simulation~\cite{ATLASGeant4} based on \texttt{GEANT4}~\cite{Geant4}. 

\begin{figure}[!ht]
\centering
\subfigure{
\includegraphics[width=0.48\columnwidth]{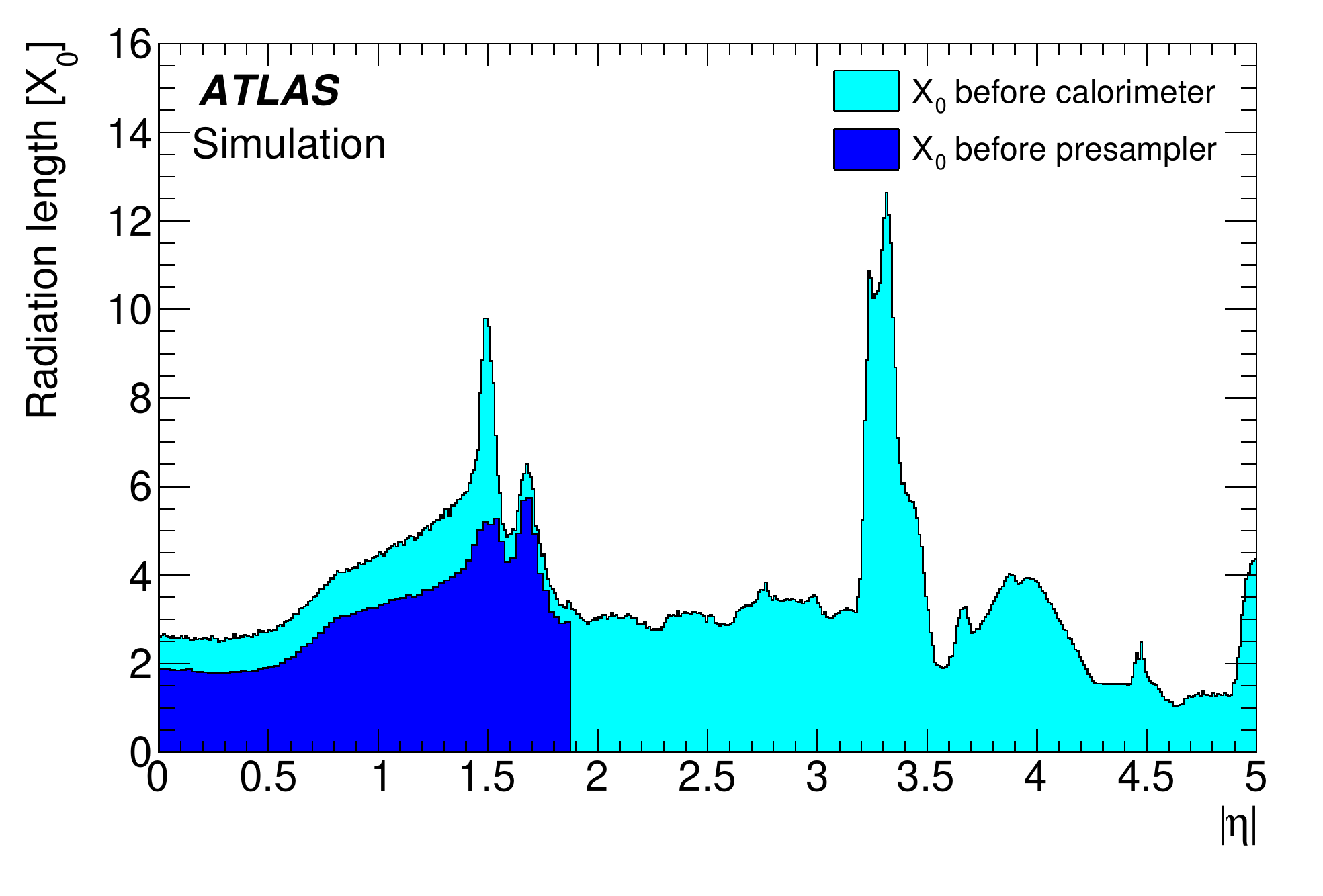} 
}
\subfigure{
\includegraphics[width=0.48\columnwidth]{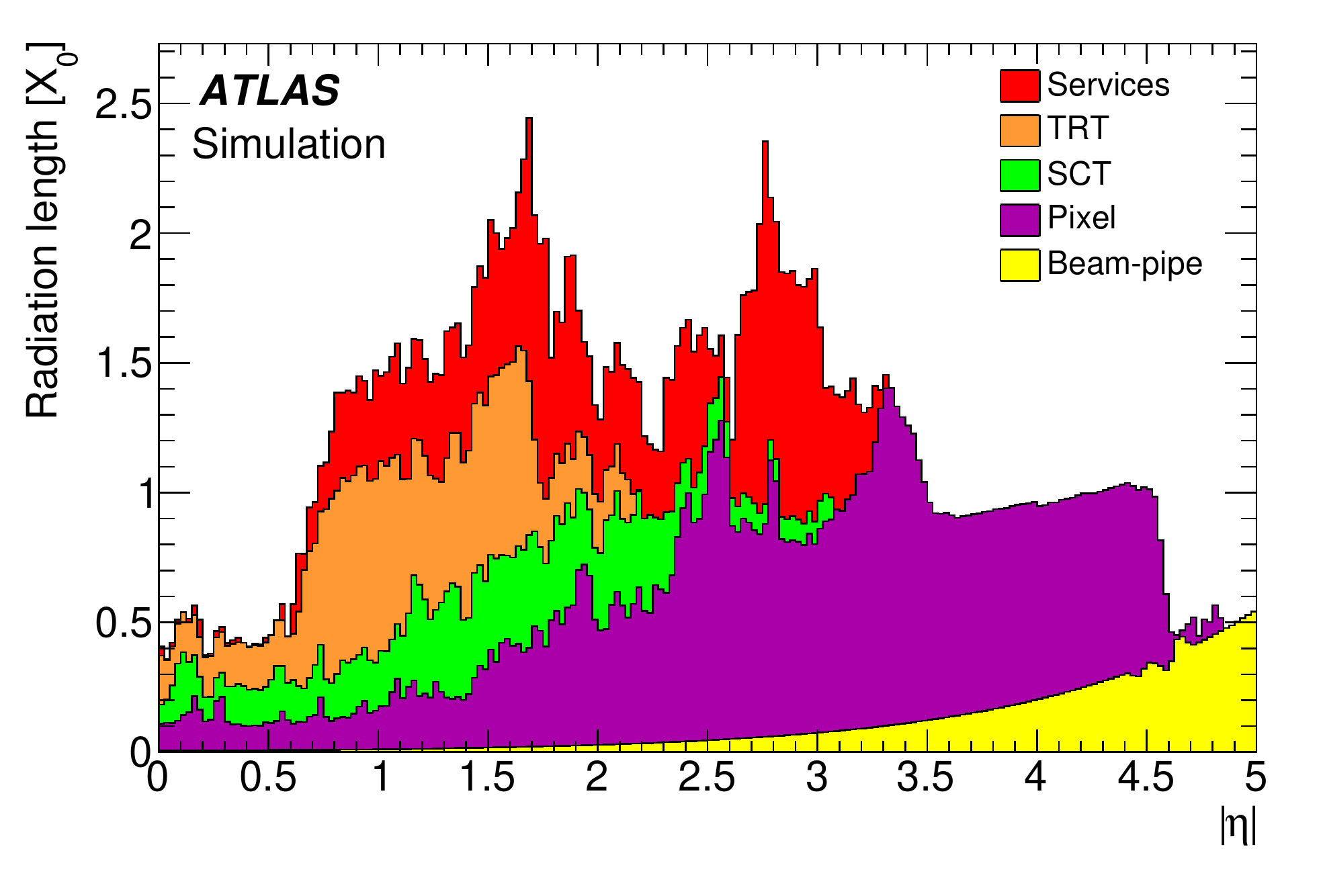} 
}
\caption{
Amount of material, in units of radiation length
$X_{0}$,  traversed by a particle as a function of $|\eta|$: material in
front of the presampler detector and the EM accordion calorimeter (left), and material
up to the ID boundaries (right). The contributions of the different detector elements, 
including the services and thermal enclosures are shown separately by filled
colour areas.  
\label{fig:geometries}}
\end{figure}

The distribution of material in front of the presampler detector and the EM accordion calorimeter as a function of $|\eta|$  is shown in the left plot of Figure~\ref{fig:geometries}. 
The contributions of the different detector elements up to the 
ID boundaries, including services and thermal enclosures,
are detailed on the right. These material distributions are used as input to the MC simulation.

The peak at $|\eta| \approx 1.5$ in the left plot of Figure~\ref{fig:geometries}, corresponding to the transition region between the barrel
and endcap EM accordion calorimeters, is due to the cryostats, the corner of the barrel
EM accordion calorimeter, the ID services and parts of the scintillator-tile hadronic calorimeter.
The sudden increase of material at $|\eta| \approx 3.2$, corresponding to the
transition between the endcap calorimeters and the forward calorimeter, is mostly due to the
cryostat that acts also as a support structure. 

\begin{figure}
\centering
\includegraphics[width=0.49\textwidth]{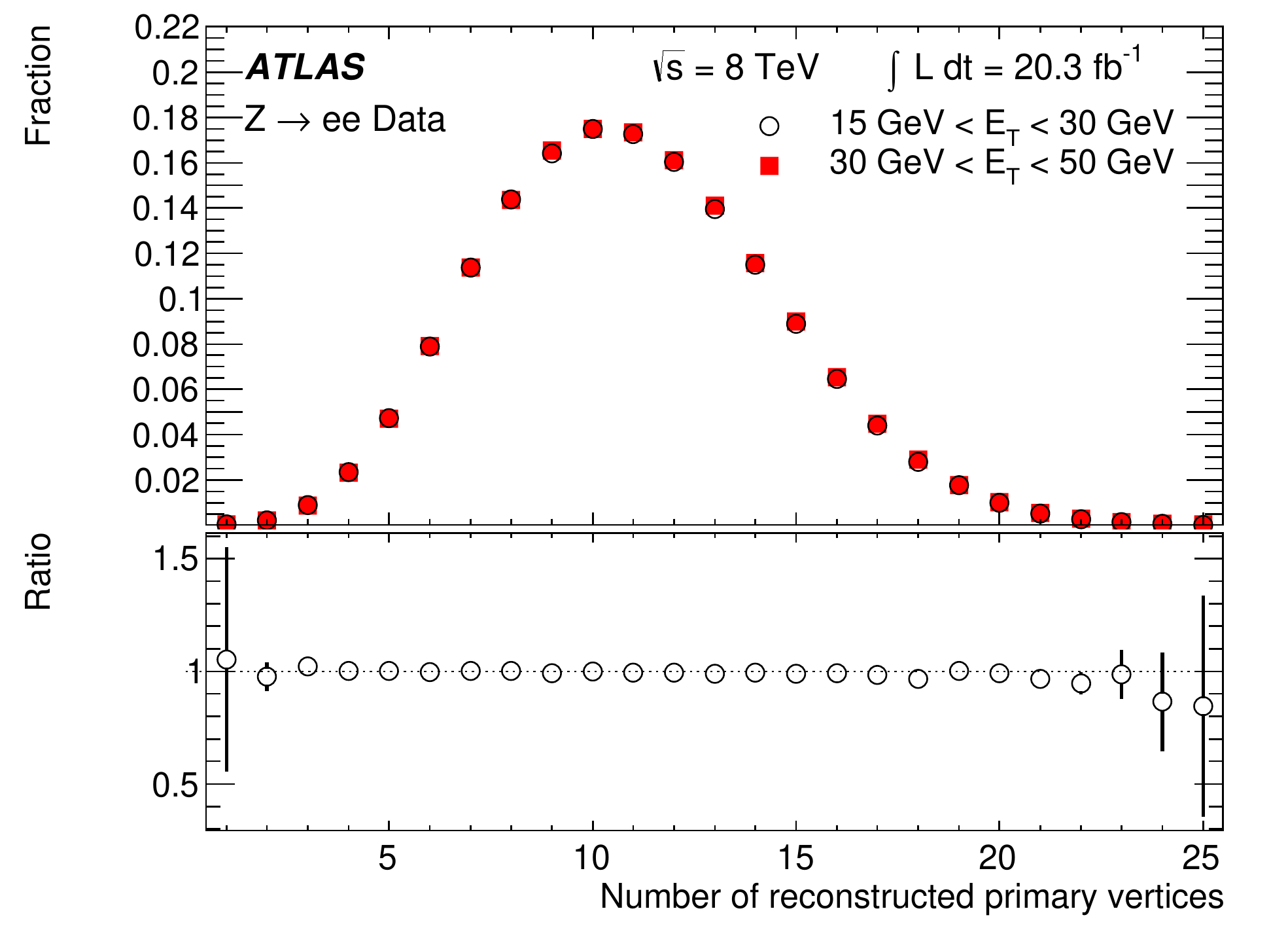}
\caption{ Number of reconstructed primary vertices in events with an electron cluster candidate with 15~\GeV~$<$~\et~$<$~30~\GeV\ (open circles) and  30~\GeV~$<$~\et~$<$~50~\GeV\ (filled squares)
in the \Zee\ data set used for the reconstruction efficiency measurement
described in Section~\ref{sec:MeasureReco}.}
\label{fig:Samples_pileup}
\end{figure}

The simulation also includes realistic modelling (tuned to the data) 
of the event pile-up from the same, previous, and subsequent bunch crossings. 
The energies of the electron candidates in simulation are smeared to match the resolution in data
and the simulated MC events are weighted to reproduce the distributions of the primary-vertex $z$-position 
and the number of vertices in data, the latter being a good indicator of pile-up. 
Figure~\ref{fig:Samples_pileup} shows the distribution of the number of primary collision vertices in events with an identified electron and an electron cluster candidate 
(with 15~\GeV~$<$~\et~$<$~30~\GeV\ and 30~\GeV~$<$~\et~$<$~50~\GeV) in the \Zee\ data set used for the reconstruction 
efficiency measurement described in Section~\ref{sec:MeasureReco}. 
The distribution does not depend on the transverse energy of the cluster of the probe electron candidate.

\section{Identification efficiency measurement}
\label{sec:MeasureID}

The efficiencies of the identification criteria (\loose, \medium, \tight, \multilepton\ and \looseLLH,
\mediumLLH, \veryTightLLH) are determined in data and in the simulated samples with respect to reconstructed electrons with associated tracks that have
at least one hit in the pixel detector and at least seven total hits in the pixel and SCT detectors (this requirement is referred to as ``track quality'' below). The
efficiencies are calculated as the ratio of the number of electrons passing a certain identification selection (numerator) to the number of electrons with a matching track satisfying the track quality requirements (denominator). 

\begin{figure}[!ht]
\centering
\includegraphics[width=0.48\textwidth]{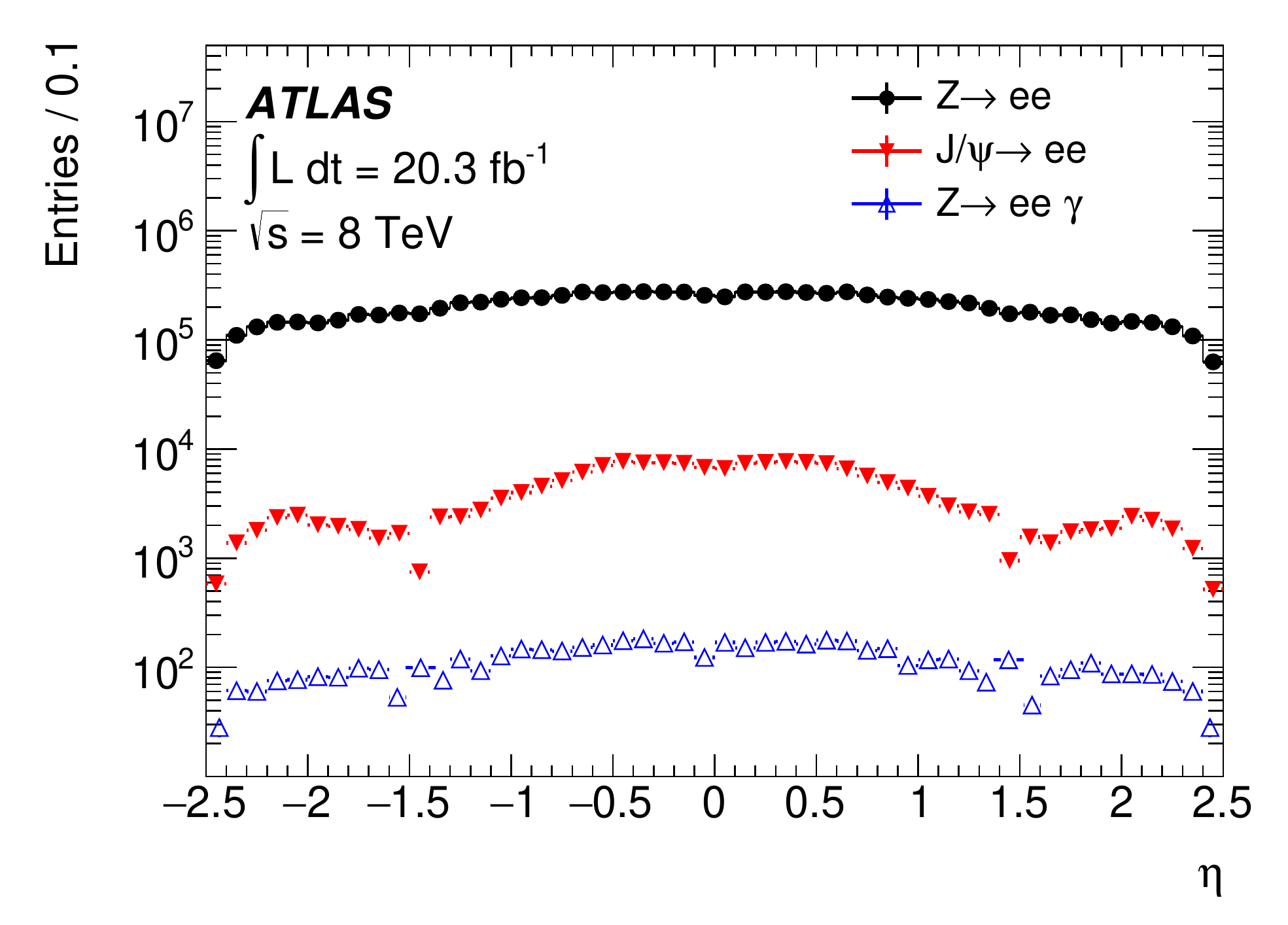}
\includegraphics[width=0.48\textwidth]{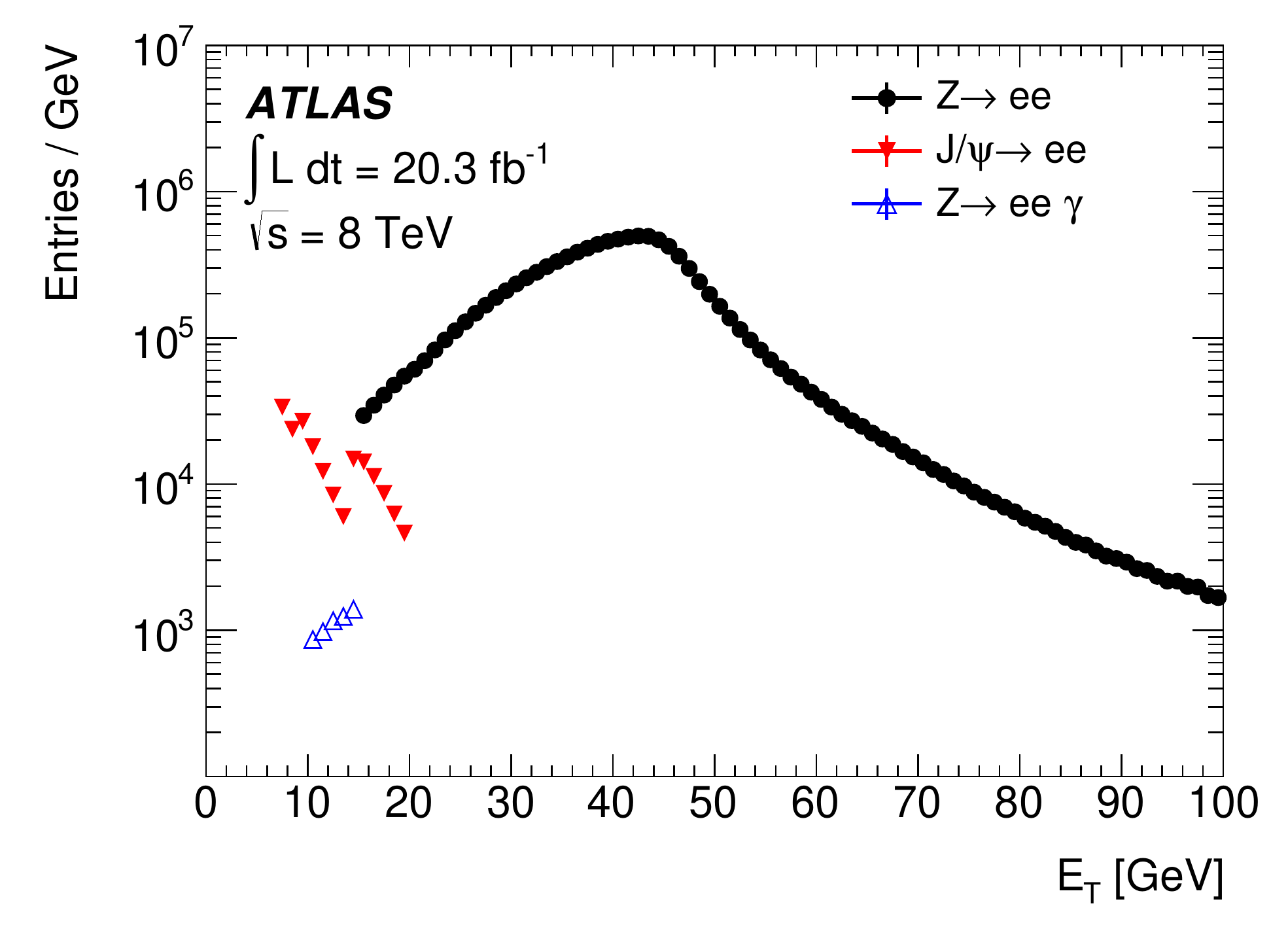}
\caption{Pseudorapidity and transverse energy distributions of probe electron candidates satisfying the \tight\ identification criterion in the \Zee\ (full circles), 
\Zeegamma\ (empty triangles) and the \Jpsiee\ (full triangles) samples. The \et\ distribution of probe electron candidates from \Jpsiee\ is discontinuous, as the sample is selected by a number of triggers with different \et\ thresholds.}
\label{fig:stats}
\end{figure}

For the identification efficiencies determined in this paper, three different decays of resonances are used, and combined in the overlapping regions as described in Section~\ref{sec:combi}: radiative decays of the $Z$ boson, \Zeegamma, for electrons with 10~\GeV~$<$~\et~$<$~15~\GeV, \Zee\ for electrons with \et~$>$~15~\GeV\ and \Jpsiee\ for
electrons with 7~\GeV~$<$~\et~$<$~20~\GeV. The distributions of the probe electron candidates passing the \tight\ identification selection are depicted in Figure~\ref{fig:stats} as a function of \eta\ (left) and \et\ (right), giving an indication of the number of events available for each of the measurements in the respective \eta\ and \et\ bin. The \et\ spectrum of probe electron candidates from \Jpsiee\ is discontinuous, as the sample is selected by a number of triggers with different \et\ thresholds
as discussed in Section~\ref{sec:JpsiMain}.

\subsection{\Tnp\ with \Zee\ events}

\Zee$(\gamma)$ decays are used to measure the identification efficiency for electrons with \et~$>$~10~\GeV. The \tnp\ method using \Zee\ decays
provides a clean sample of electrons, especially when the probe electron candidates have \et~$>$~25~\GeV.  For lower transverse
energies, background subtraction becomes important. Two different distributions are used to discriminate signal
electrons from background: the invariant mass of the \tnp\ pair is used in the \Zmass\ method and the isolation distribution of the probe electron candidate is used in the \Ziso\ method. 

Probe electrons with \et\ between 10~\GeV\ and 15~\GeV\ are selected from \Zee$\gamma$ decays in which an electron has lost much of its energy due to final-state radiation (FSR). At low  \et, this topology has less background than \Zee\ decays. The invariant mass in these cases is computed from three objects: the tag electron, the probe electron and a photon. 

\subsubsection{Event selection}\label{sec:eventsel_ZTPID}

Events are selected using a logical OR between two single-electron triggers, one with an \et\ threshold of 24~\GeV\ requiring medium
identification and one with an
\et\ threshold of 60~\GeV\ and loose identification requirements.\footnote{The electron identification selection in
the trigger is looser than or equivalent to the corresponding analysis requirements.}

Events are required to have at least two reconstructed electron candidates in the central region of the detector, $|\eta|$~$<$~2.47, with opposite charges (see Section~\ref{sec:MisID} for the measurement of the charge misidentification). The tag electron candidate is required to have a transverse energy \et~$>$~25~\GeV, be matched to a trigger electron within $\Delta R$~$<$~0.15 and be outside the transition region between barrel
and endcap of the EM calorimeter, 1.37~$<$~$|\eta|$~$<$~1.52. 
Furthermore, it has to pass the \tight\ identification requirement (\medium\ for \Zeegamma). 
The probe electron candidates must have 
\et~$>$~10~\GeV\ and satisfy the track quality criteria. 
The invariant mass of the tag--probe (tag--probe--photon for \Zeegamma) system is required to be within $\pm$15~\GeV\ of the \Z\ mass. 
About 15.5 million probe electron candidates are selected for further analysis.

For the \Zeegamma\ method, in addition to the tag and the probe electron candidates, a photon is selected passing \tight\ photon identification requirement \cite{photonID2011} and fulfilling \et(probe)~+~\et(photon)~$>$~30~\GeV.
Requirements are placed on the angular distance between the photon and the electron candidates to avoid double counting of objects:
$\Delta R$(tag--photon)~$>$~0.4 and  $\Delta R$(probe--photon)~$>$~0.2. The reason for the asymmetry between tag and probe electron requirements is an isolation requirement with a cone size of 0.4 which is applied to the tag electron as one of the variations for assessing the systematic uncertainties. Furthermore, FSR photons from the probe electron tend to be closer to the probe electron than to the tag electron. Further
requirements are placed on the tag--probe and tag--photon invariant mass to select events with FSR: $m$(tag+photon)~$<$~80~\GeV,
$m$(tag+probe)~$<$~90~\GeV. All possible tag--probe--photon combinations are used.
About 13~000 probe electron candidates with a transverse energy of 10~\GeV~$<$~\et~$<$~15~\GeV\ are selected integrated over the full $|\eta|$~$<$~2.47 range.
\subsubsection{Background estimation and variations for assessing the systematic uncertainties of the \Zmass\ method}\label{sec:Zmass_bkgsub}

The invariant mass of the tag-and-probe pair (and the photon in the case of \Zeegamma) is used as the discriminating variable between
signal electrons and background.

In order to form background templates, reconstructed electron candidates with an associated track, satisfying track quality criteria, are chosen as probes. In addition, identification and isolation requirements are inverted to minimize the contribution of signal electrons. A study was performed on data and simulated samples to test the shape biases of possible background templates due to the inversion of selection requirements 
and contamination from signal electrons, and the least-biased templates were chosen. 
The remaining signal electron contamination in the background templates is estimated using simulated events.

The normalization of the background template is determined by a sideband method:
for the denominator (defined at the beginning of Section~\ref{sec:MeasureID}), the templates are normalized to the invariant-mass distribution above the \Z\ peak (120~\GeV~$<$~\mee~$<$ 250~\GeV\ for \Zee\ and 100~\GeV~$<$~$m_{ee\gamma}$~$<$ 250~\GeV\ for \Zeegamma).

\begin{figure}
\centering
\includegraphics[width=0.45\textwidth]{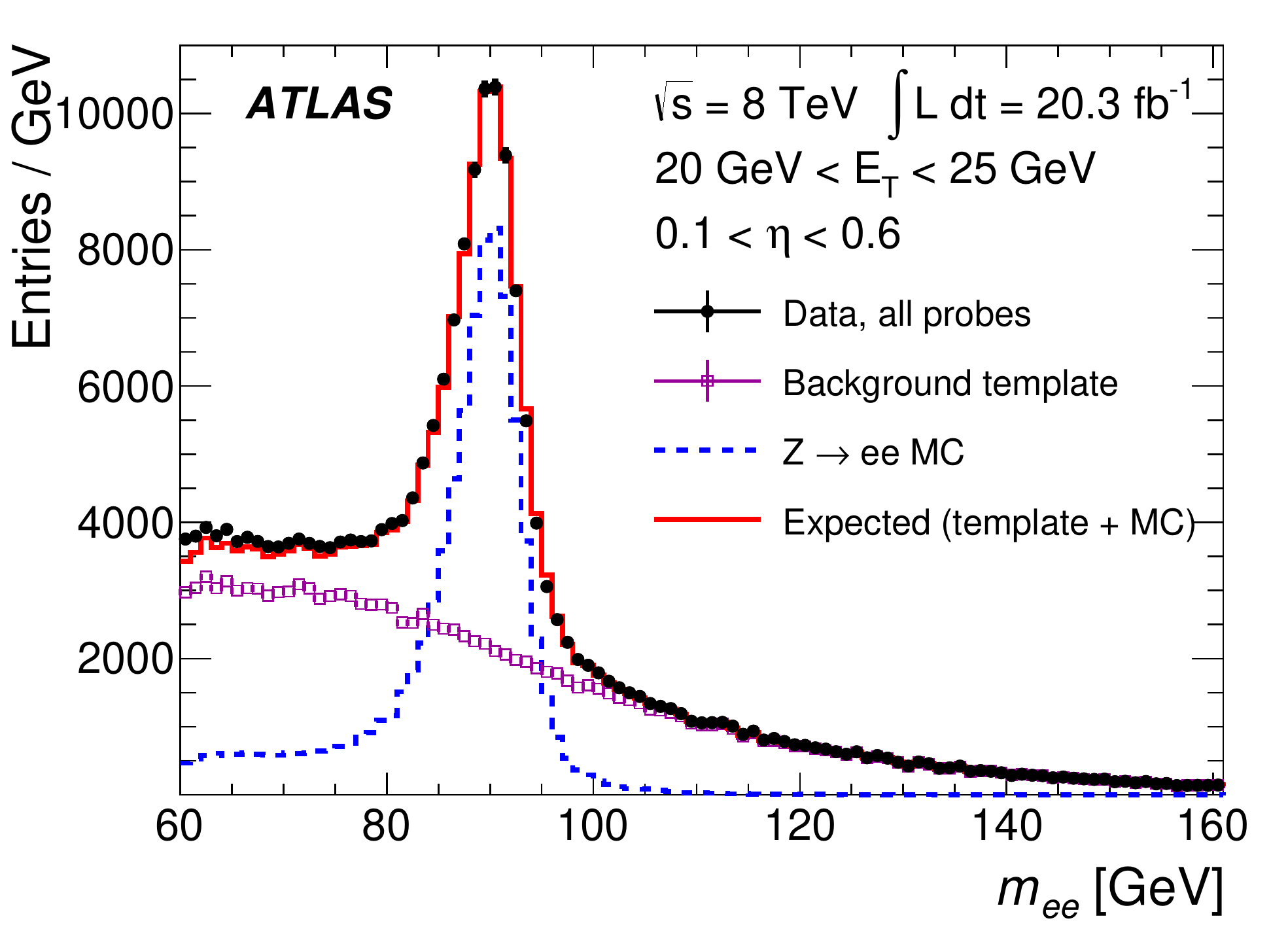}
\includegraphics[width=0.45\textwidth]{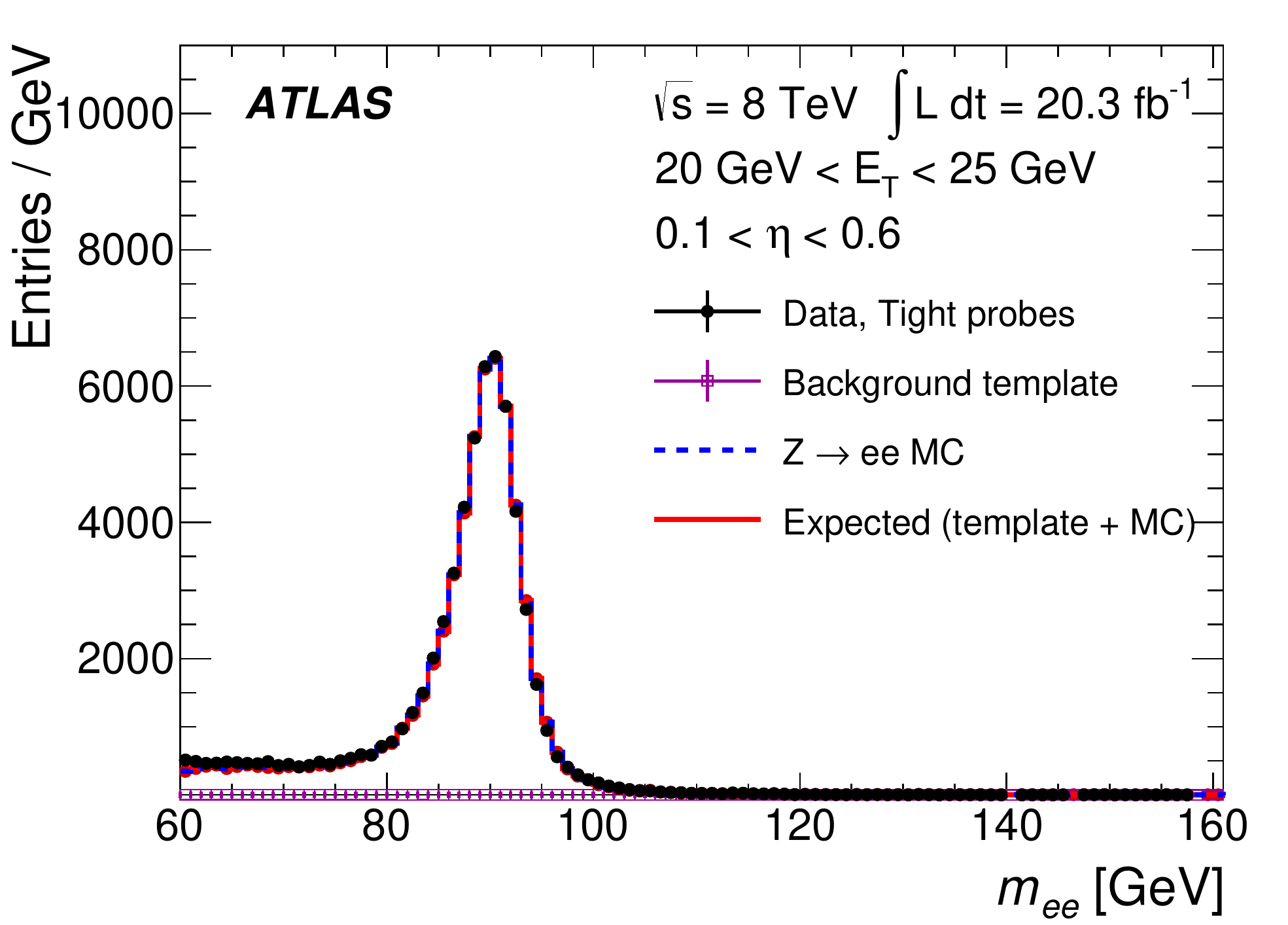}
\caption{  Illustration of the background estimation using the \Zmass\ method in the 20~\GeV~$<$~\et~$<$~25~\GeV, 0.1~$<$~\eta~$<$~0.6~bin, at reconstruction+track-quality level (left) and for probe electron candidates passing the cut-based \tight\ identification (right). The background template is normalized in the range 120~\GeV~$<$~\mee~$<$ 250~\GeV. The tag electron passes cut-based \medium\ and isolation requirements. The signal MC simulation is scaled to match the estimated signal in the \Z-mass window. }
\label{fig:BkgMass}
\end{figure}

\begin{figure}
\centering
\includegraphics[width=0.45\textwidth]{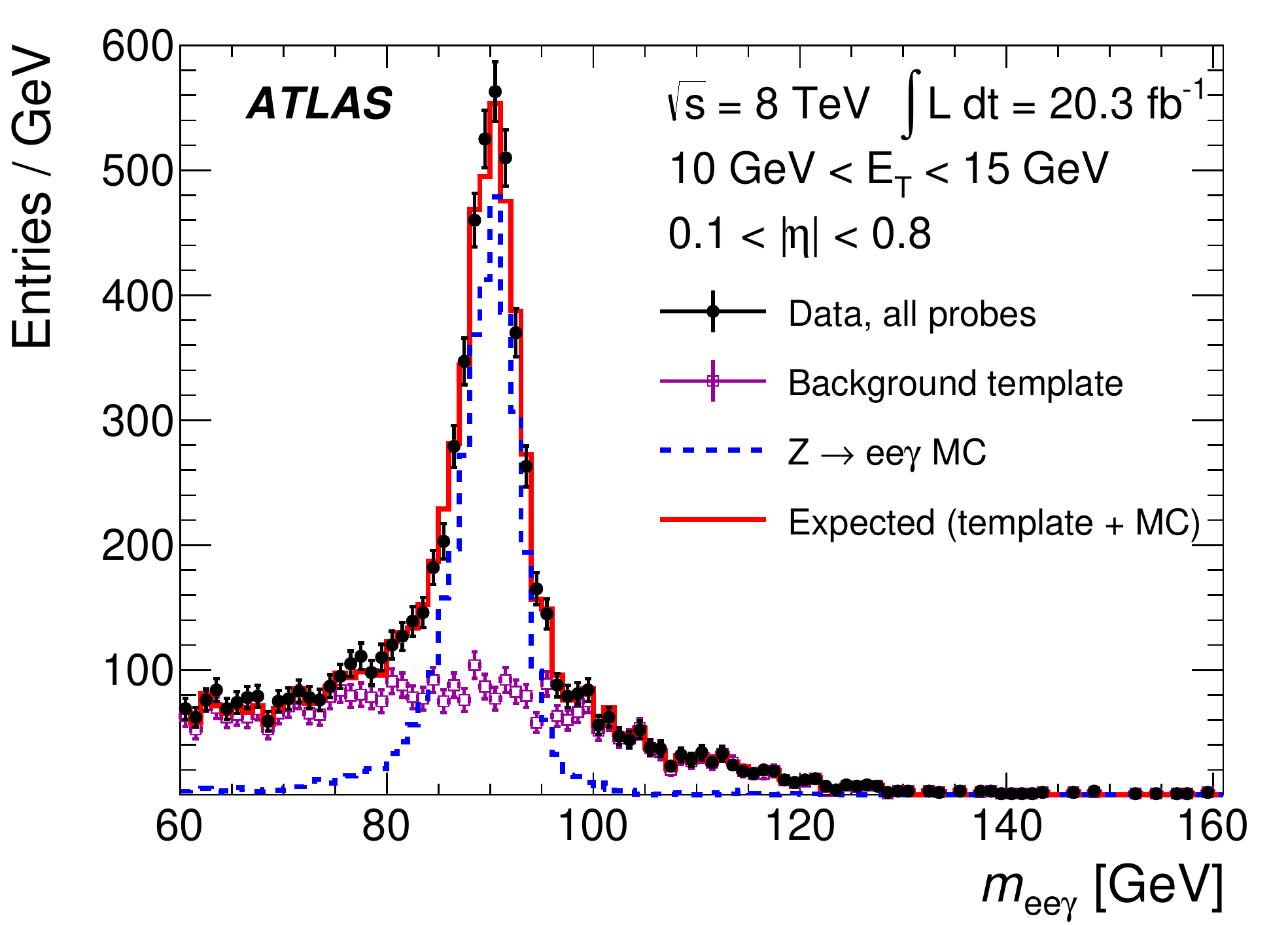}
\includegraphics[width=0.45\textwidth]{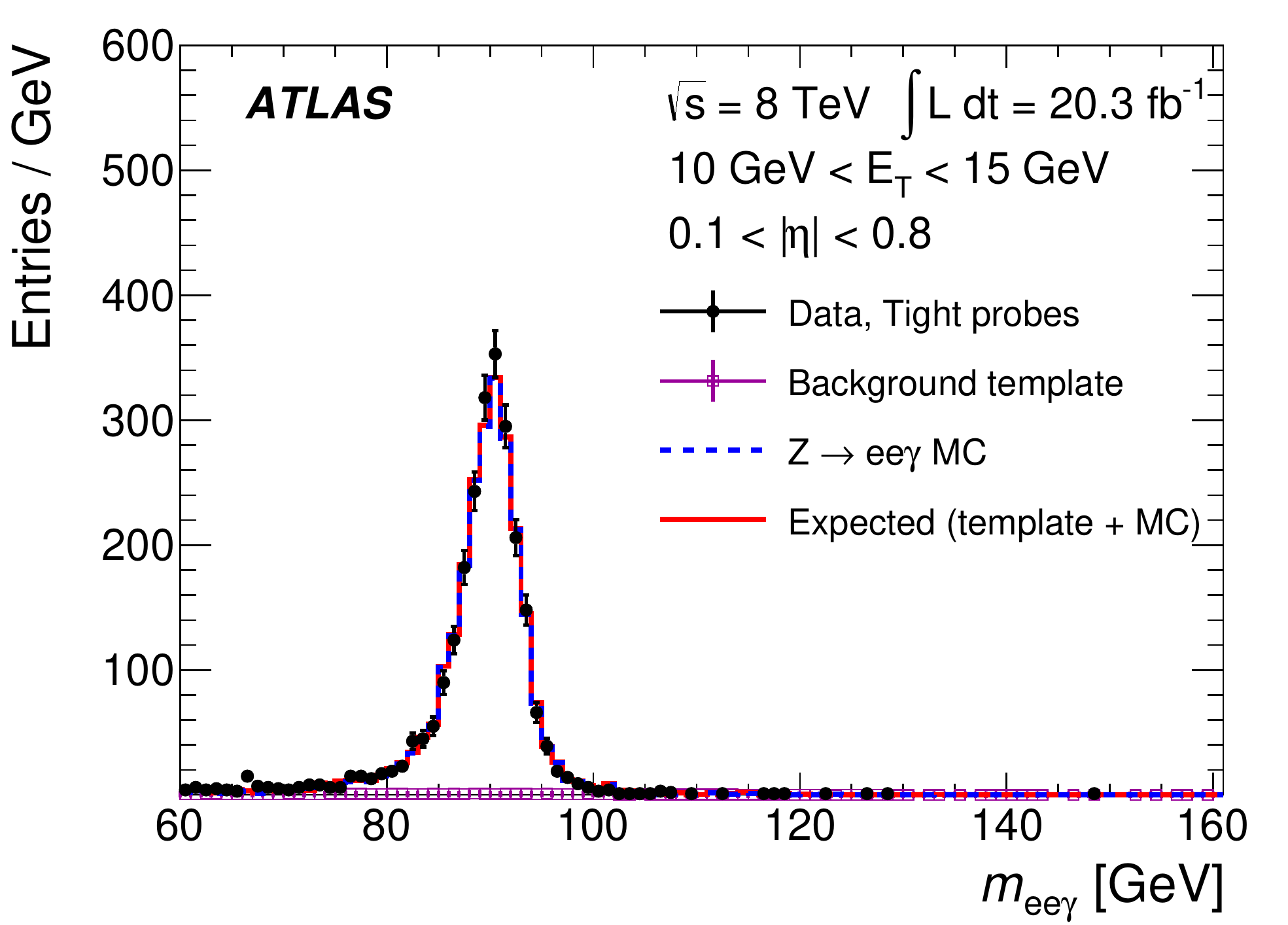}
\caption{  Illustration of the background estimation using the \Zeegamma\ method in the 10~\GeV~$<$~\et~$<$~15~\GeV, 0.1~$<$~$|\eta|$~$<$~0.8~bin, at reconstruction+track-quality level (left) and for probe electron candidates passing the cut-based \tight\ identification (right). The background template is normalized in the range 100~\GeV~$<$~\mee~$<$~250~\GeV. The tag electron passes cut-based \medium\ and isolation requirements. The signal MC simulation is scaled to match the estimated signal in the \Z-mass window. }
\label{fig:BkgZeeGamma}
\end{figure}

Care is taken to remove the small contribution of signal electrons in the tails of the distribution of all probes before normalizing the background template to them. \Tight\ probe electrons and \tight\ data efficiencies are used to perform this subtraction, except for the \tight\ efficiency extraction, for which the MC efficiency is used.
For the numerator, the same templates are used as in the denominator, but they are normalized to the same-sign invariant-mass distribution (all numerator requirements are imposed on the probe). The normalization is done in the same ranges as in the denominator. The same-sign distribution is used as reference because it has less signal contamination than the opposite-sign distribution, an effect that is more important in the numerator. Figure~\ref{fig:BkgMass} shows the \Zee\ tag-and-probe invariant-mass distribution in one example bin for both numerator and denominator, including the normalized background template and the MC \Zee\ prediction. Figure~\ref{fig:BkgZeeGamma} shows the same for the \Zeegamma\ invariant-mass distribution.

In order to assess systematic uncertainties, efficiency measurements based on the following variations of the analysis are considered. The mass window is changed from 15~\GeV\ to 10~\GeV\ and 20~\GeV\ around the \Z\ mass, 
the tag electron requirement is varied by applying a requirement on the calorimetric isolation variable
and, in
the \Zee\ case, by loosening the identification requirement to \medium. Furthermore, for \et~$<$~30~\GeV, two normalization
regions, below  and above the \Z\ peak are used. The normalization range below the peak is 60~\GeV~$<$~\mee~$<$~70~\GeV. For \et~$>$~30~\GeV, the number of events in the low-mass region is too small for a reliable normalization, so instead two different background template selections are considered.
All possible combinations of these variations are produced and taken into account as described in Section~\ref{sec:SysUncert}.

\subsubsection{Background estimation and variations for assessing the systematic uncertainties of the \Ziso\ method}
\label{sec:Ziso_bkgsub}

In this approach, the calorimeter isolation
distribution $E_\mathrm{T}^\mathrm{cone0.3}$ of the probe electron candidates is used as the 
discriminating variable. 

\begin{figure}
\centering
\includegraphics[width=0.45\textwidth]{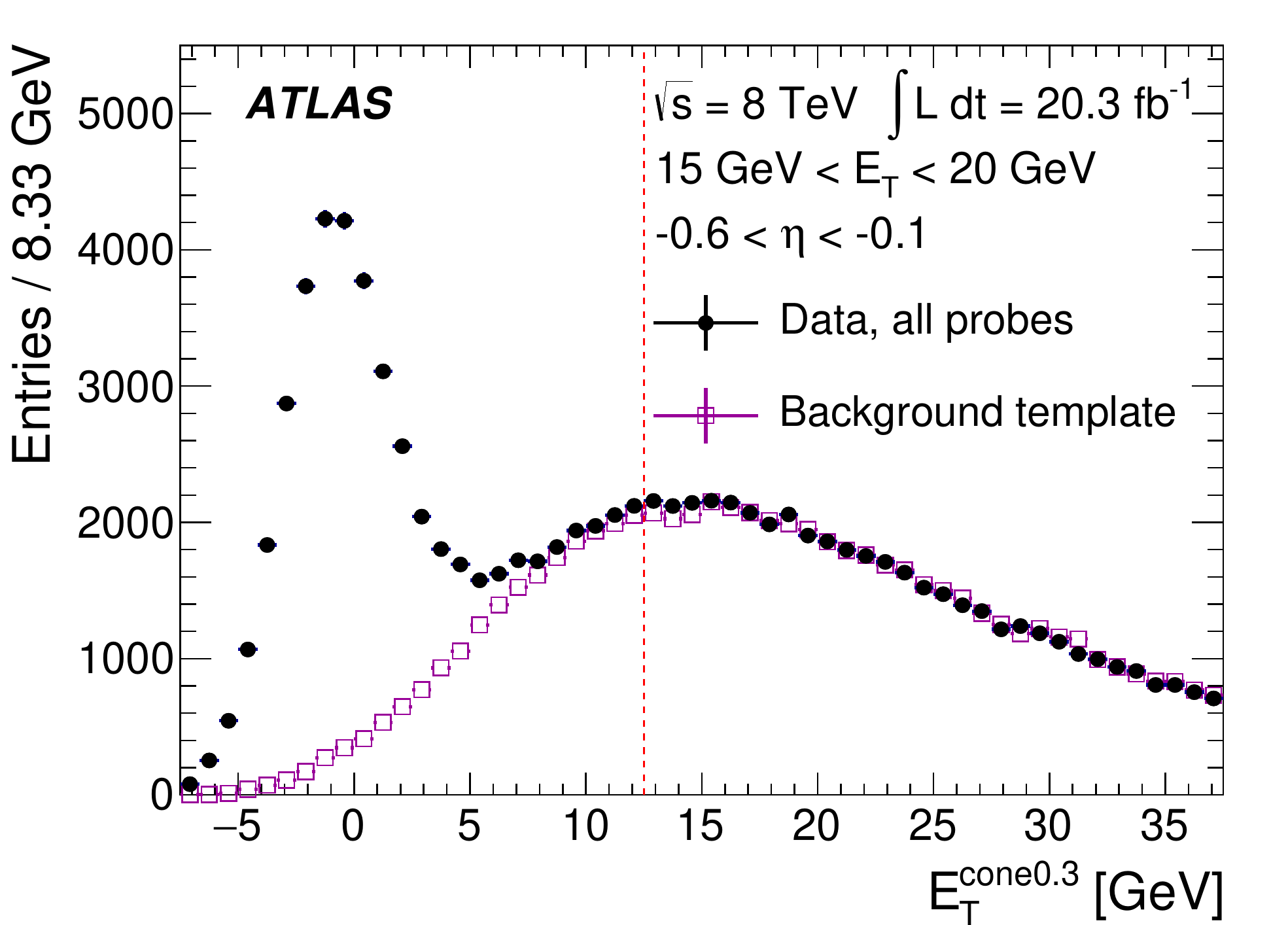}
 \includegraphics[width=0.45\textwidth]{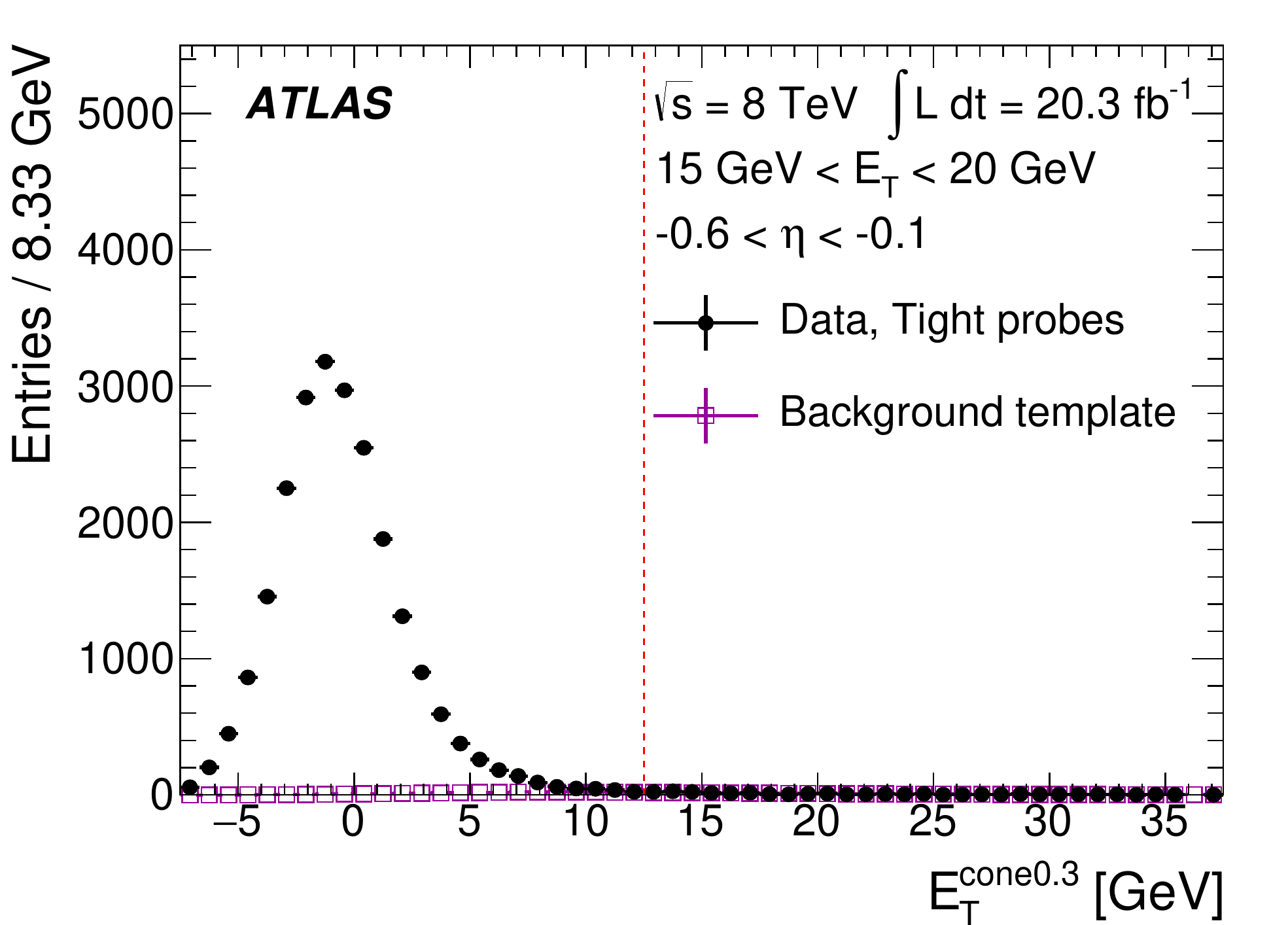}
 \caption{
   Illustration of the background estimation using the \Ziso\ method in the 15~\GeV~$<$~\et~$<$~20~\GeV, $-$0.6~$<$~\eta~$<$~-0.1~bin, at reconstruction+track-quality level (left) and after applying the cut-based \tight\ identification (right). The tag electrons are selected using the cut-based \tight\ identification and a \Z-mass window of 15~\GeV\ is applied. The threshold chosen for the sideband subtraction is $\et^\mathrm{cone0.3}$~$=$~12.5~\GeV.}
\label{fig:BkgIso}
\end{figure}

The background templates are formed as subsets of all probe electron candidates used in the denominator of the identification efficiency calculation. The probes 
for the background template are required to be reconstructed as electrons with a matching track that satisfies track quality criteria; however, they
are required to fail some of the identification requirements, namely the requirements on \wstot\ and \TRTHighTHitsRatio. A study was performed on possible background templates and the bias due to the inversion of selection requirements and contamination from signal electrons. The
least-biased templates were chosen. As illustrated in Figure~\ref{fig:BkgIso}, the background templates are normalized to the
isolation distribution of the probe electron candidates using the background dominated tail region of the isolation distribution.

To assess the systematic uncertainty of the efficiency, the parameters of the measurement are varied.  
The threshold for the sideband subtraction is chosen between $\et^\mathrm{cone0.3}$~$=$~10~\GeV\ and~$=$~15~\GeV. 
As in the \Zmass\ case, the mass window is changed from 15~\GeV\ to 10~\GeV\ and 20~\GeV\ around the \Z\ mass, 
the tag electron requirement is varied by applying a requirement on the calorimetric isolation variable, $E_\mathrm{T}^\mathrm{cone0.4}$~$<$~5~\GeV.

In addition, different identification requirements are inverted to form two alternative templates 
and an alternative probe electron isolation distribution $E_\mathrm{T}^\mathrm{cone0.4}$ with a larger isolation cone size ($\Delta R = 0.4$)  is used as the discriminant.
As in the \Zmass\ case, all possible combinations of these variations are considered.

For the \Zmass\ and \Ziso\ methods together, there are in total 90 variations, which are treated as variations of the same measurement in order to estimate the systematic uncertainty due to the background estimation method.

\subsection{\Tnp\ with \Jpsiee\ events}
\label{sec:JpsiMain}

\Jpsiee\ events are used to measure the electron identification efficiency for 7~\GeV\ $< \et <$ 20~\GeV. At such low energies, the probe  sample suffers from a significant background fraction, 
which can be estimated using the reconstructed dielectron invariant mass
($m_{ee}$) 
of the selected tag-and-probe pairs. Furthermore, the \JPsi\ sample is
composed of two contributions. In prompt production, the \JPsi\ meson is
produced directly in the proton--proton collision via strong interaction or from the decays of directly produced
heavier charmonium states. The electrons from the decay of prompt \JPsi\
particles are expected to be isolated and therefore to have identification
efficiencies close to those of isolated electrons from other physics processes
of interest in the same transverse energy range, such as Higgs boson decays. In
non-prompt production, the \JPsi\ meson originates from $b$-hadron decays and
its decay electrons are expected to be less isolated. 

Experimentally, the two production modes can be distinguished by measuring the
displacement of the \Jpsiee\ vertex with respect to the primary vertex. 
Due to
the long lifetime of $b$-hadrons, electron-pairs from non-prompt \JPsi\ production have a measurably displaced vertex, while prompt decays occur at the
primary vertex. To reduce the dependence on the \JPsi\ transverse momentum,
the variable used in this analysis to discriminate between prompt and
non-prompt production, called pseudo-proper time~\cite{PseudoProperTime}, is defined as

\begin{equation}
\label{eq:tau}
\tau = \frac{ L_{xy} \cdot m^{J/\psi}_{\mathrm{PDG}} }{ p^{J/\psi}_\mathrm{T} }.
\end{equation}

Here, $L_{xy}$ measures the displacement of the $J/\psi$ vertex with respect to\ the primary vertex in the transverse plane, while $m^{J/\psi}_{\mathrm{PDG}}$ and
$p^{J/\psi}_\mathrm{T}$ are the mass~\cite{PDG} 
and the reconstructed transverse
momentum of the $J/\psi$ particle.

Two methods have been developed to measure the electron efficiency using \Jpsiee\
decays. The short-$\tau$ method, already used in Refs.~\cite{Aad:2010Paper,Aad:2011Paper},
considers only events with short pseudo-proper
time, selecting a subsample dominated by prompt $J/\psi$ production. The remaining
non-prompt contamination is estimated using MC simulation and the measurement of the non-prompt fraction in $\Jpsi \rightarrow \mu\mu$
events~\cite{ATLASJpsiXsec}.  The $\tau$-fit method, used in Ref.~\cite{Aad:2011Paper}, utilizes the full $\tau$-range and extracts the non-prompt fraction by fitting the pseudo-proper time
distribution both before and after applying the identification requirements.

\subsubsection{Event selection}

Events are selected by five dedicated \Jpsiee\ triggers. These require tight 
trigger electron identification\footnote{The tight electron identification selection applied in
the \JPsi\ trigger is looser than the corresponding analysis requirements. In particular, no selection is
applied to \deltaphi, $E/p$ and isConv.} and an electron \et\ above a threshold
for one of the two trigger objects, while only requiring an
EM cluster above a certain \et\ threshold for the other. 

Events with at least two electron candidates with $\et>5$~\GeV\
and $|\eta|<2.47$ are considered. 

The tag 
electron candidate must be matched to a tight trigger electron
object within $\Delta R < 0.005$ and satisfy the cut-based \tight\ identification selection. 
To further clean the tag electron sample an isolation criterion is applied in 
most of the analysis variations.
The other
electron candidate, the probe, needs to satisfy the track quality criteria.
It is also required to match an
EM trigger object of the \Jpsiee\ triggers within $\Delta R < 0.005$
and have a transverse energy that is at least 1~\GeV\ higher than the corresponding trigger
threshold. To ensure that the measured efficiency corresponds to well-isolated
electrons an isolation requirement is imposed on the probe electron candidate as well. The isolation criterion has
less than 1\% effect on the identification efficiency in simulated events.    It
is further required that the tag and probe electron candidates are separated by
$\Delta R_\mathrm{tag-probe} > 0.2$ to prevent one electron from affecting the
identification of the other.  The pseudo-proper time of the reconstructed \JPsi\
candidate is restricted to $-$1~ps $< \tau < 3$~ps in the $\tau$-fit method and
typically to  $-$1~ps $< \tau < 0.2$~ps in the short-$\tau$ method. The negative values of the pseudo-proper times are due to the finite resolution of $L_{xy}$.
At this stage no requirement is made on the charge of the electrons 
and all possible tag-and-probe pairs are considered.
About 700~000 probe electron candidates are selected for $\et=7$--$20$~\GeV, of which about 190~000 pass 
the \tight\ selection, within the range of  $-$1~ps $< \tau < 3$~ps and integrated over $|\eta|<2.47$.
 
\subsubsection{Background estimation and variations for assessing the systematic uncertainties}

\begin{figure}[!ht]
\centering

\includegraphics[width=0.49\textwidth]{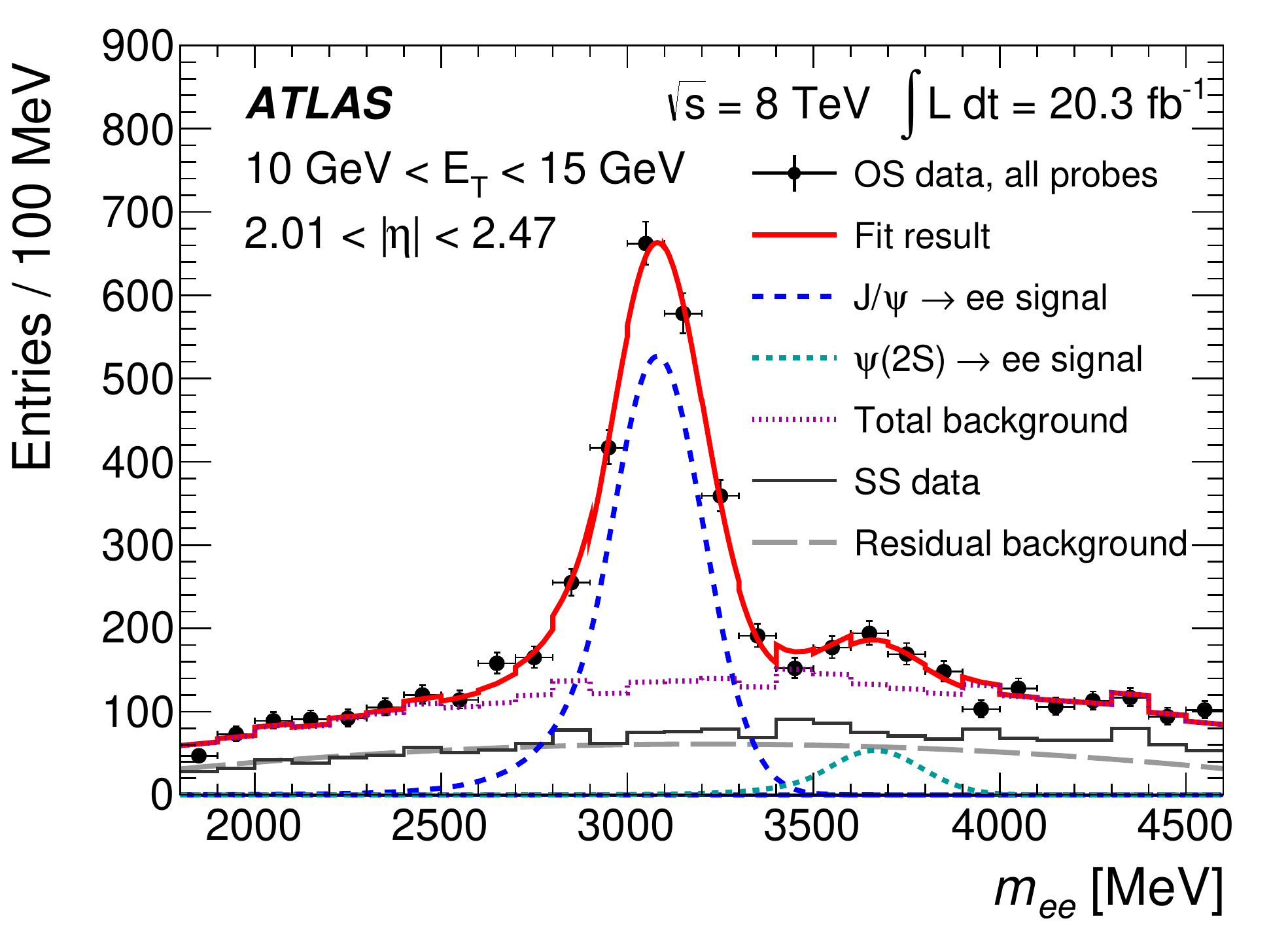}
\includegraphics[width=0.49\textwidth]{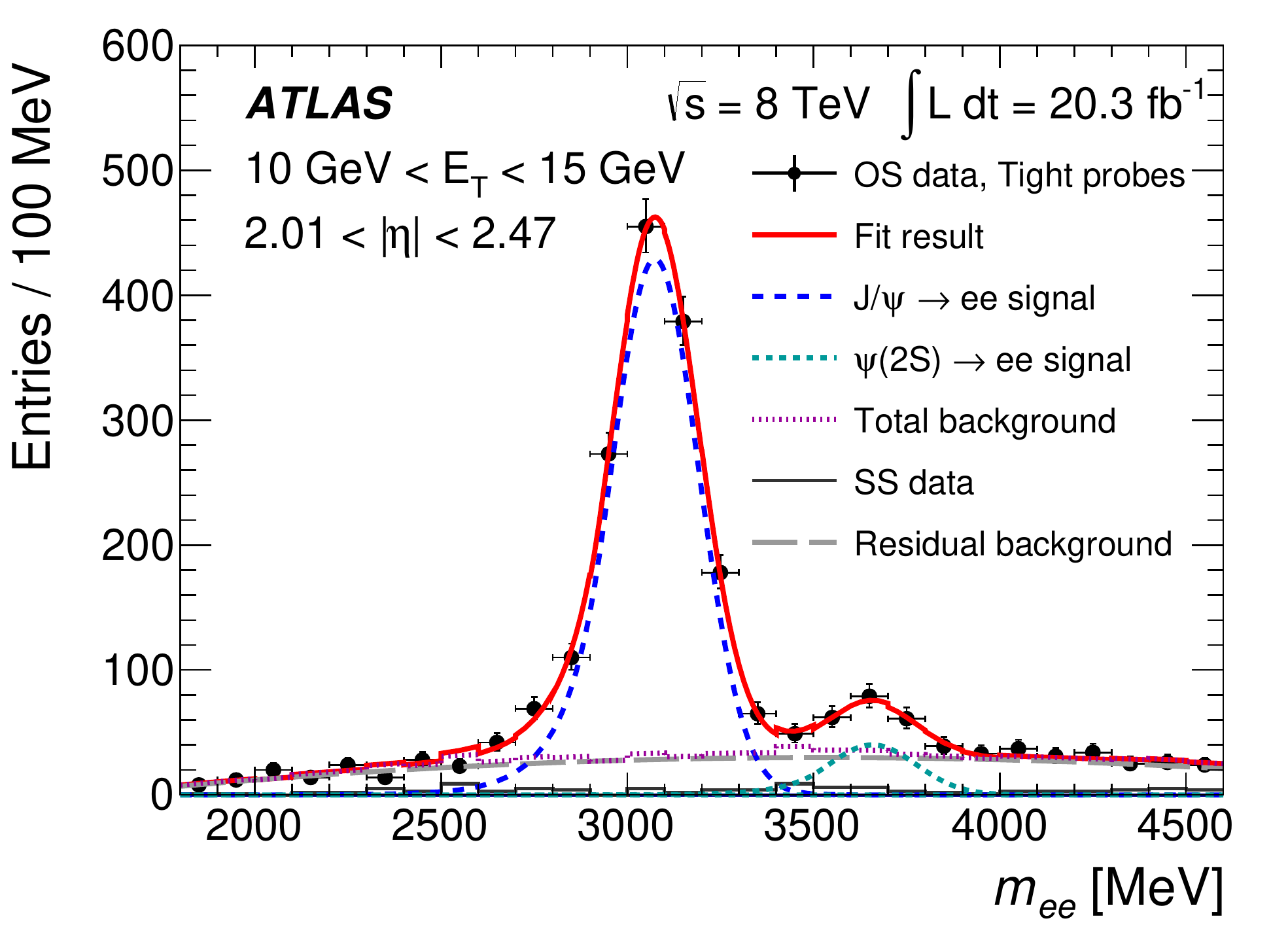}
\caption{ 
The figure demonstrates the background subtraction as carried out
in the short-$\tau$ method. Shown is the dielectron invariant-mass fit
for all probe electron candidates having a good track quality (left) and for probe electron candidates 
passing the cut-based \tight\ identification (right) for 10~\GeV $<$ \ET\ $<$
15~\GeV\ and 2.01 $<$ $|\eta| <$ 2.47.   A track
isolation requirement of $p_\mathrm{T}^\mathrm{cone0.2}/E_\mathrm{T} < 0.15$ is placed on the probe electron candidate. The
pseudo-proper time is required to be $-$1~ps $< \tau <0.2$~ps. 
Dots with error bars represent
the opposite-sign (OS) pairs for data, the fitted \JPsi\ signal is
shown by the dashed blue and the $\psi$(2S) by the dashed light blue
lines (both modelled by a Crystal-Ball function). A background
fit is carried out using the sum of the same-sign (SS) distribution (solid grey) 
from data and a Chebyshev polynomial of 2nd order describing the residual
background (dashed grey). The sum of the two background
contributions is depicted as a purple dotted line.}
\label{fig:jpsi_massFit0}
\end{figure}

\begin{figure}[!ht]
\centering
 \includegraphics[width=0.49\textwidth]{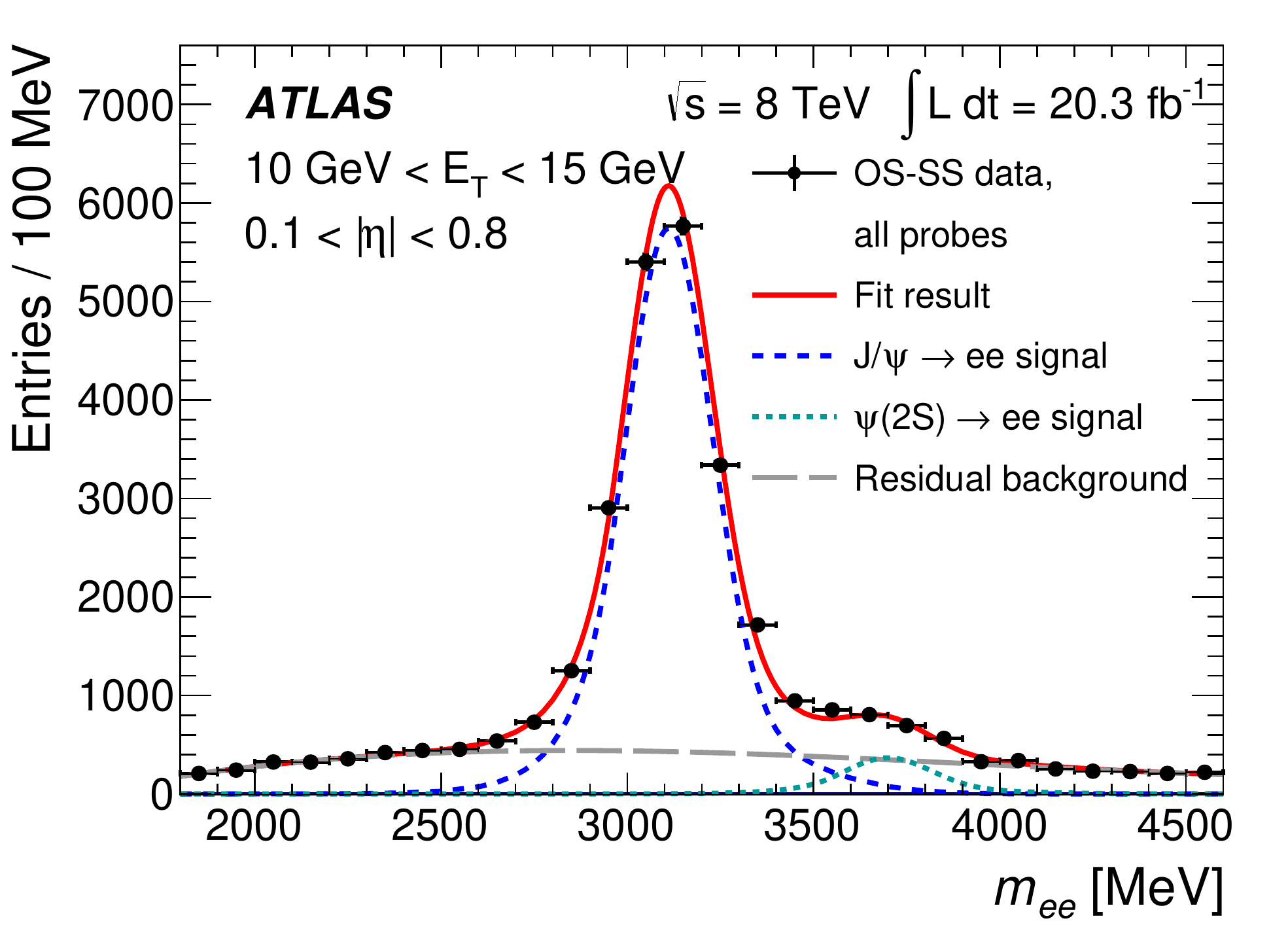}
\includegraphics[width=0.49\textwidth]{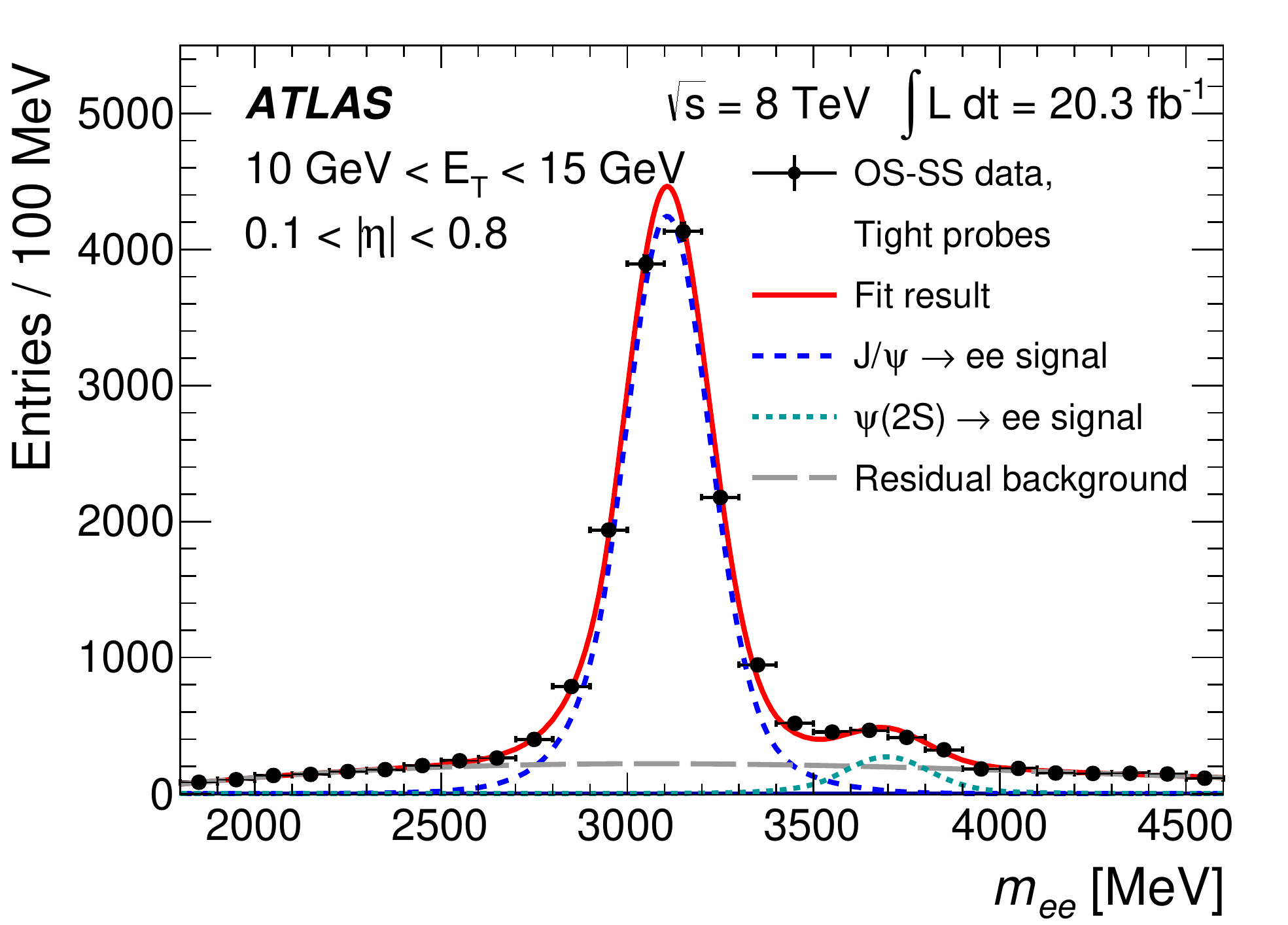}

\caption{ 
Illustration of the background determination for the \JPsi\ analysis, in the $\tau$-fit method. 
The dielectron invariant-mass fit for all probe electron candidates passing track-quality requirements (left) and for probe electron candidates passing the
cut-based \tight\
identification (right) for 10~\GeV\ $<\et<15$~\GeV\ and $0.1<\eta<0.8$ is shown.
A track isolation requirement of
$p_\mathrm{T}^\mathrm{cone0.2}/E_\mathrm{T} < 0.15$ is placed on both the tag and
the probe electron candidates. The pseudo-proper time is required to be $-$1~ps $<$ $\tau$ $<$ 3~ps.  Dots with error bars represent the OS minus SS data, 
the fitted $J/\psi$ signal 
is shown by the dashed blue and the $\psi$(2S) by the dashed light blue lines 
(both modelled by a Crystal-Ball + Gaussian function). 
The residual background (Chebyshev polynomial of 2nd order) is
shown by the dashed grey line. 
 }
\label{fig:jpsi_massFit}
\end{figure}

The invariant mass of the \tnp\ pair is used to discriminate between signal electrons and 
background. The most important contribution to the background, even after requiring the tag-and-probe pair to have opposite-sign (OS) charges, comes from random
combinations of two particles. This can be
evaluated -- assuming charge symmetry --  using the mass spectrum of same-sign (SS) charge
pairs. The remaining background is small and can be described
using an analytical model. For this, the invariant-mass distribution of the two electron candidates is fitted with the sum of three contributions: \JPsi, $\psi$(2S) and background,
typically in the range of 1.8~\GeV~$<$~$m_{ee}$~$<$~4.6~\GeV. 
To model the \JPsi\ component, a Crystal-Ball~\cite{CrystalBall1,CrystalBall2} function is used. In the $\tau$-fit method to better describe
the tail, a Crystal-Ball + Gaussian function is used instead. The $\psi$(2S) is modelled with the same shape except
for an offset corresponding to the mass difference between the \JPsi\ and
$\psi$(2S) states. Finally the residual background is modelled by a Chebyshev
polynomial (as variation by an exponential function) with the parameters determined from a combined signal + background fit to the data. The background estimated using SS pairs is 
either added to the residual background in the binned fit (see Figure~\ref{fig:jpsi_massFit0} for the short-$\tau$ method) or 
subtracted explicitly before performing the unbinned fit (see Figure~\ref{fig:jpsi_massFit} for the $\tau$-fit method).
The number of \JPsi\ candidates is counted within 
a mass window of 2.8~\GeV~$<$~$m_{ee}$~$<$~3.3~\GeV.

\begin{figure}[!ht]
\centering
 \includegraphics[width=0.49\textwidth]{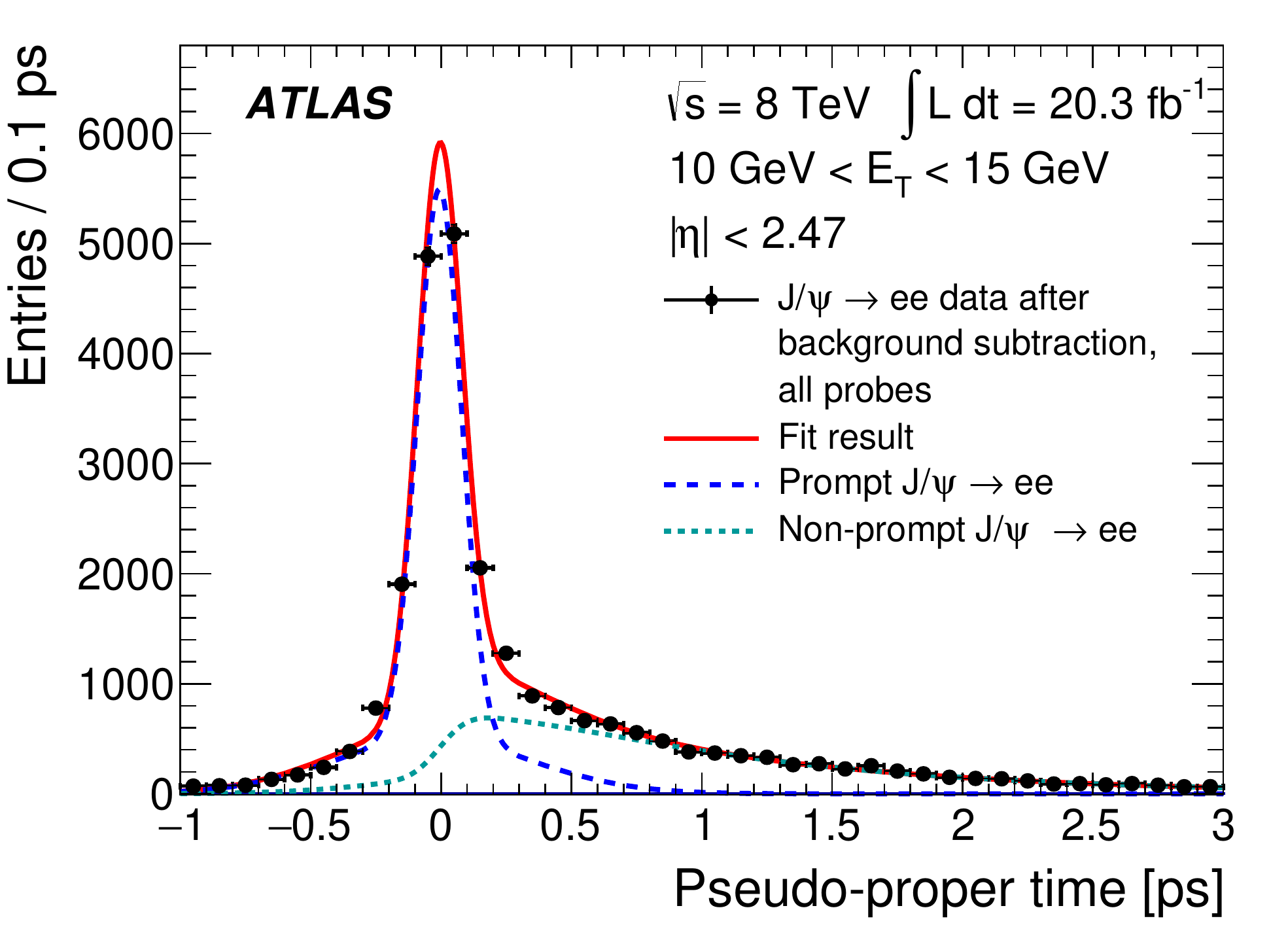}
\includegraphics[width=0.49\textwidth]{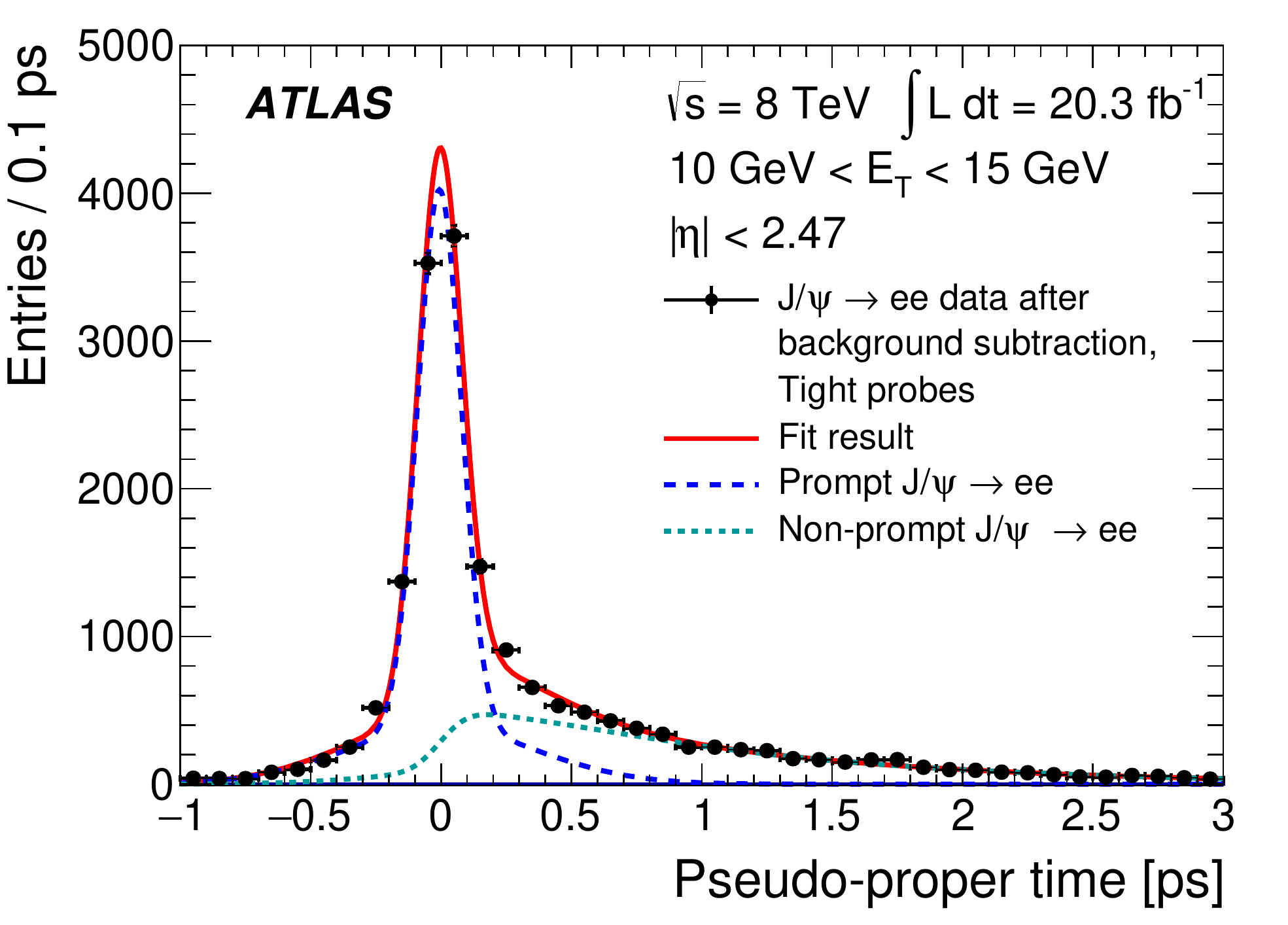}
\caption{ 
Pseudo-proper time fit for all probe electron candidates passing reconstruction+track-quality requirements (left) and for probe electron candidates passing the \tight\ identification (right) for 10~\GeV\ $<\et\ <15$~\GeV, 
integrated over $|\eta|<2.47$. A calorimetric isolation requirement of 
$E_\mathrm{T}^\mathrm{cone0.2}/E_\mathrm{T} < 0.2$
is placed on the probe electron candidate. 
Dots with error bars represent the OS minus SS data with the residual 
background subtracted using the reconstructed dielectron mass distribution sidebands.
The prompt signal component is shown by the dashed blue line (sum of two Gaussian functions) 
and the non-prompt signal component is shown by the light blue dashed
line (exponential decay function convolved with the sum of two Gaussian functions). 
 }
\label{fig:jpsi_fit_timeFit}
\end{figure}

In the $\tau$-fit method, the prompt component is then extracted by an unbinned fit of the pseudo-proper time distribution
in the range $-$1~ps $<$ $\tau$ $<$ 3~ps,
after correcting the contribution for the estimated
background by subtracting the $\tau$ distribution in the mass sidebands 2.3~\GeV~$<$ $m_{ee}$ $<$ 2.5~\GeV\ and 
4.0~\GeV\ $<$ $m_{ee}$ $<$ 4.2~\GeV\ normalized to the estimated background within the signal mass window as given by the $m_{ee}$ fit.
The non-prompt component is modelled by an exponential 
decay function convolved with the sum of two Gaussian functions, while the shape of the prompt component is
described by the sum of the same Gaussian functions describing the detector resolution, as shown in 
Figure~\ref{fig:jpsi_fit_timeFit}. 

In the short-$\tau$ method, strict requirements on $\tau$ are made, requiring it to be below 0.2 or 0.4 ps.
The resulting non-prompt contamination is below $\sim$20\%, decreasing with decreasing probe electron \et.
The measured efficiency is compared to the prediction of the MC simulation after mixing the simulated prompt and non-prompt \Jpsiee\ samples
according to the ATLAS measurement of the non-prompt \JPsi\ fraction in the dimuon final state at 
$\sqrt{s}=7$~\TeV~\cite{ATLASJpsiXsec}.

 Systematic uncertainties arise predominantly from the background estimation and the probe electron definition. 
They are estimated by varying the \tnp\ selection (such as the isolation and the $\tau$ requirements),
the fit parameters (background and signal shapes, fit window and sideband definitions) and 
the size of the mass window (changed by $\pm$40\%) for signal counting after the mass fit.
In total, 186 variations were considered in each (\et, $|\eta|$) bin, 
using the two methods,  
to determine the efficiency and its uncertainty.

\subsection{Combination}\label{sec:combi}

To calculate the final results for the identification efficiency, the data-to-MC correction factors are combined. The following measurements are used in the different \et\ bins:
\begin{itemize}
\item $7$--$10$~\GeV: \Jpsiee, 
\item $10$--$15$~\GeV: \Jpsiee\ and \Zeegamma,
\item $15$--$20$~\GeV: \Jpsiee\ and \Zee,
\item $20$--$25$~\GeV\ and bins above: \Zee.
\end{itemize} 

Only the two \et\ bins $10$--$15$~\GeV\ and $15$--$20$~\GeV\ allow a combination of independent measurements, 
which is done using a program originally developed for the HERA experiment
\cite{Aaron:2009bp} and used in Ref.~\cite{Aad:2011Paper}. It performs a $\chi^2$ fit over all bins, separately for the bins below and above 20~\GeV, adjusting the input values taking into account correlations of the
systematic uncertainties in $\eta$ and \ET\ bins.

Both the $\chi^2$ (ranging
from 3.4 to 12.3 for 12 degrees of freedom, depending on the identification selection) and the pulls of the combination indicate good
agreement for the measurements in the $10$--$15$~\GeV\ and $15$--$20$~\GeV\ bins.

\subsection{Results}
\label{sec:main_id_results}

The combined data efficiencies are derived by applying the combined data-to-MC efficiency ratios to the MC efficiency prediction from simulated \Zee\ decays. 
Similarly, when comparing the results of different efficiency measurements, the measured data-to-MC efficiency ratios are used to correct the \Zee\ MC sample.

The measured efficiencies for the various identification criteria are presented as functions of the electron $\eta$, \et\ and the
number of reconstructed primary collision vertices in the event. 
The latter is a measure of the amount of activity due to overlapping collisions 
which affects the reconstructed electrons, for example by making the calorimeter shower shapes more background-like due to nearby particles.
The efficiency dependence in bins of primary vertices is only measured for
electrons with \et~$>$~15~\GeV\ using \Zee\ events with the \Zmass\ method, 
as the \Jpsiee\ sample size is not large enough. 

\begin{figure}
\centering
\includegraphics[width=0.49\textwidth]{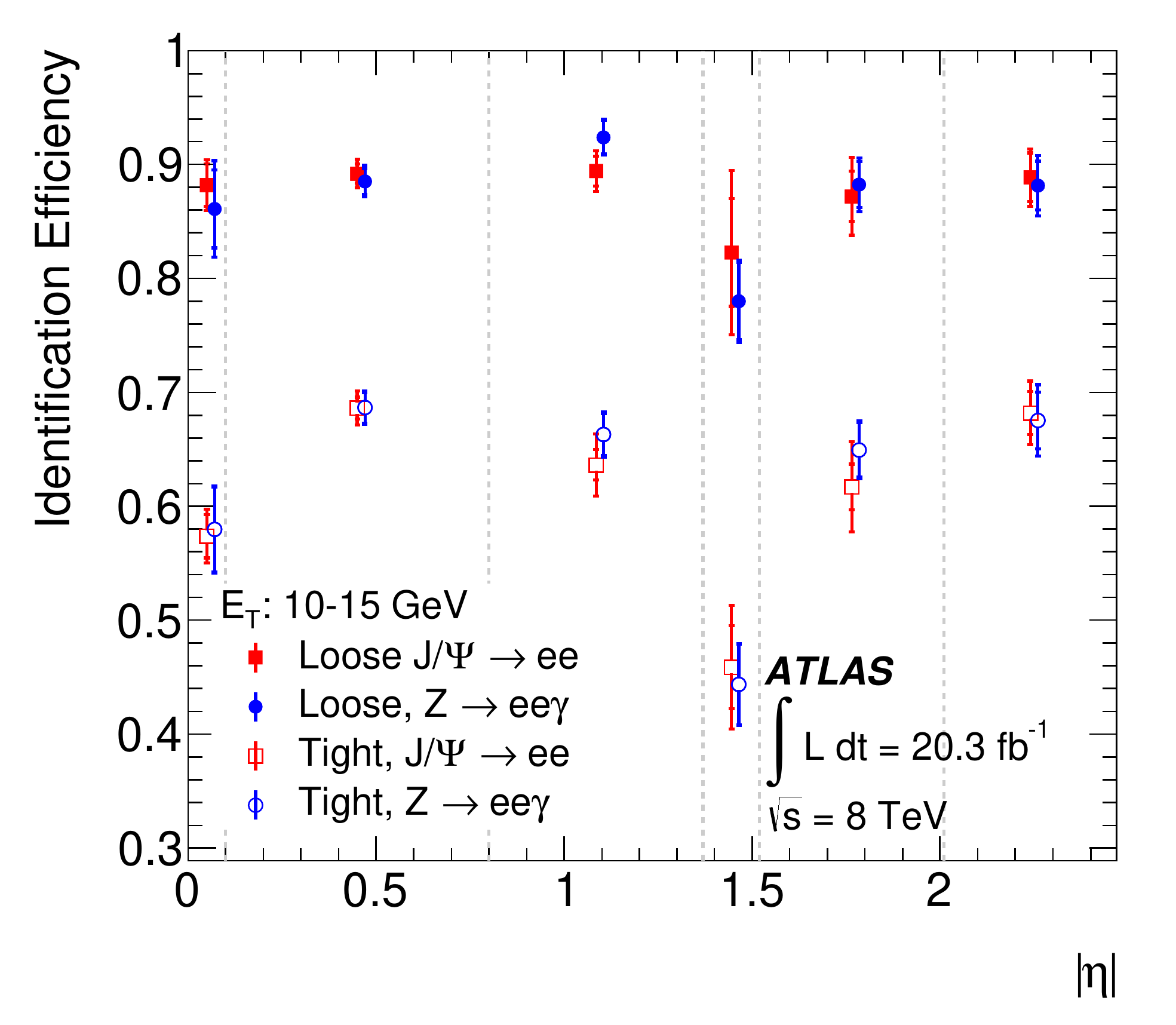}
\includegraphics[width=0.49\textwidth]{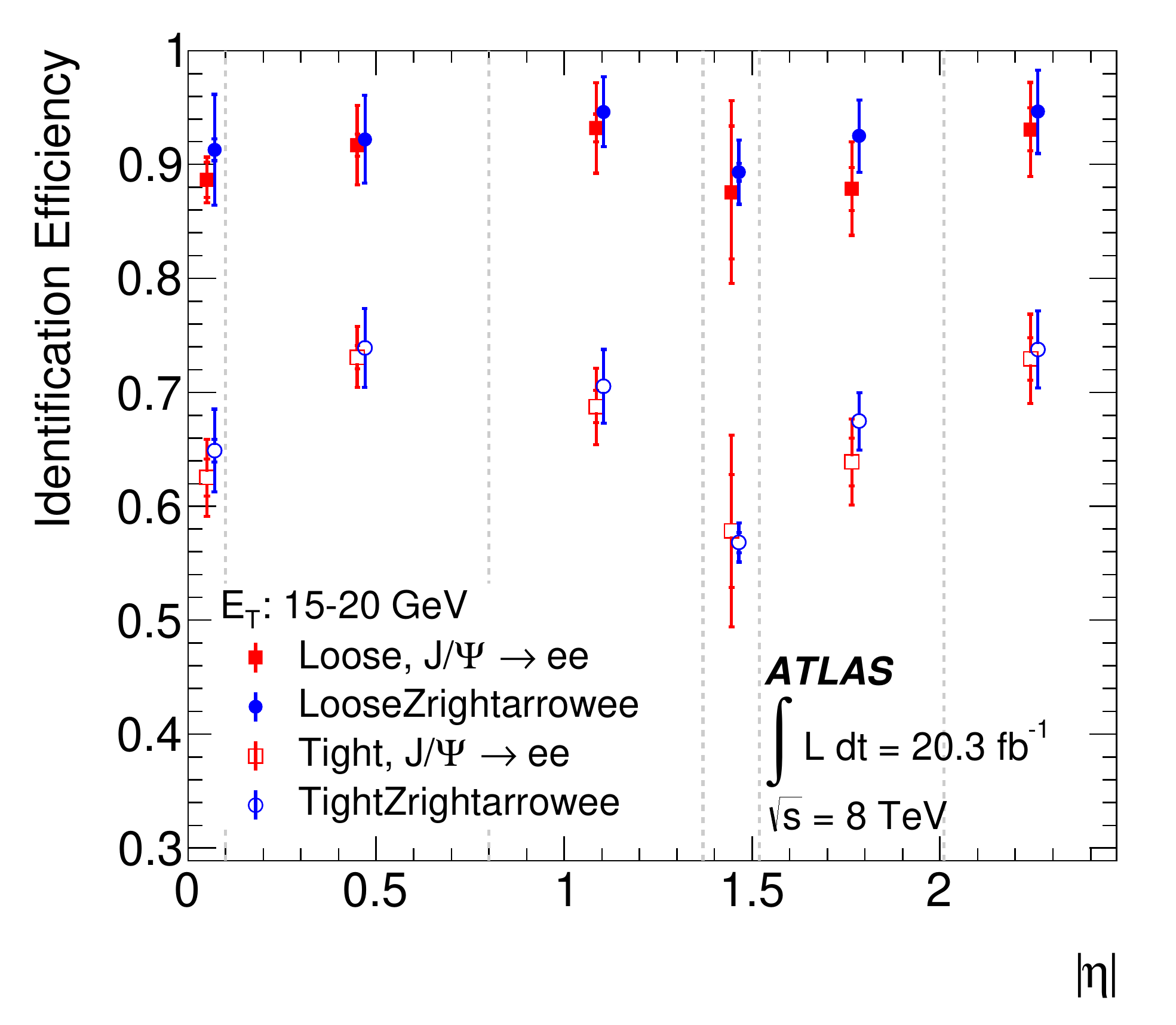}
\includegraphics[width=0.49\textwidth]{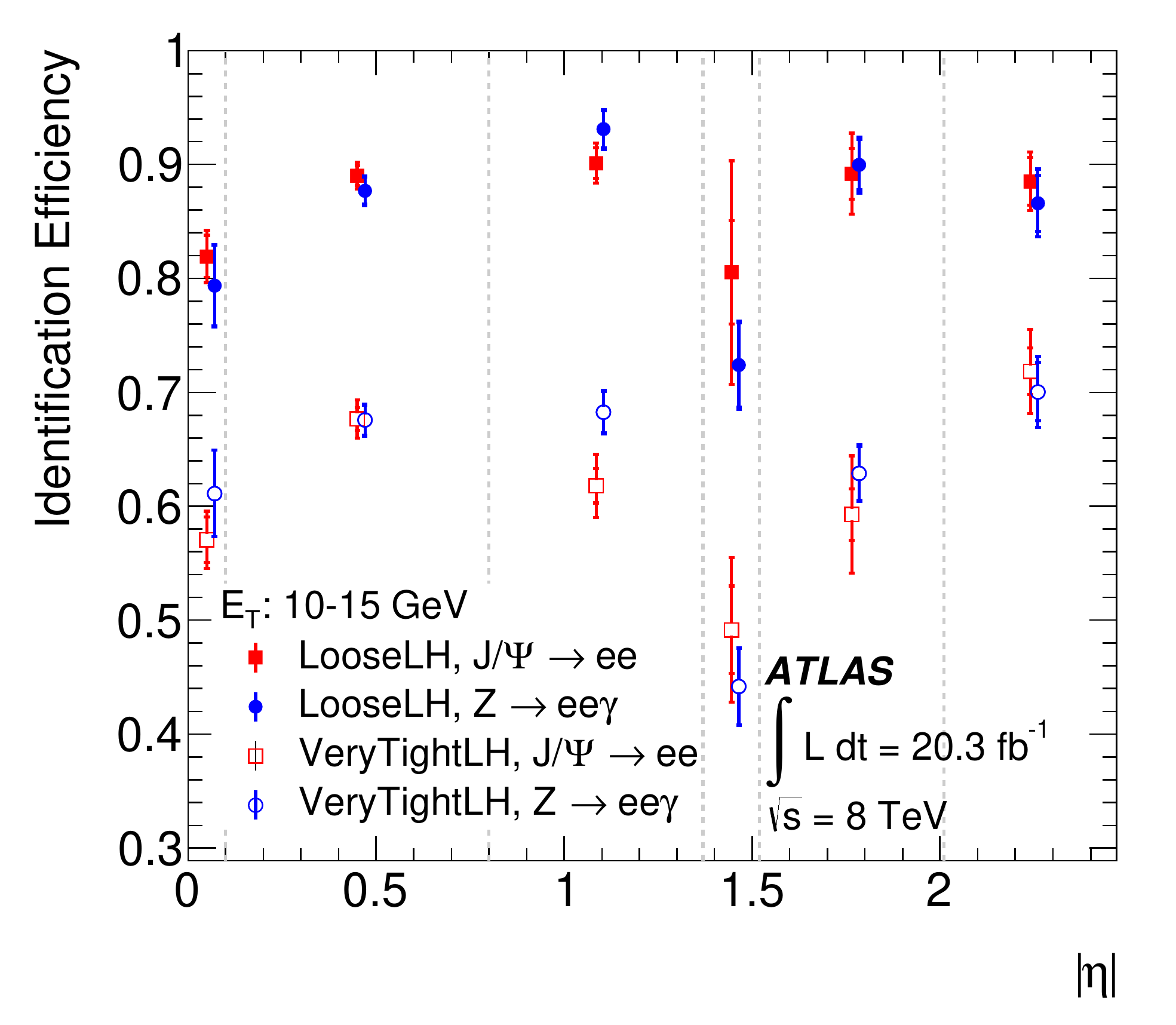}
\includegraphics[width=0.49\textwidth]{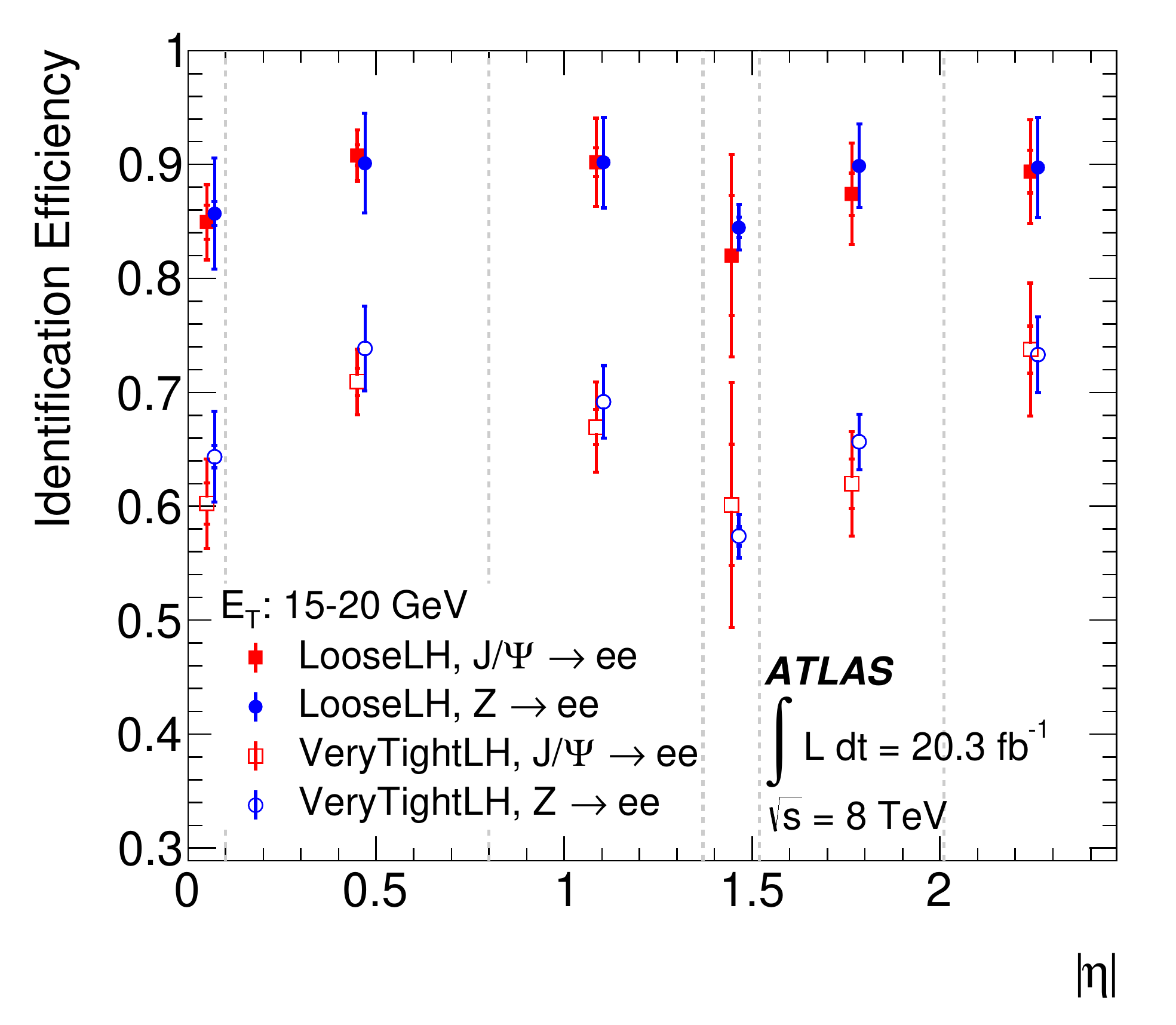}
\caption{  Measured identification efficiency as a function of $|\eta|$ for $\et=10$--$15$~\GeV\ (left) and 
$\et=15$--$20$~\GeV\ (right)
for the cut-based \loose\ and \tight\ selections (top) 
and for \looseLLH\ and \veryTightLLH\ (bottom).
The data efficiency is derived by applying the measured data-to-MC efficiency ratios, 
determined with either the \JPsi\ or the  \Z\ methods,
to the prediction of the MC simulation from \Zee\ decays. 
The uncertainties are statistical (inner error bars) and statistical+systematic (outer error bars). The dashed lines indicate the bins in which the efficiencies are calculated. For better visibility, the measurement points are displayed as slightly shifted with respect to each other.}
\label{fig:Results_ZJPsi}
\end{figure}

Figure~\ref{fig:Results_ZJPsi} shows a comparison 
between efficiencies computed for \Zee\ decays
in the two \et\ bins in which different measurements overlap. 
The methods agree well.

\begin{figure}
\centering
\includegraphics[width=0.49\textwidth]{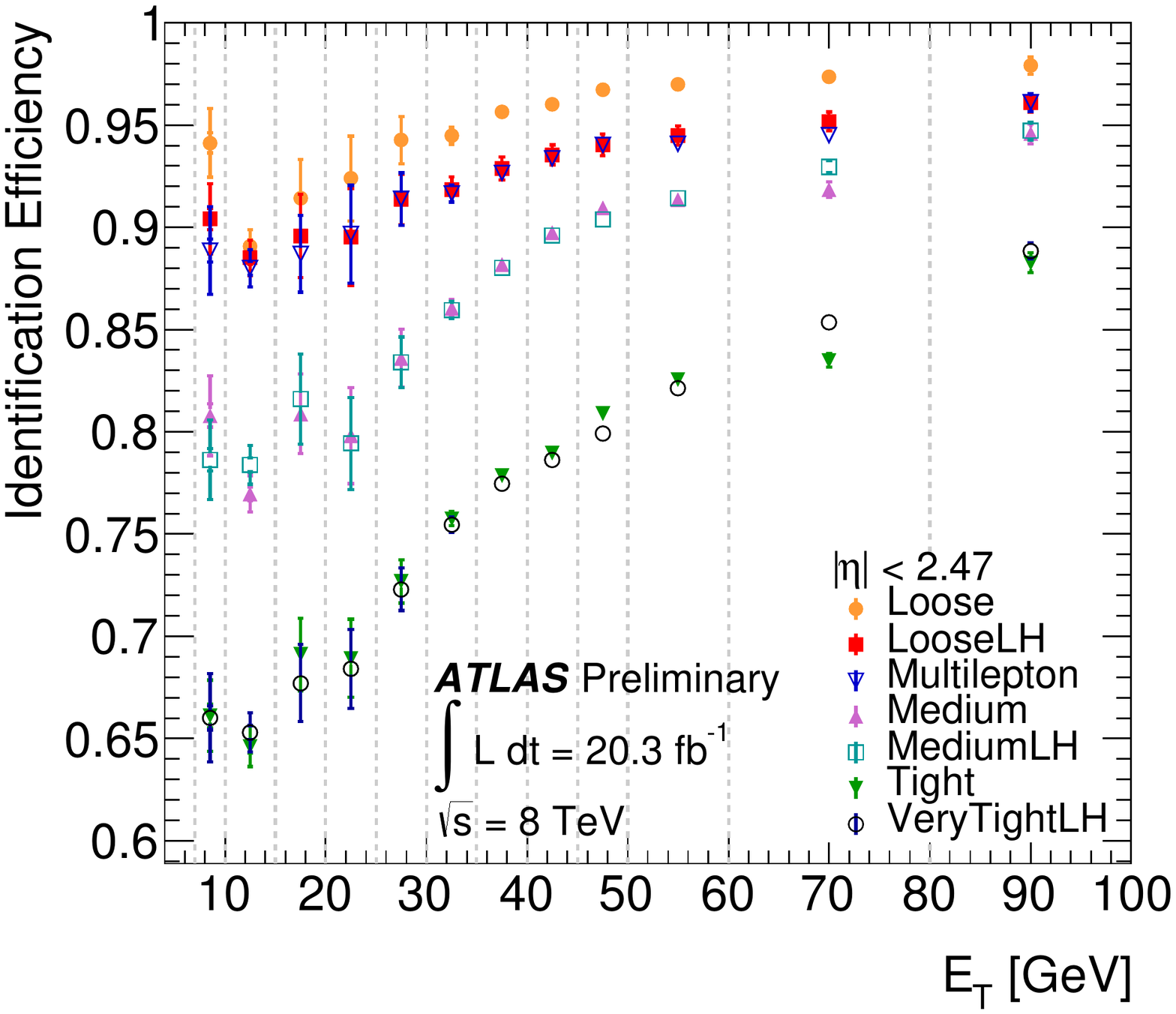}
\includegraphics[width=0.49\textwidth]{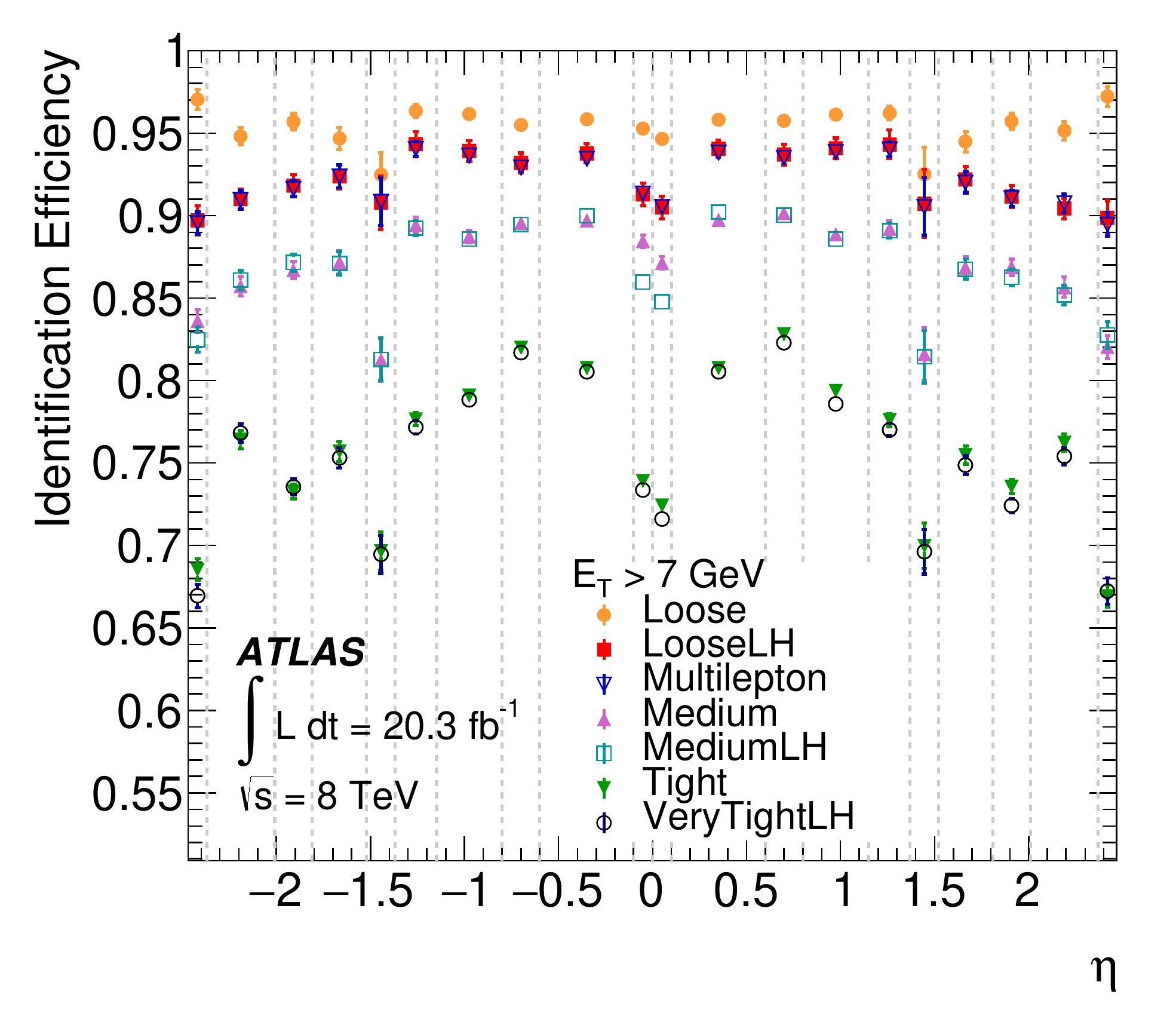}
\includegraphics[width=0.49\textwidth]{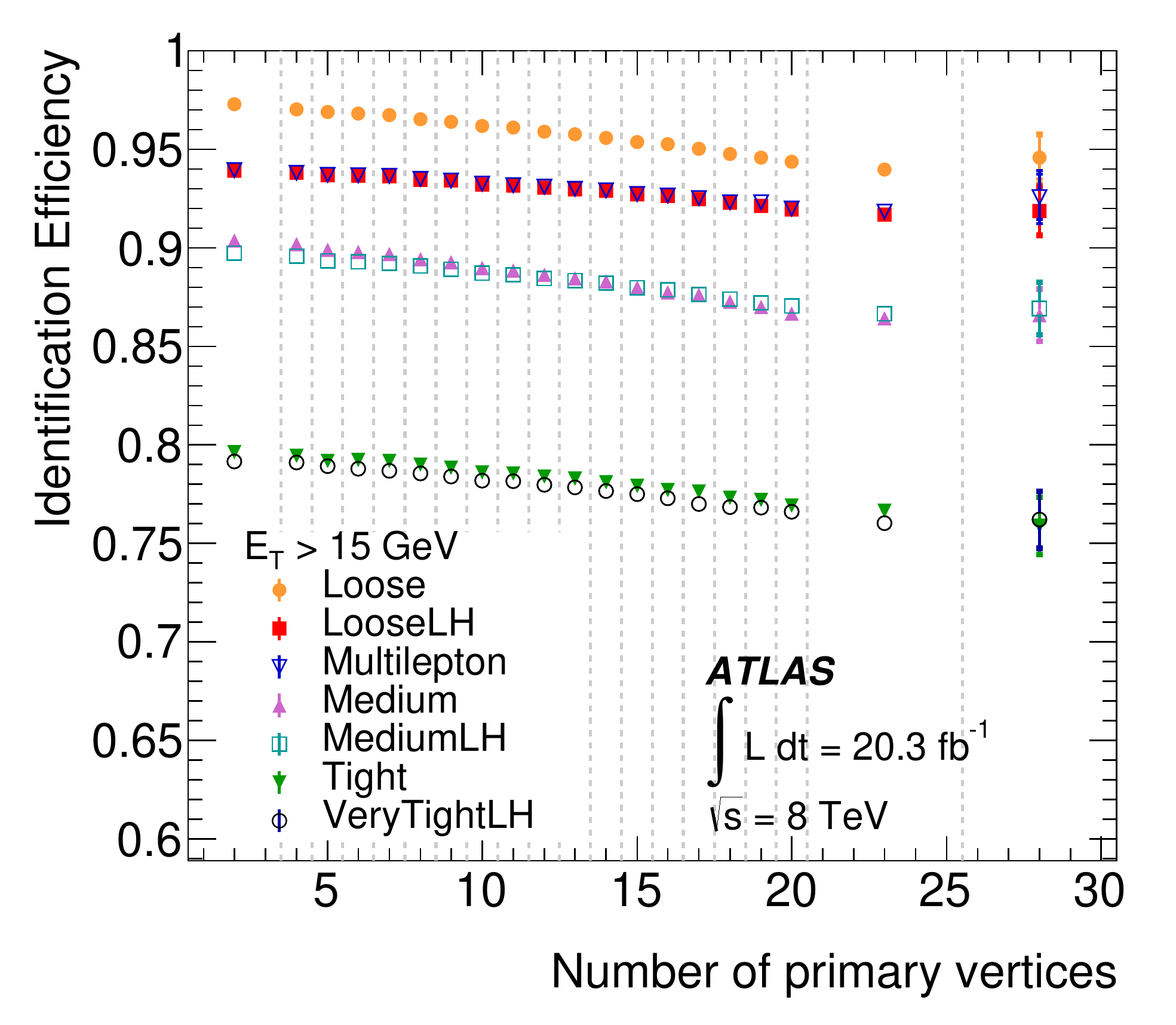}
\caption{  Measured identification efficiency  for the various cut-based and LH selections 
as a function of \et\ (top left), $\eta$ (top right) and the number of reconstructed primary vertices (bottom).
The data efficiency is derived from the measured data-to-MC efficiency ratios and the prediction of the MC simulation from \Zee\ decays. 
The uncertainties are statistical (inner error bars) and statistical+systematic (outer error bars). 
The last bin in \et\ and number of primary vertices includes the overflow.  The dashed lines indicate the bins in which the efficiencies are calculated.}
\label{fig:Results_IDoverlays_eteta}
\end{figure}

The efficiencies integrated over \et\ or $\eta$, as well as the dependence on the number of primary vertices is shown in Figure~\ref{fig:Results_IDoverlays_eteta}. These distributions assume the (\et, $\eta$) distribution of electrons from \Zee\ decays and treat the total uncertainties as fully correlated between bins, as done for most analyses. 

\begin{figure}
\centering
\includegraphics[width=0.49\textwidth]{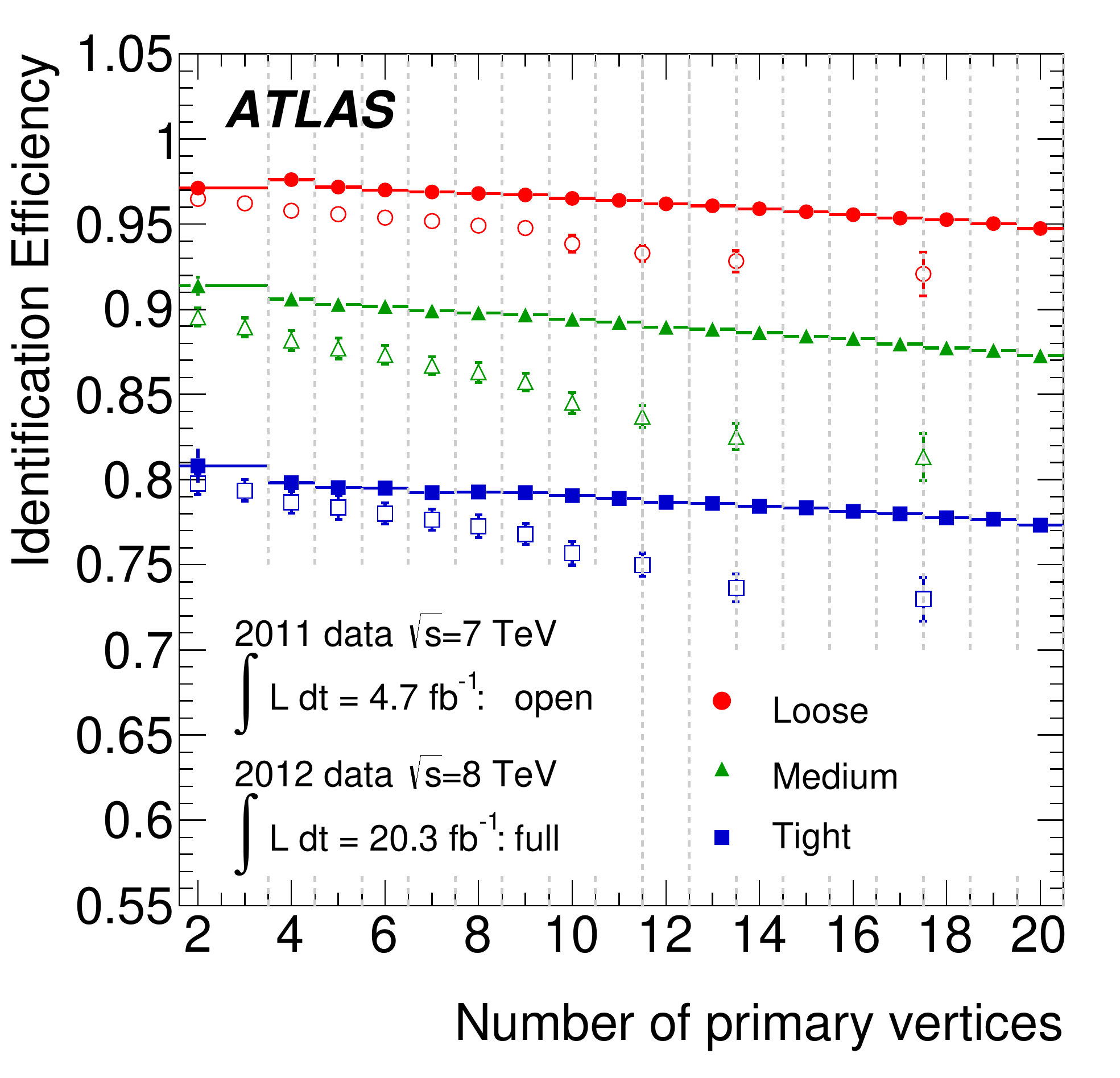}
\caption{  Identification efficiency for the various cut-based selections measured with 2011 and 2012 data as a function of the number of reconstructed primary vertices.}
\label{fig:Results_IDoverlays_nvtx20112012}
\end{figure}

With tighter requirements on more variables, the overall identification efficiency decreases, while the dependence on \et\ and $\eta$ increases, as expected. The efficiency of the cut-based \multilepton\ selection shows less variation with the number of primary vertices than the cut-based \loose\ selection, as
it relies less on the pile-up-sensitive variables \reta\ and \rhad. Overall, the 2012 update of the cut-based menu (see Section~\ref{sec:cutbasedID}) has been successful: the efficiencies and rejections could be kept at values similar to those in 2011, while the remaining pile-up dependence is small (variation below 4\% for 1 to 30 vertices). The improvement of the 2012 menu regarding the pile-up robustness of the requirements is demonstrated in Figure~\ref{fig:Results_IDoverlays_nvtx20112012}, where the efficiencies for the cut-based \loose, \medium\ and \tight\ selections as a function of the number of reconstructed primary vertices are compared for 2011 and 2012. 

The \looseLLH\  is tuned to match the efficiencies
of the cut-based \multilepton\ selection, while the (\mediumLLH) \veryTightLLH\ is tuned to match those of the cut-based (\medium) \tight\ selection. The efficiency figures show that this
tuning is successful in almost all bins. While the efficiencies match, the background rejection of the LH selections is
better. 
The background efficiencies  
are reduced by a factor of about two when comparing the cut-based identification to the corresponding LH selections (see Section~\ref{sec:Rejection}).  

The efficiencies as a function of \ET\ and \eta, as presented in Figure~\ref{fig:Results_IDoverlays_eteta}, show some well-understood features. 
The identification efficiencies in general rise as a function of \ET\ because electrons with higher \ET\ are better separated from the background in many of the discriminating variables.
For the lowest ($7$--$10$~\GeV) as well as for the highest (above 80~\GeV) \ET\ bin, a significant and somewhat discontinuous increase in the identification efficiency is observed. This is
explained by the fact that at very low and very high \ET\ some requirements are relaxed.  For the high \ET\ bin the $E/p$ requirement is removed, because the measurement of the electron's track momentum is less precise for high-\pT\ tracks and can therefore not safely be used to distinguish electrons from backgrounds. It was checked that the data-to-MC correction factor measured for electrons above 80~\GeV\ is applicable to electrons even at \ET\ greater than 400~\GeV\ using the \Ziso\ method. Within the large statistical uncertainties, data-to-MC correction factors binned in \ET\ for the high-\ET\ region were found to agree with the combined data-to-MC correction factor above 80~\GeV\ that is presented in this paper.
The lowest \et\ bin ($7$--$10$~\GeV) was tuned separately from the other bins, 
choosing the signal efficiency to be a few percentage points higher. This leads to higher background contamination.

The shape of the identification efficiency distributions as a function of \eta\ is mainly due to features of the detector design and the selection optimization procedure that is typically based on the signal-to-background ratio. 
A small gap between the two calorimeter half-barrels and in the TRT around $|\eta| \approx 0$ explains 
the slight drop in efficiency. 
Another, larger drop in efficiency is observed for $1.37<|\eta|<1.52$, where the transition region between the barrel and endcap calorimeters is situated. At high $|\eta|$ the efficiencies are lower due to the larger amount of material in front of the endcap calorimeters.

\begin{figure}
\centering
\includegraphics[width=0.49\textwidth]{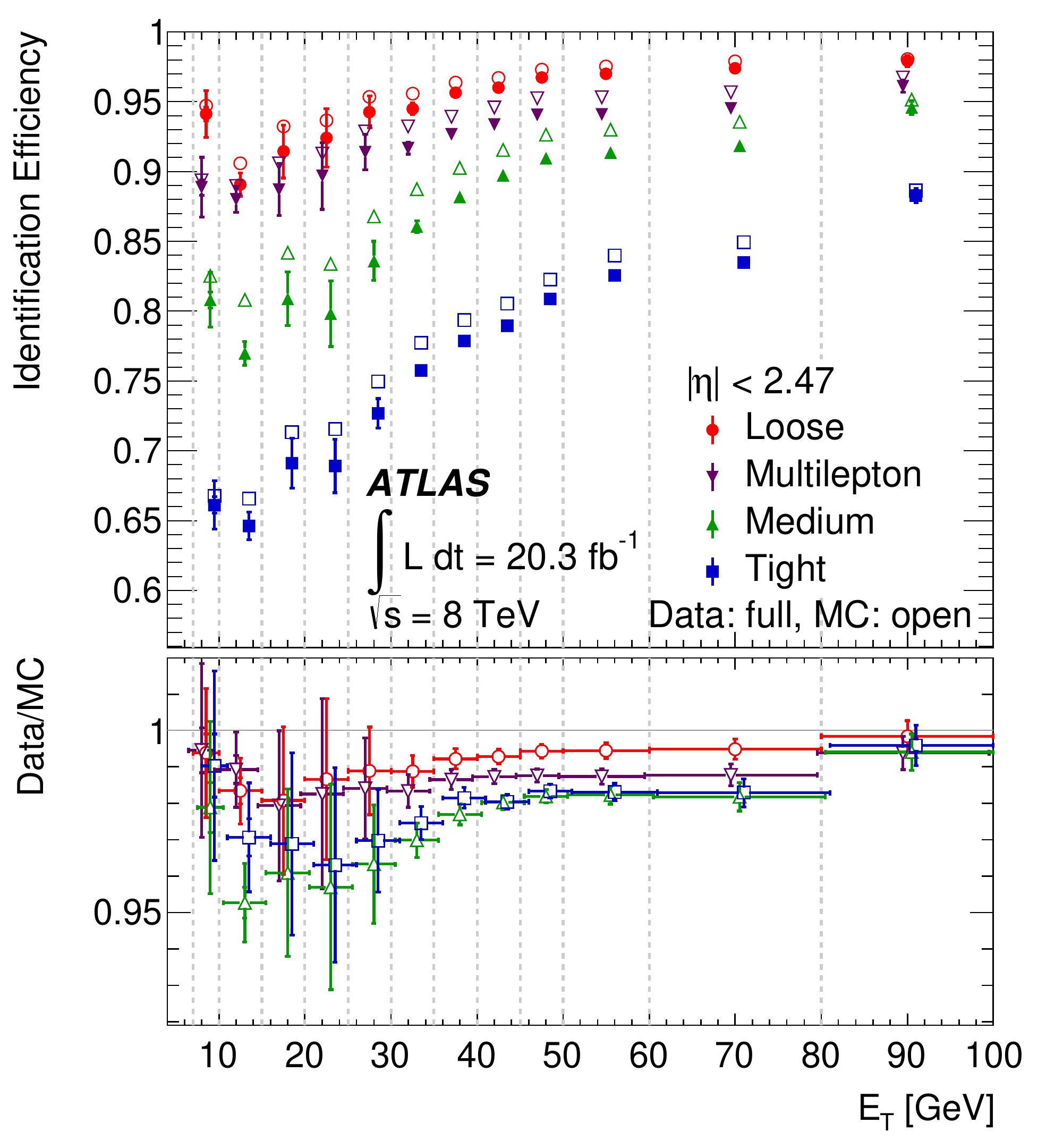}
\includegraphics[width=0.49\textwidth]{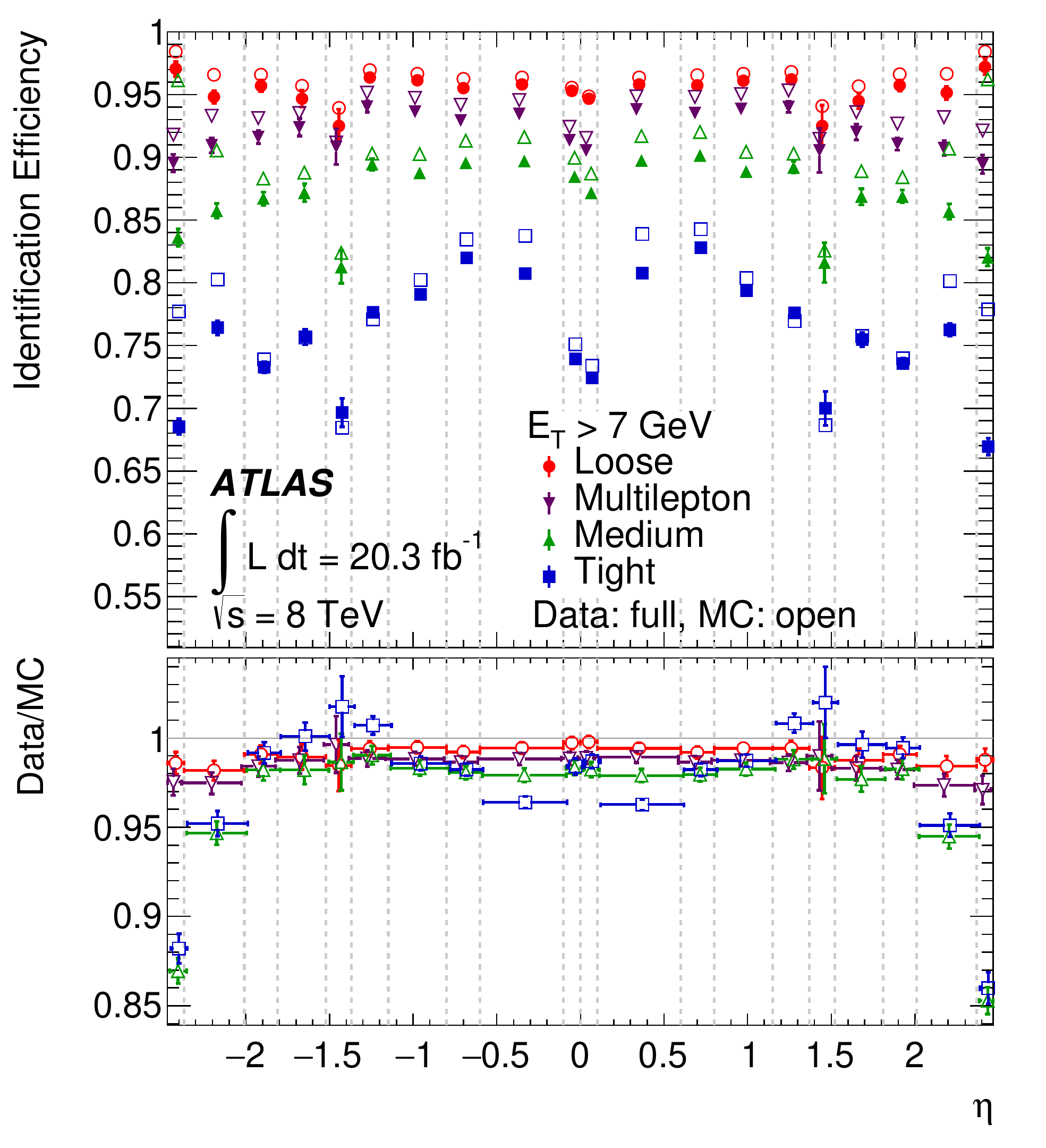}
\includegraphics[width=0.49\textwidth]{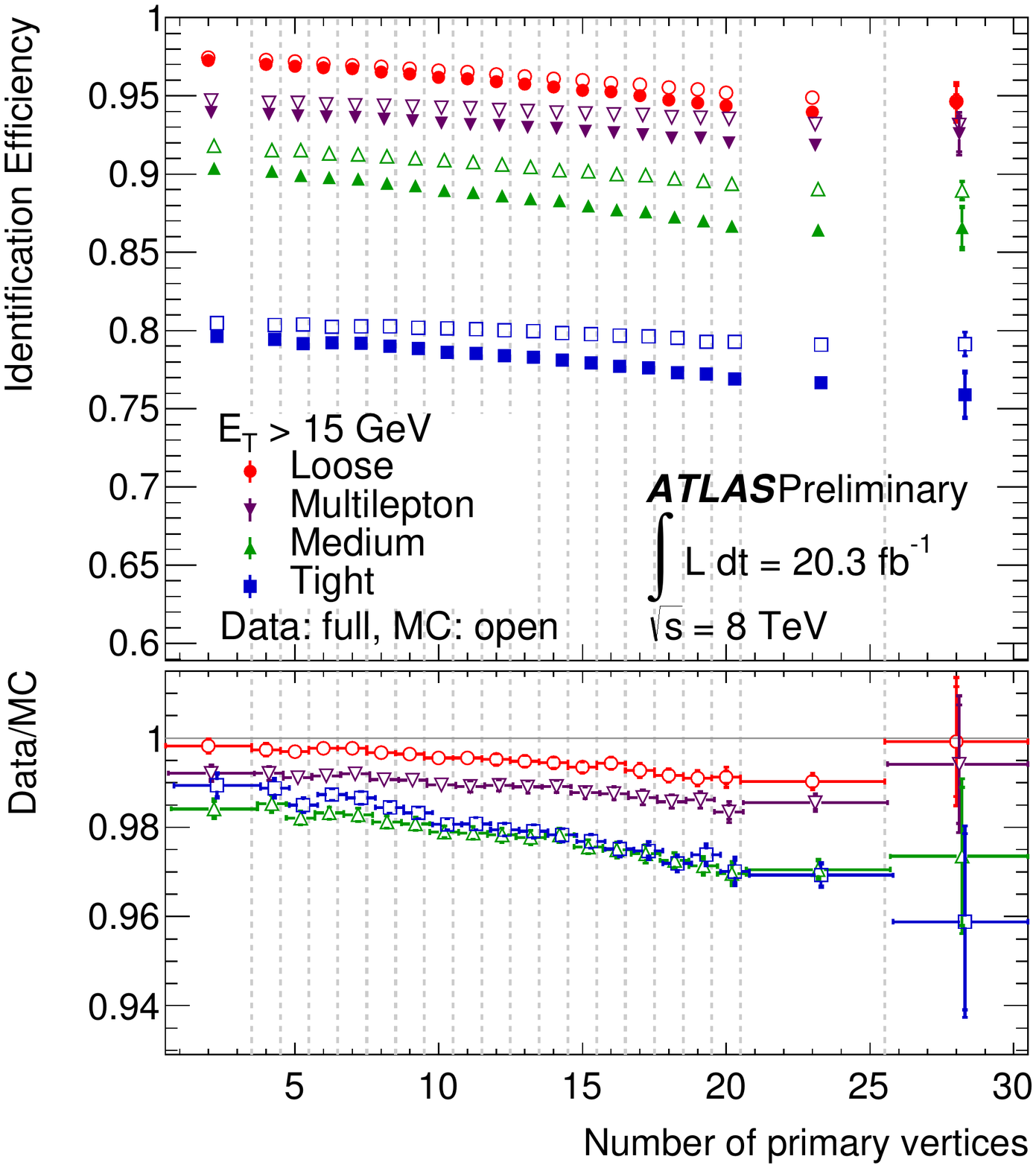}
\caption{Identification efficiency in data
as a function of \et\ (top left), $\eta$ (top right) and the number of reconstructed primary vertices (bottom) 
for the cut-based 
\loose, \multilepton, \medium\ and \tight\ selections, compared to predictions of the MC simulation for electrons from \Zee\
decay. 
The lower panel shows the data-to-MC efficiency ratios.
The data efficiency is derived from the measured data-to-MC efficiency ratios and the prediction of the MC simulation for electrons from \Zee\
decays. The last bin in \et\ and number of primary vertices includes the overflow. 
The uncertainties are statistical (inner error bars) and statistical+systematic (outer error bars).   The dashed lines indicate the bins in which the efficiencies are calculated.}
\label{fig:Results_MCDataoverlaysLooseTight_eteta}
\end{figure}

\begin{figure}
\centering
\includegraphics[width=0.49\textwidth]{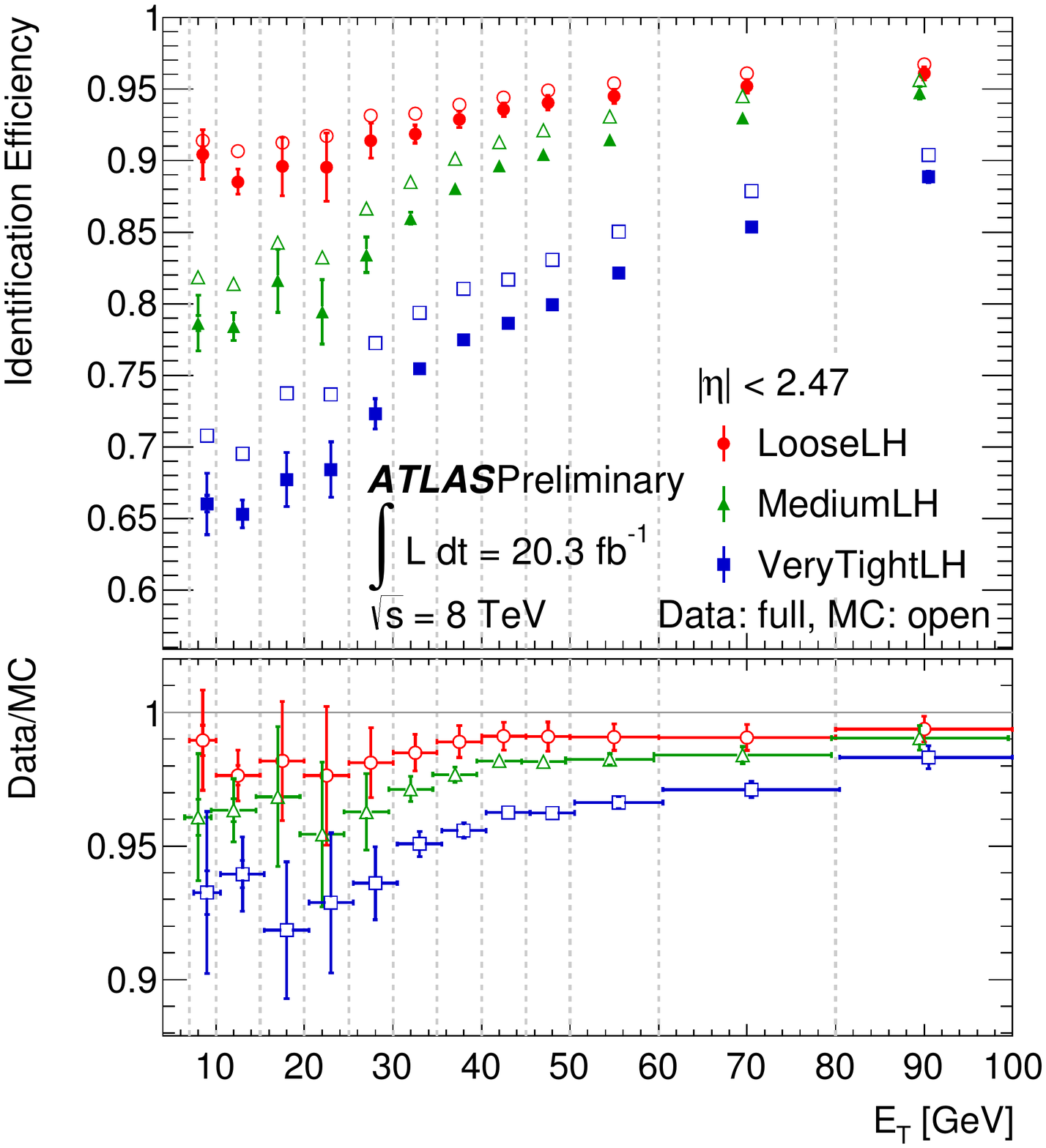}
\includegraphics[width=0.49\textwidth]{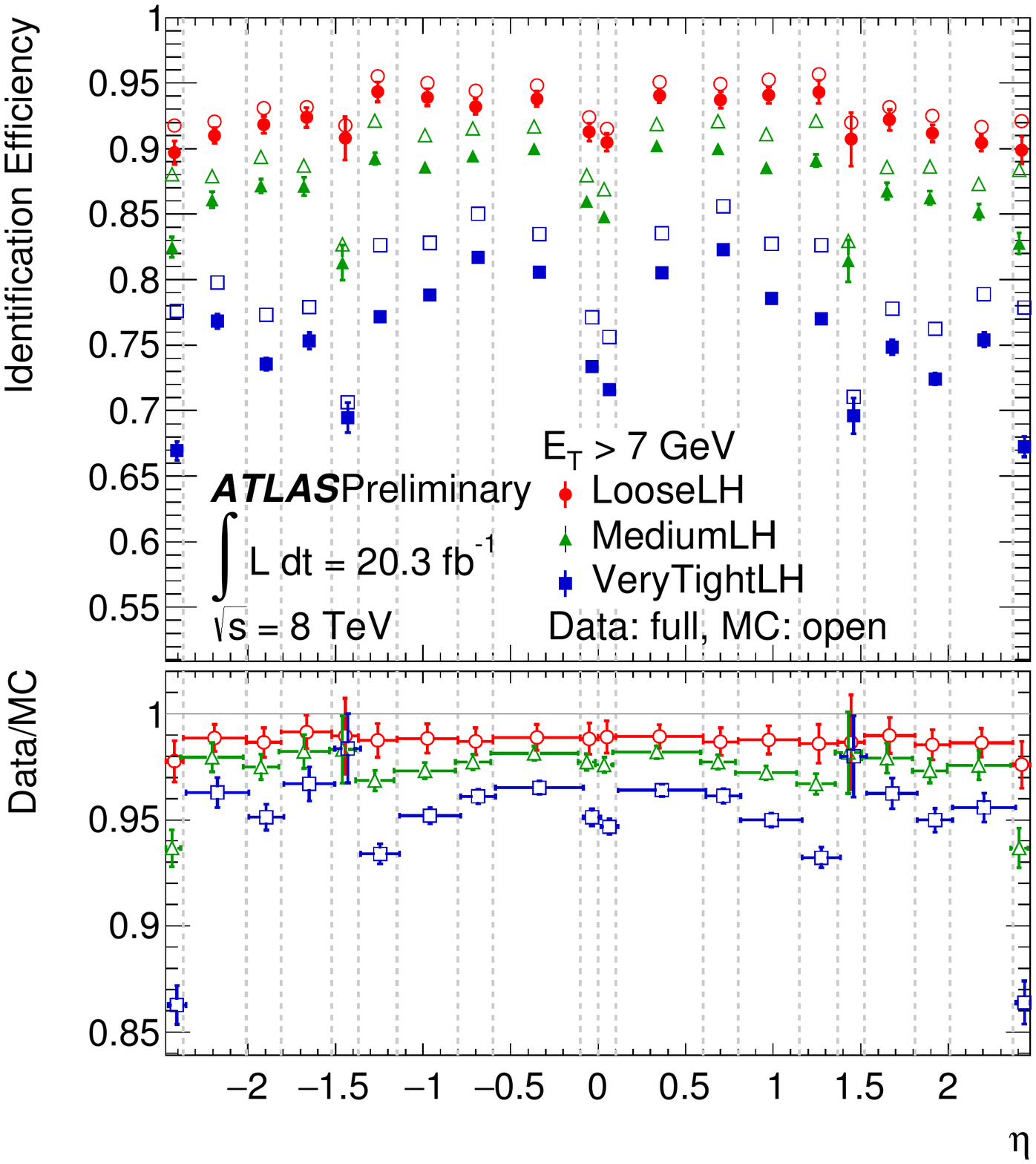}
\includegraphics[width=0.49\textwidth]{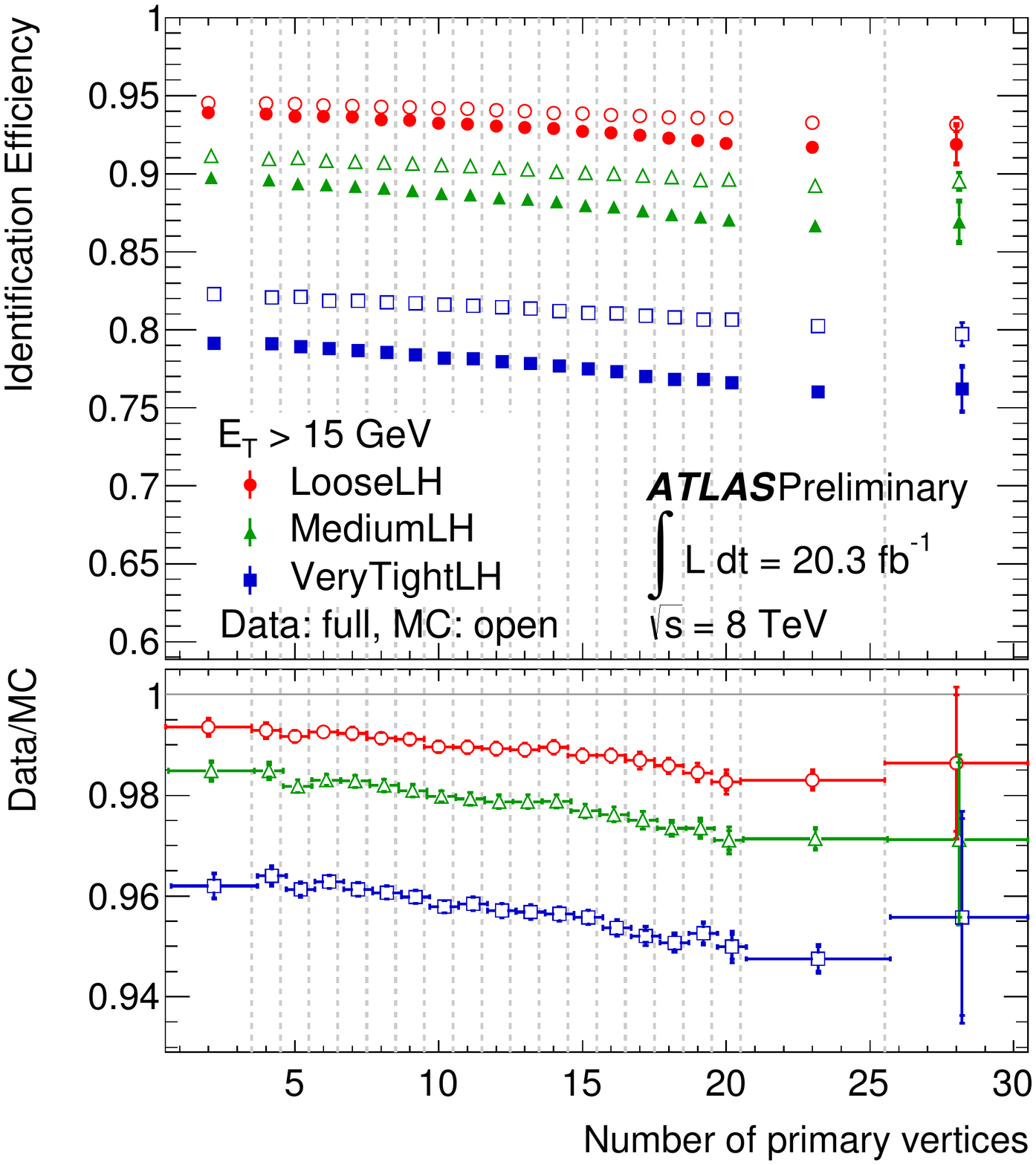}
\caption{  Identification efficiency in data
as a function of \et\ (top left), $\eta$ (top right) and the number of reconstructed primary vertices (bottom) 
for \looseLLH, \mediumLLH\ and \veryTightLLH\ selections, compared to predictions of the MC simulation for electrons from \Zee\
decay.  
The lower panel shows the data-to-MC efficiency ratios.
The data efficiency is derived from the measured data-to-MC efficiency ratios and the prediction of the MC simulation for electrons from \Zee\
decays. The last bin in \et\ and number of primary vertices includes the overflow. 
The uncertainties are statistical (inner error bars) and statistical+systematic (outer error bars). The dashed lines indicate the bins in which the efficiencies are calculated.}
\label{fig:Results_MCDataoverlaysLooseLLHVeryTightLLH_eteta}
\end{figure}

 Figures~\ref{fig:Results_MCDataoverlaysLooseTight_eteta} and
\ref{fig:Results_MCDataoverlaysLooseLLHVeryTightLLH_eteta} show the identification efficiencies when integrated over \et\ or $\eta$,
and as a function of the number of reconstructed primary vertices. 
These figures depict in their lower panels the data-to-MC correction factors.  
As can be seen, the correction factors are close to one, with cut-based selections showing better data--MC agreement than the LH. Only for low \et\ or high values of \eta, corrections reaching 10\% have to be applied for the more stringent selection criteria. The combined statistical and systematic uncertainties in the data-to-MC correction factors range from 0.5\% to 10\%, with the highest uncertainties found at low \et, and in the transition region of the calorimeter, 1.37~$<|\eta|<$~1.52. 
At low \et, a large contribution to the uncertainties is statistical in nature and can be considered uncorrelated between bins when propagating the uncertainties to the final results of analyses (in the presented figures the uncertainties are treated as fully correlated between bins).

As discussed in Ref.~\cite{Aad:2010Paper}, the difference between identification efficiencies in data and MC simulation can be
traced back to differences in the distribution of the variables used in the identification, particularly the shower shape
variables and the TRT high-threshold hit ratio \TRTHighTHitsRatio, the latter being defined only for $|\eta|<2$.
The distributions of the lateral shower shapes are not well modelled by the \texttt{GEANT4}-based simulation of the detector: in comparison to
predictions of the MC simulation, most shower shapes in data are wider and centred at values closer to the background distributions. 
These effects lead to higher efficiencies in MC simulation. 
\TRTHighTHitsRatio, on the other hand, is underestimated in the simulation for $|\eta|$~$>$~1, leading to higher efficiencies in data
than in the simulation. These two effects cancel each other, as can be seen in Figure~\ref{fig:Results_MCDataoverlaysLooseTight_eteta}, where the data and MC efficiency
values of the cut-based \tight\ selection are quite close to each other for 1~$<|\eta|<$~2.

Figures~\ref{fig:Results_MCDataoverlaysLooseTight_eteta} and \ref{fig:Results_MCDataoverlaysLooseLLHVeryTightLLH_eteta} show that the data has a more significant dependence on pile-up than predicted by simulation. For the cut-based \multilepton\ and 
\loose\ selections, the data-to-MC ratio is almost constant as a function of the number of primary vertices, while it decreases for the cut-based \medium\ and \tight\ selections as well as the LH selections by about 2\% from 1 to 30 primary vertices. 
This effect is primarily caused by the mismodelling in MC simulation of the $R_\mathrm{Had(1)}$, \wstot\ and \TRTHighTHitsRatio\ variables. The \TRTHighTHitsRatio\ variable is sensitive to the pile-up conditions due to higher occupancies in events with many vertices, which can lead to hit overlaps in the TRT straws increasing the chance of passing the high threshold. The effect is not well modelled by the simulation, independent of the modelling of the pile-up itself. 
Both the $R_\mathrm{Had(1)}$ and \wstot\ variables, as well as additional energy deposits from pile-up particles, are not well modelled by the \texttt{GEANT4} simulation of the calorimeter, leading to differences as a function of pile-up between data and MC simulation.
The pile-up profile of the collision data analyses which use the results of these efficiency measurements is very close to the pile-up profile of the efficiency measurements presented here. The data-to-MC correction factors will therefore adjust the MC efficiencies in the collision data analyses for the residual pile-up dependence.

In general, the mismodelling of the distributions affects cut-based and LH selections differently. For cut-based selections, a
mismodelling in MC simulation is reflected in the efficiency only if it occurs around the cut value. In the case of the LH, a
mismodelling anywhere in the distribution can affect the efficiency. 
The harder the requirement on the discriminant of the LH, the larger the effect of the differences between data and MC
distributions on the data-to-MC correction factors, as can be seen in Figures~\ref{fig:Results_MCDataoverlaysLooseTight_eteta} and \ref{fig:Results_MCDataoverlaysLooseLLHVeryTightLLH_eteta}.

\section{Identification efficiency for background processes}
\label{sec:Rejection}

The three main categories of electron background (in descending order of abundance after electron reconstruction) are light-flavour
hadrons, electrons from conversions and Dalitz decays (referred to as background electrons in the following), and non-isolated electrons from heavy-flavour decays. 
The background efficiencies of the different identification selections were studied using both MC simulation and data.

\subsection{Background efficiency from Monte Carlo simulation}

The efficiencies of the different identification selections for backgrounds were studied using MC simulation of  
all relevant $2\rightarrow2$ QCD processes filtered at particle level to mimic a level-1 EM
trigger requirement.
The sample is enriched in electron backgrounds, with electrons from $W$ and \Z\ decays excluded at particle level using generator-level simulation information. 
Furthermore, the sample is required to pass a set of electron and photon triggers without identification criteria, to allow better comparison with data-driven measurements.
The estimated background efficiency and the composition of the background are shown in Table~\ref{tab:bkg_composition_table1} for reconstructed electron candidates passing track quality requirements with transverse energies between 20~\GeV\ and 50~\GeV. The quoted uncertainties are statistical only.
The composition of this background-enriched  sample is categorized according to
simulation information: non-isolated electrons from heavy-flavour decays,
electrons from conversions and Dalitz decays, and hadrons. No explicit isolation requirement is applied. In analyses of collision data, the background efficiencies translate to background from multijet processes of typically $2$--$10$\% for leptonic $W$ and semileptonic $t\bar{t}$ decays, where the cut-based \tight\ identification and some moderate isolation requirements have been applied. For a typical selection for a $Z$ cross-section measurement that relies on the cut-based \medium\ identification, the multijet background is below 0.5\% in the $Z$ mass peak region.  

\def\zee{\ensuremath{Z\rightarrow ee}}

\begin{table}[!h]
\begin{center}
{\scriptsize
\begin{tabular}{|l|c|c|c|c|c|c|c|c|c|}
\hline
  \multicolumn{9}{|c|}{$20$ \GeV $<\et<50$ \GeV} \\\hline
Selection    & Data efficiency [\%] &  MC efficiency [\%] & \multicolumn{3}{c|}{Background composition [\%]} & \multicolumn{3}{c|}{MC efficiency [\%]} \\   
     & \zee\ signal &  Background & \multicolumn{3}{c|}{} & \multicolumn{3}{c|}{for background categories} \\   
     &  (prompt iso $e$) & (prompt $e$ excluded) & non-iso $e$ & bkg $e$ & hadron & non-iso $e$ & bkg $e$ & hadron \\\hline
Track quality        & 100            & 100         & \hphantom{0}1.1 & 16.1 & 82.8   & 100             & 100          & 100                \\\hline
\loose\ requirements         & 95.7 $\pm$ 0.2 & 4.76 $\pm$ 0.04 & \hphantom{0}7.4 & 48.4 & 44.2   & 32.5$\pm$0.8 & 14.3$\pm$0.2 & 2.54$\pm$0.03  \\\hline
\multilepton\ requirements   & 92.9 $\pm$ 0.2 & 1.64 $\pm$ 0.02 & 22.5 & 34.5 & 43.0  & 34.2$\pm$0.8 & \hphantom{0}3.51$\pm$0.08 & 0.85$\pm$0.02   \\\hline
\medium\ requirements        & 88.1 $\pm$ 0.2 & 1.11 $\pm$ 0.02 & 25.8 & 50.5 & 23.7  & 26.5$\pm$0.8 & \hphantom{0}3.46$\pm$0.08 & 0.32$\pm$0.01   \\\hline
\tight\ requirements         & 77.5 $\pm$ 0.2 & 0.46 $\pm$ 0.01 & 54.5 & 29.9 & 15.6  & 23.0$\pm$0.7 & \hphantom{0}0.85$\pm$0.04 & 0.086$\pm$0.006   \\\hline
\looseLLH          & 92.8 $\pm$ 0.2 & 0.94 $\pm$ 0.02 & 40.2 & 42.0 & 17.9 & 34.8$\pm$0.8 & \hphantom{0}2.44$\pm$0.07 & 0.20$\pm$0.01   \\\hline
\mediumLLH         & 87.8 $\pm$ 0.3 & 0.51 $\pm$ 0.01 & 48.8 & 40.6 & 10.7 & 23.1$\pm$0.7 & \hphantom{0}1.29$\pm$0.05 & 0.066$\pm$0.005   \\\hline
\veryTightLLH     & 77.0 $\pm$ 0.3 & 0.29 $\pm$ 0.01 & 63.7 & 28.9 & \hphantom{0}7.4  & 16.9$\pm$0.7 & \hphantom{0}0.51$\pm$0.03 & 0.026$\pm$0.003   \\\hline

\end{tabular}
}
\vskip0.5truecm
\caption{
Background efficiency of different identification selections taken from a MC simulation  containing all relevant
$2\rightarrow2$ QCD processes. The reconstructed electron candidates are required to have transverse energies between 20~\GeV\ and 50~\GeV\ and electrons from $W$ and \Z\ decays are removed at particle level. 
Furthermore, the sample is required to pass a set of electron and photon triggers without identification criteria, to allow better comparison with data-driven measurements.
The composition of the sample is categorized according to
MC simulation information: non-isolated electrons from
heavy-flavour decays, background electrons from photon conversions and Dalitz decays, and hadrons.
The background efficiency for each category is also quoted.
The efficiency
is always quoted with respect to reconstructed electron candidates passing the track quality requirement.
For completeness, the isolated electron efficiency for \zee\ decays, measured from data, is also given. 
The uncertainties are statistical only.
}
\label{tab:bkg_composition_table1}
\end{center}
\end{table} 

After applying the looser cut-based selections, the background generally consists of hadrons and background
electrons in similar fractions, with a small contribution of electrons from heavy-flavour decays. As the cut-based selections
get tighter, heavy-flavour decays begin to dominate the remaining background, followed by background
electrons.
In contrast, the \looseLLH\ selection retains significantly less hadronic background than its cut-based
counterpart; instead, non-isolated and background electrons dominate in this regime. After the \veryTightLLH\
selection, hadrons are highly suppressed and the sample is dominated by non-isolated electrons. To suppress these further, in many analyses isolation and tighter impact parameter requirements are added to the electron identification selection.

To estimate absolute background efficiencies, it is necessary to determine the efficiency for background objects to pass the denominator requirement of the relative efficiencies listed in Table~\ref{tab:bkg_composition_table1}. 
An unfiltered MC sample consisting of minimum-bias, single- and double-diffractive events is used. The numerator consists of reconstructed electron candidates passing the trigger and track quality requirements with transverse electron energy $\et>20$~\GeV. The denominator is defined as the numerator plus any object reconstructed as a hadronic jet using the anti-$k_t$ jet reconstruction algorithm \cite{antikt}, with a radius parameter $R =$ 0.4, and transverse jet energy $E_{\mathrm{T},\mathrm{jet}}>20$~\GeV. 
Jets overlapping with reconstructed electron candidates within a $\Delta R$ of 0.4 are removed to prevent double-counting. Reconstructed objects
matched to simulated electrons from $W$ and $Z$ decays are also removed from the calculation. Using this methodology, it is found that
$8.89\%\pm0.16\%$ (stat.) of the simulated jets built from hadrons, photon conversions or heavy-flavour decays are reconstructed as electrons with \et\ $>$ 20~\GeV\ and pass trigger 
and track quality requirements.
The efficiencies in Table~\ref{tab:bkg_composition_table1} can be multiplied by this number to obtain absolute background efficiencies for jets with \et\ $>$ 20~\GeV.

\subsection{Background efficiency ratios measured from collision data}
Studying the electron backgrounds in MC simulation can give an approximate estimate of the background efficiency. However, the description of the MC simulation has several limitations:
misidentification efficiencies depend on the tails of the distributions of many discriminating variables, which are typically more susceptible to
mismodelling than the core of the distribution. Furthermore, a small deviation in shape can lead to a large data-to-MC 
efficiency correction factor due to the low fraction of candidates in the tails. 
A data-driven estimate of the background efficiency is therefore essential. In this section, the ratio of background efficiencies from cut-based and LH menus is determined using data.

An inclusive background sample is selected by a set of electron and photon triggers with different \ET\ thresholds
and no identification requirement.
To prevent contamination from isolated electrons from $W$ and \Z\ decays,
the reconstructed electron candidate (matched to the trigger electron) is rejected if it forms a pair with an invariant mass of $40$--$140$~\GeV\ with an electron candidate
passing the \medium\ requirement.
Likewise, the electron candidate is also rejected if there is significant missing transverse momentum in the event
($\met > 25$~\GeV \footnote{The \met\ is the magnitude of the negative vectorial sum of the transverse momenta from calibrated objects, such as identified electrons, muons, photons, hadronic decays of tau leptons, and jets. Clusters of calorimeter cells not associated with any object are also included.}), or if the transverse mass calculated using \met\ is compatible 
with $W$-boson production ($m_\mathrm{T}>40$~\GeV). 
In order to remove the residual true electron contamination, these kinematic requirements are furthermore applied to simulated \Zee\ and \Wen\ samples; the surviving events are scaled to the corresponding integrated luminosity and subtracted from the data yields before the background efficiency calculation.

\begin{figure}[!ht]
\centering
\includegraphics[width=0.49\textwidth]{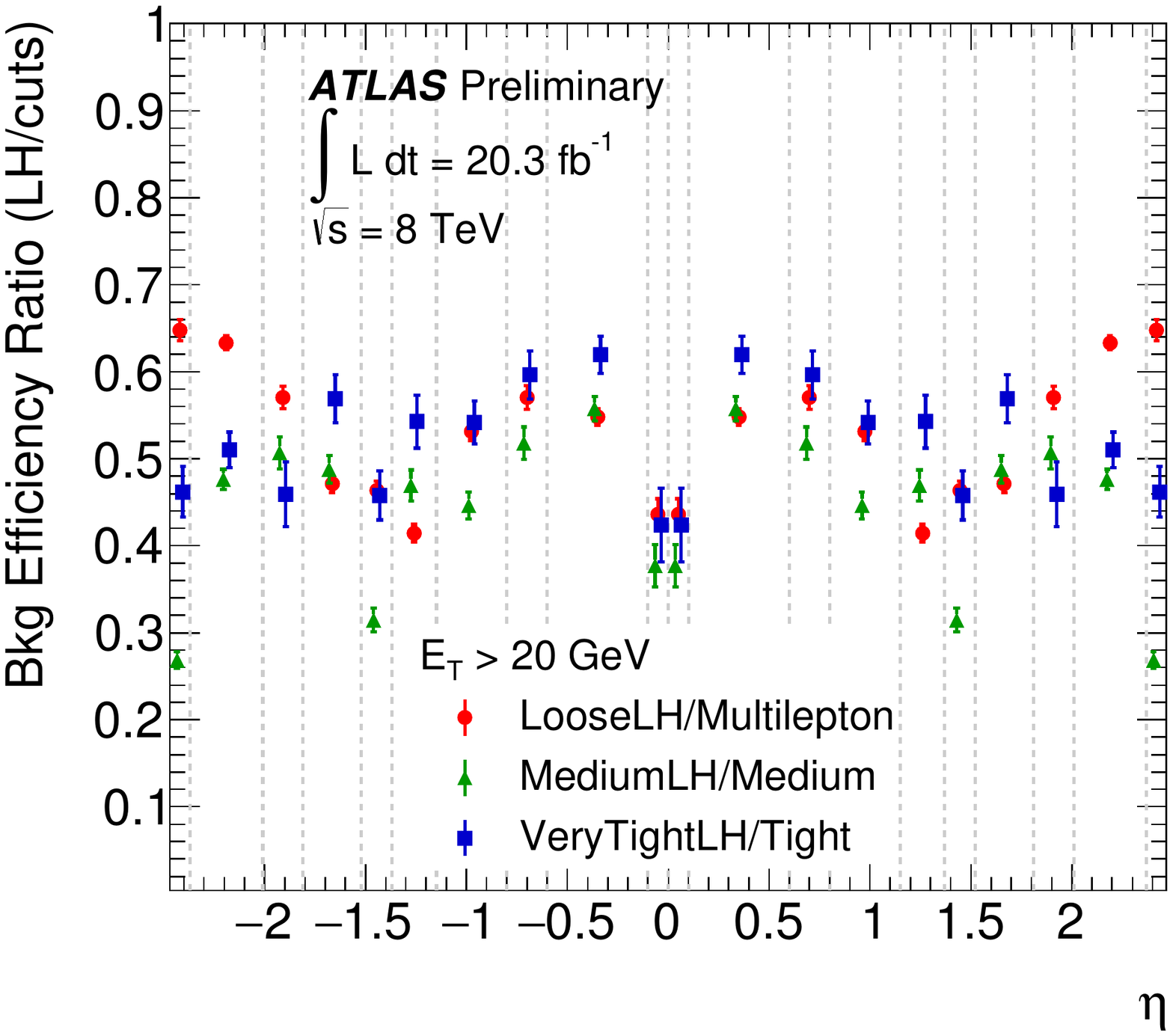}
\includegraphics[width=0.49\textwidth]{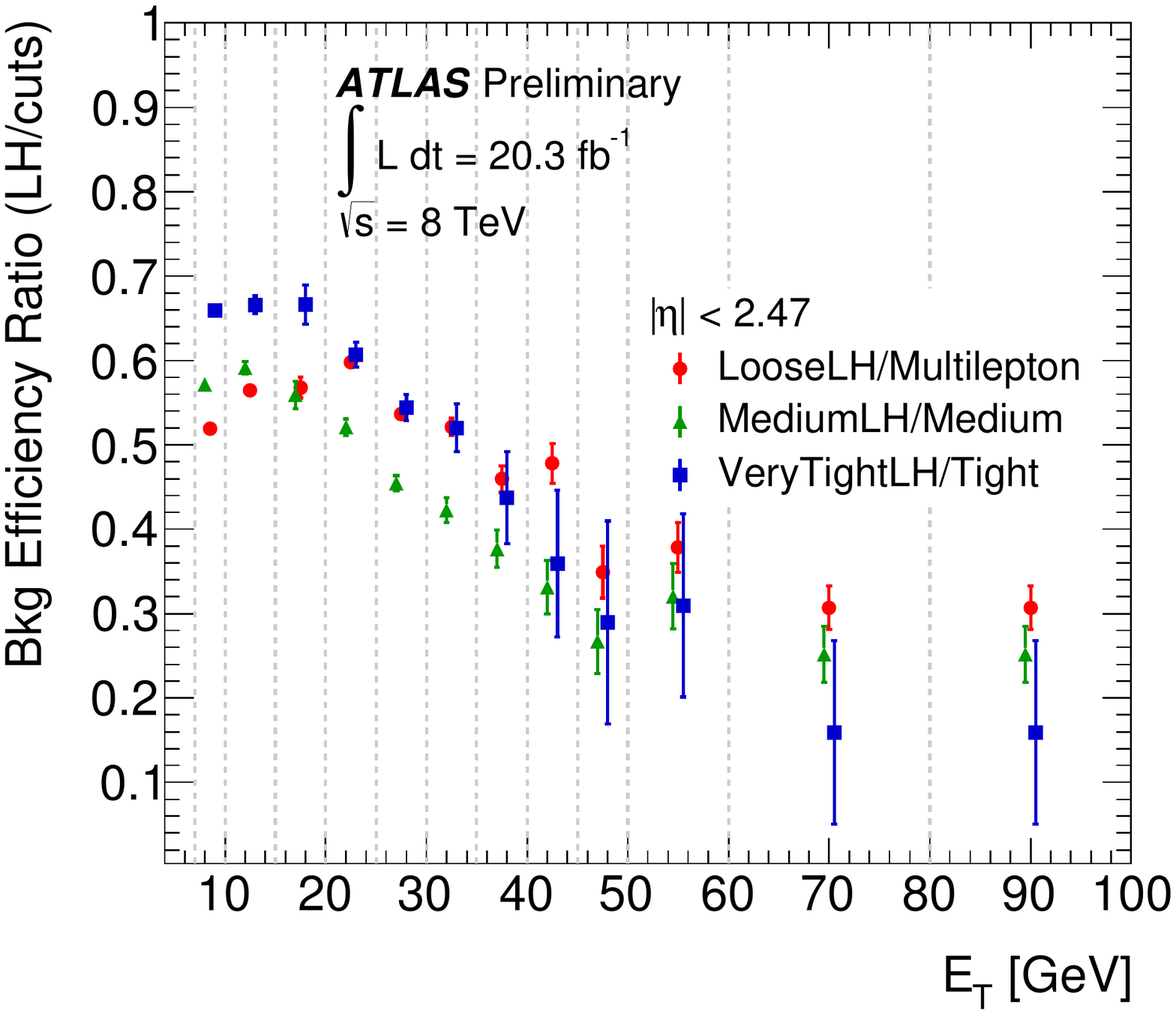}
\caption{  Ratio of background efficiencies for a LH to that of 
the closest-efficiency cut-based selections as a function of $\eta$ (left) and \et\ (right), 
as obtained using an inclusive background sample (see text). 
The uncertainties are statistical as well as systematic: a systematic uncertainty of 21\% is assigned to the subtraction of signal events using the simulation; this uncertainty is dominated by the mismodelling of the missing transverse momentum.}
\label{fig:RejectionRatio1}
\end{figure}

The background sample is dominated by light-flavour hadrons, followed by 
photon conversions and a small fraction of heavy-flavour decays. The ratio of the background efficiency for a LH to that for 
the closest-efficiency cut-based selection is shown in Figure~\ref{fig:RejectionRatio1}. It can be seen that the LH selections let through only about $40$--$60$\% of the background
compared to the cut-based selections, while it is shown in Section~\ref{sec:main_id_results} that they retain approximately the same signal electron efficiency.
These results cannot be directly compared to those derived from MC simulation 
and given in Table~\ref{tab:bkg_composition_table1}, as the composition of the samples might differ.
Nonetheless, the data-driven and MC-based estimates show the same trend when comparing the background rejection of cut-based and LH selections.

\section{Determination of the charge misidentification probability}
\label{sec:MisID}

Charge misidentification occurs if an isolated prompt electron is reconstructed with a wrong charge assignment. The misidentification is mostly caused by the emission of bremsstrahlung at a small angle with a subsequent conversion of the emitted photon and the mismatching of one of the conversion tracks to the cluster of the original electron. In addition, for high \et\ and therefore increasingly straight tracks, charge misidentification can be caused by a failure to correctly determine the curvature of the track matched to the electron. 
For electrons with transverse energies of $\et<300$~\GeV, the causes of charge misidentification are predominantly conversions combined with inefficiencies in matching the correct track to the electron.

Various physics analyses such as measurements of same-sign $WW$ scattering~\cite{Aad:2014zda} or $Z$ polarization~\cite{Aad:2016izn} as well as searches for supersymmetry in final states with two same-sign leptons~\cite{Aad:2014pda} rely on correct charge assignment. Therefore the measurement of the charge misidentification rate and its description in MC simulation is crucial.

In the range of \et\ for which the $Z$ decays yield a sufficiently large sample, and which is used by most analyses, the charge misidentification probability is dominated by material effects, rather than the precision of the measurement of the track curvature, as studies using MC simulation have shown. Therefore the charge misidentification rate is determined as a function of $\eta$ rather than \et\ using electrons with \et\ greater than 15~\GeV. 

\begin{figure}
\center
\includegraphics[scale=0.4]{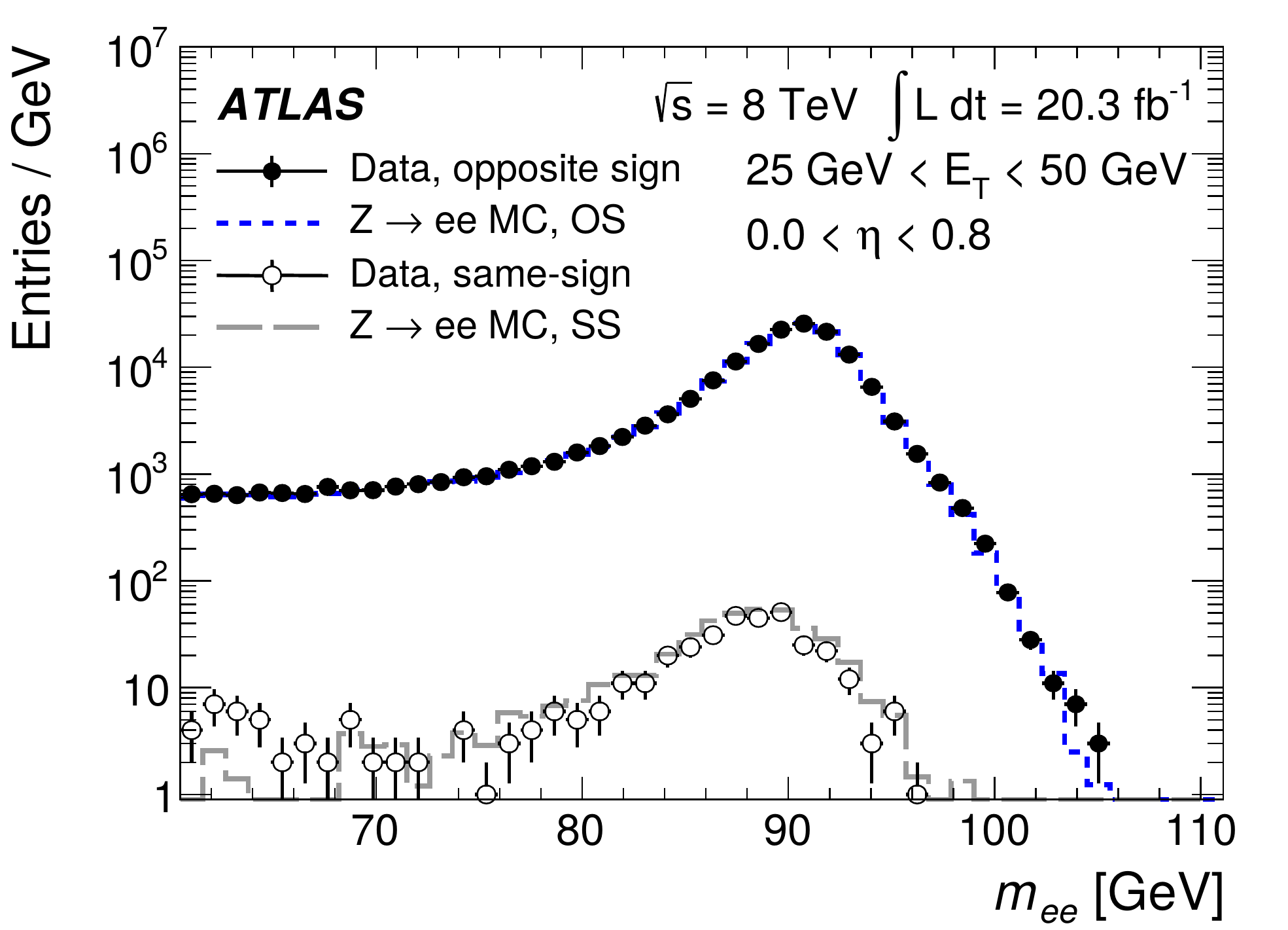}
\includegraphics[scale=0.4]{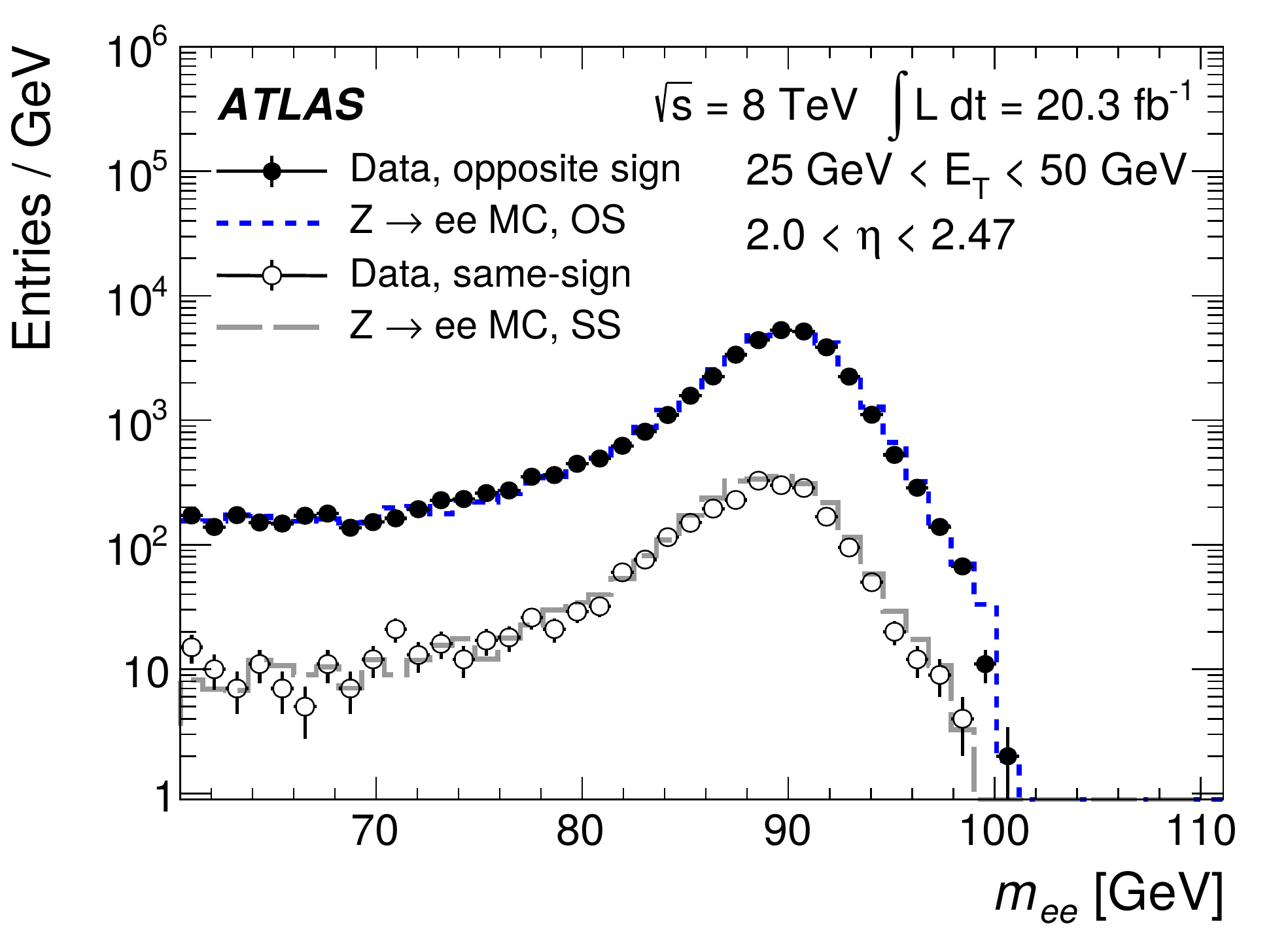}      
\caption{Distribution of the invariant mass \mee\ of the selected opposite-sign (OS) or same-sign (SS) electron pairs in data and MC simulation for 25~\GeV~$<$~\et~$<$~50~\GeV\ in the 0.0~$<$~\eta~$<$~0.8~bin (left) and in the 2.0~$<$~\eta~$<$~2.47~bin (right). Tag and probe electron candidates are required to pass the cut-based \tight\ identification and a track isolation requirement of $p_\mathrm{T}^\mathrm{cone0.2}/E_\mathrm{T} < 0.14$.}
\label{fig:onezcf}
\end{figure}

The event selection described in Section~\ref{sec:eventsel_ZTPID} is applied to select a sample of \Zee\ events, except for the opposite-charge requirement. Additionally, both the tag and probe electron candidates are required to satisfy certain identification and isolation criteria. Figure~\ref{fig:onezcf} shows the \mee\ distribution of the selected OS and SS electron pairs for a representative selection. 

The probability for an electron to be charge misidentified in a certain bin $i$ in \eta\ and \et\ is referred to as $\epsilon_i$. The probabilities $\epsilon_i$ in the different regions are statistically independent. The average number of SS events $N^{\mathrm{SS}}_{ij}$ that is expected for a pair of electrons in the bins $i$ and $j$ follows from the number of total events $N^{\mathrm{OS+SS}}_{ij}$, where no charge requirement is applied, using the respective charge misidentification probabilities $\epsilon_{i,j}$ as:

\begin{equation}
N^{\mathrm{SS}}_{ij} = N^{\mathrm{OS+SS}}_{ij}[(1-\epsilon_i)\epsilon_j + (1-\epsilon_j)\epsilon_i].\label{eq:nevents}
\end{equation}

$N^{\mathrm{SS+OS}}_{ij}$ is taken from data after background subtraction. A likelihood function can be constructed using a Poissonian approximation of the probability to observe a specific number of SS events $n^{\mathrm{SS,obs}}_{ij}$ in data if the electrons are reconstructed in the bins $i$ and $j$:

\begin{equation}
L = \prod_{i,j}L_{ij}  =\prod_{i,j} \frac{(N^{\mathrm{SS}}_{ij} + N^{\mathrm{SS, bkg}}_{ij})^{n^\mathrm{SS,obs}_{ij}}\times e^{N^{\mathrm{SS}}_{ij}+ N^{\mathrm{SS, bkg}}_{ij}}}{
n^\mathrm{SS,obs}_{ij}}!\label{eq:ssposs}
\end{equation}

The likelihood function is maximized to estimate the charge misidentification probabilities $\epsilon_i$ in each bin $i$. 

As in the other efficiency measurements, the backgrounds originate from hadronic jets as well as from photon conversions, Dalitz decays and semileptonic heavy-flavour hadron decays. The backgrounds for total and same-sign candidate events are estimated by extrapolating linearly the number of events from equally sized sidebands of the invariant-mass distributions above and below the $Z$ mass peak to the signal region. 
As an estimate of the uncertainties, the measurement is performed by varying the invariant-mass window from 15~\GeV\ to 10~\GeV\ and 20~\GeV\ around the \Z\ mass, the width of the sidebands used in the background subtraction is changed to be 20, 25, or 30~\GeV. All variations have very small effects on the measured rates. The average value of these variations is taken as the measured value, the RMS as the systematic uncertainty. The uncertainty returned by the minimization is accounted for as a statistical uncertainty. 

The charge misidentification rate is determined for three representative sets of requirements applied in analyses:
\begin{itemize}
 \item \textbf{\Medium}: \medium\ identification requirements.
 \item \textbf{\Tight\ + isolation}: \tight\ identification requirements combined with selection criteria for the track isolation of $p_\mathrm{T}^\mathrm{cone0.2}/E_\mathrm{T} < 0.14$.
 \item \textbf{\Tight\ + isolation + impact parameter}: \tight\ identification combined with calorimetric and track isolation criteria of $E_\mathrm{T}^\mathrm{cone0.3}/E_\mathrm{T} < 0.14$ and $p_\mathrm{T}^\mathrm{cone0.2}/E_\mathrm{T} < 0.07$ and in addition requirements on the track impact parameters of $|z_0| \times \sin\theta <0.5$ mm and $|d_0|/\sigma_{d_0}<$ 5.0.
\end{itemize}
  
\begin{figure}
\center
\includegraphics[scale=0.6]{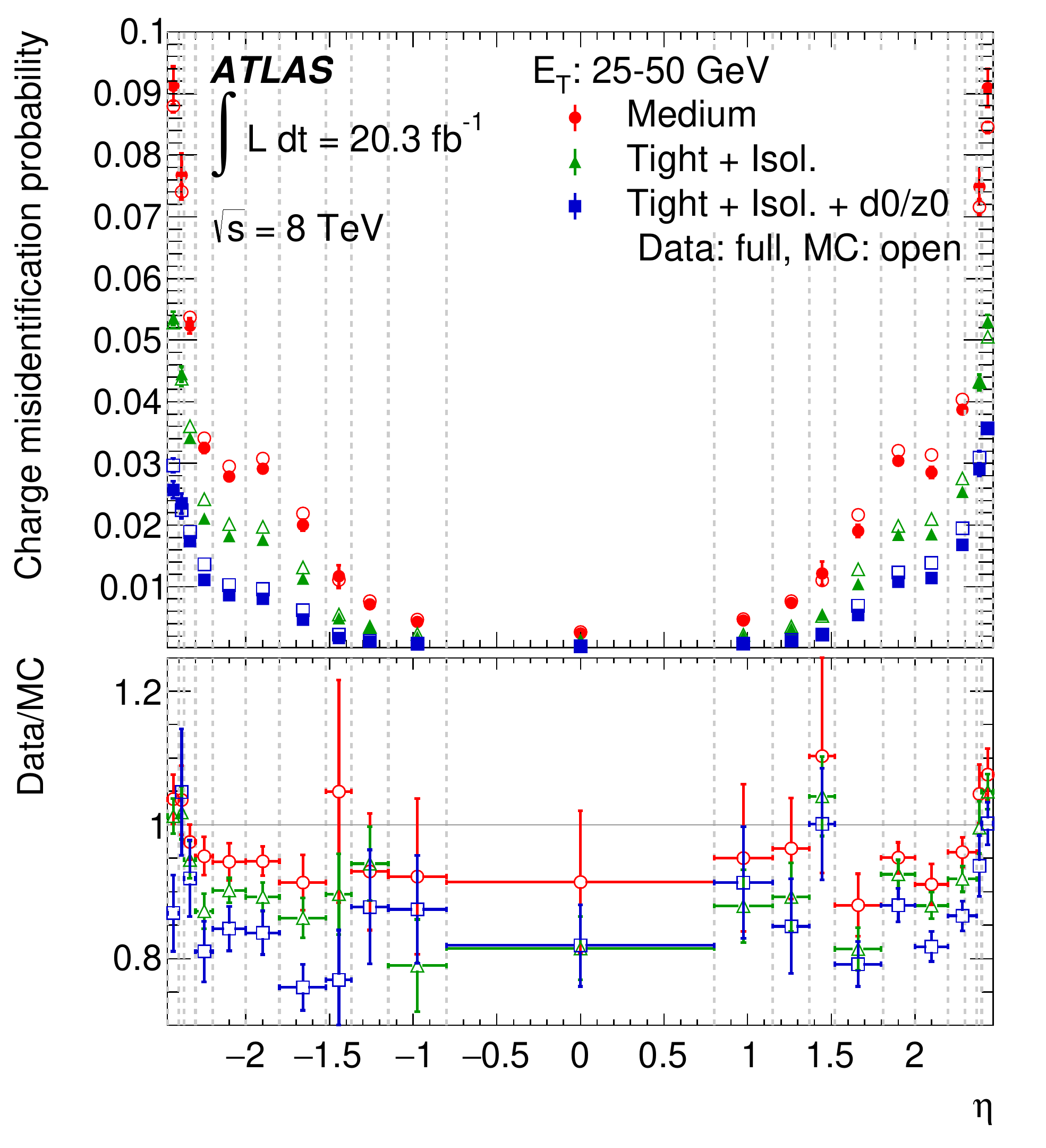}
\caption{
Charge misidentification probability in data as a function of $\eta$ for three different sets of selection requirements (\Medium, \Tight\ + Isolation and \Tight\ + Isolation + impact parameter), compared to the expectation of the MC simulation as measured on a sample of electron pairs from \Zee\ decays. The lower panel shows the data-to-MC charge misidentification probability ratios. The uncertainties are the total uncertainties from the sum in quadrature of statistical and systematic uncertainties. The dashed lines indicate the bins in which the efficiencies are calculated.
}
\label{fig:chargemis}
\end{figure}

Figure~\ref{fig:chargemis} shows the charge misidentification probability for the three working points as determined by the measurement in data and MC simulation. Since the charge misidentification probability is correlated with the amount of bremsstrahlung and thus with the amount of traversed material, the probabilities are quite low in the central region of the detector but can reach almost 10\% for high values of $|\eta|$. The energy in a cone around the electron can be indicative of energy deposited by bremsstrahlung. Equally, large values of the track impact parameters can mean that the track matched to the electron is not a prompt track from the primary vertex but from a secondary interaction or bremsstrahlung and a subsequent conversion. Thus, tighter selection criteria, in particular requirements on the isolation or track parameters, can decrease the charge misidentification probability by a factor of up to four, depending on the additional selection requirements.

\section{Reconstruction efficiency measurement}\label{sec:MeasureReco}

\subsection{\Tnp\ with \Zee\ events}
 Electrons are reconstructed from EM clusters that are matched to
tracks in the ID, as described in Section~\ref{sec:Reconstruction}.  
The tracks are required to satisfy the track quality criteria, i.e. to 
have at least one hit in the pixel detector and in total at least seven
hits in the pixel and SCT detectors. 
The measurement of the efficiency to detect an energy cluster in the EM calorimeter using the
sliding-window algorithm is very challenging in data and not performed here. In MC simulation, it is found to be
above 99\% for \et~$>$~15~\GeV\ as discussed in Section~\ref{sec:Reconstruction}. EM clusters form the starting point of the reconstruction efficiency measurement. 

The reconstruction efficiency is defined as the ratio of the number of electrons reconstructed as a 
cluster matched to a track satisfying the track quality criteria (numerator) to the number of clusters with or without a matching track (denominator). This reconstruction efficiency is measured using a \tnp\ analysis
which is very similar to the \Zmass\ method introduced in Section~\ref{sec:MeasureID}.  
 In comparison to the measurement of the identification efficiency, the probe definition is relaxed to include all EM clusters. 
The background estimation is adapted to include the contribution
of EM clusters with no associated track.
The measurement is only performed for probe electron candidates with \et~$>$~15~\GeV, as the background contamination of the sample becomes too high at lower \et.

\subsubsection{Event selection}

The general event selection as well as the criteria for the tag electron are identical to the ones used in the \Zmass\ method,
described in Section~\ref{sec:eventsel_ZTPID}.

Each event is required to have at least
one tag electron candidate and one probe, which in this case is an EM cluster.   
In order to veto EM clusters from converted photons, 
no other cluster within $\Delta R$~$=$~0.4 of a reconstructed electron candidate is considered. 
No requirement on the charge of the tag and the probe electron candidates 
is applied, since there is no charge associated with EM clusters unless they are matched to a track.

\subsubsection{Background estimation and variations for assessing the systematic uncertainties}

The background estimation for the numerator of the reconstruction efficiency (electrons passing the reconstruction requirements) follows that of the \Zmass\ method described in
Section~\ref{sec:Zmass_bkgsub}. However, for the denominator (all reconstructed EM clusters) an additional contribution from photon candidates must be determined separately.
The total background at the denominator level is the sum of two contributions: background to electrons reconstructed as a cluster with and without an associated track. The background estimation for these two contributions is explained below.

\paragraph{Background estimate for electrons reconstructed as clusters with no associated track}
\label{sec:Reco_Bkg_notrk}

\begin{figure}[h!]
\centering
\includegraphics[width=0.8\columnwidth]{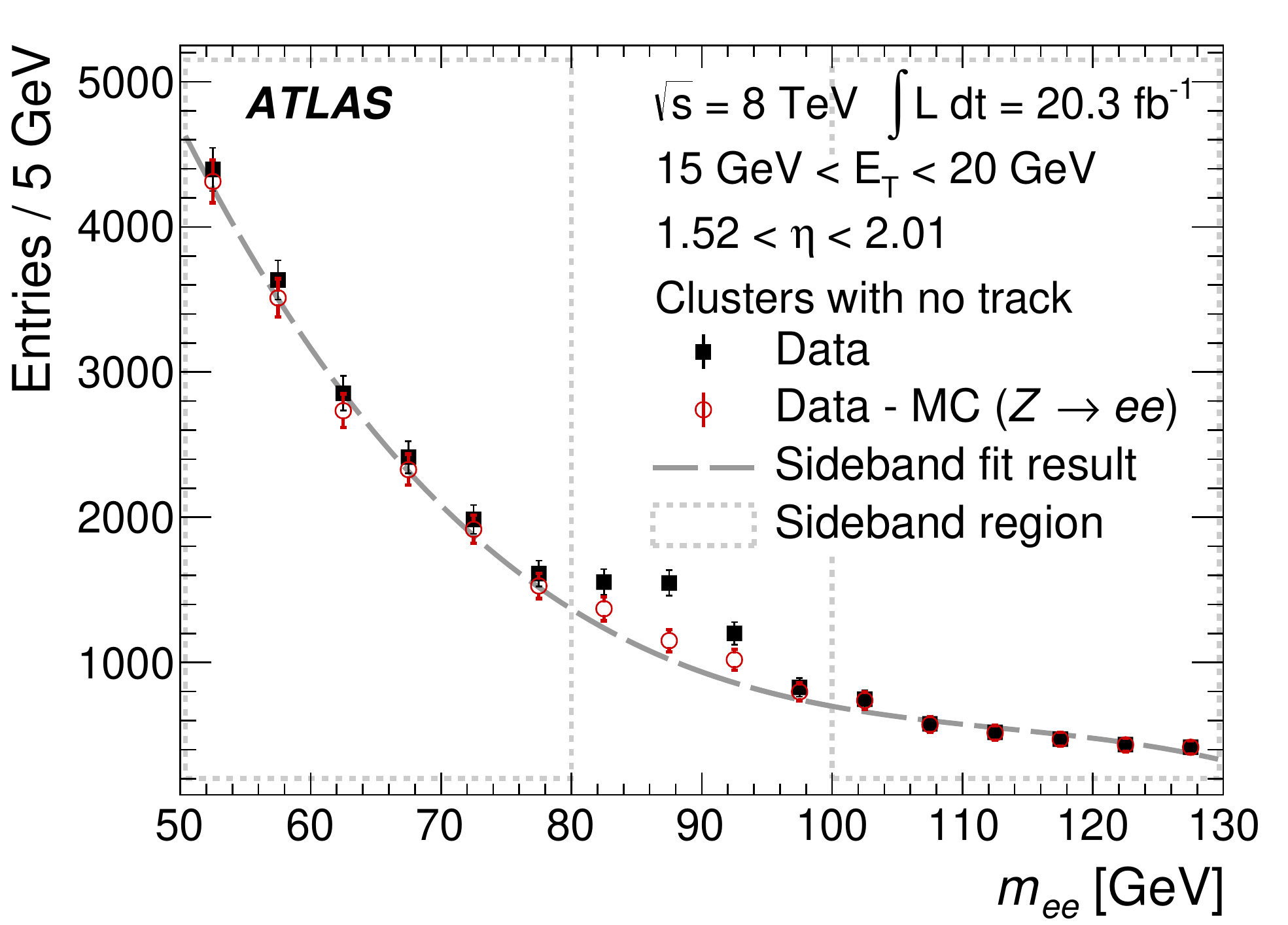} 
\caption{Estimate of the background to the selected EM clusters with no associated track 
for 15~\GeV~$<$~\ET~$<$~20~\GeV\ and 1.52~$<$~\eta~$<$~2.01. 
A polynomial fit (shown by a dashed dark grey line) is carried out in the sideband region (indicated by dashed light grey boxes) 
of the invariant-mass distribution of data events from which genuine electrons have been subtracted using MC simulation
(the data are shown by filled squares before the subtraction of the prediction of the MC simulation and by open circles afterwards). 
In the signal region, defined as the events with an invariant mass of $80$~\GeV\ to $100$~\GeV, the fit result is used to obtain a data-driven estimate, which is compared to the data minus the prediction of the MC simulation. Only statistical uncertainties are shown for the data minus the MC prediction; the systematic uncertainty in the scaling of the MC simulation and description of the MC simulation of the inefficiency to match an electron with a track is $10$--$20$\%. 
}
\label{fig:main_ZeeRecoPhotonFit}
\end{figure}

Electrons reconstructed as EM clusters but not matched to any track are 
interpreted as photons. In
order to estimate the photon background, which, unlike the signal electrons, has a smoothly falling invariant-mass shape, a
third-order polynomial is fitted to the invariant-mass distribution of the selected electron--photon pairs (corresponding
to the tag and the probe electron candidates). The fit is carried out using the two sideband regions above and below the \Z\ mass peak, as
illustrated in Figure~\ref{fig:main_ZeeRecoPhotonFit}. Residual
signal electron contamination in the background-dominated sideband regions is subtracted using MC simulation before the fit.
Systematic uncertainties in the scaling of the MC simulation and description of the MC simulation of the inefficiency to match an electron with a track are $10$--$20$\% and are not shown in Figure~\ref{fig:main_ZeeRecoPhotonFit}. These uncertainties explain the small  difference in the signal region between the data minus the MC prediction and the polynomial fit to the sidebands. The prediction of the MC simulation enters only in the subtraction of the very small residual signal in the sideband regions used to perform the polynomial fit. The resulting uncertainty in the measured reconstruction efficiency is negligible.
 
\paragraph{Background estimate for electrons reconstructed as clusters with an associated track}
\label{sec:Reco_Bkg_withtrk}

The method to estimate the background to EM clusters with an associated track is almost the same as for the
identification efficiency measurement, described in Section~\ref{sec:Zmass_bkgsub}: 
A background template is selected in data by inverting identification selection criteria for the probes and normalized to the data in a control region of the invariant-mass distribution of the \tnp\ pair. 

The backgrounds in the signal region are determined separately for clusters with tracks satisfying or not satisfying the track quality selection criteria. Therefore, the track quality selection criteria must be satisfied (not satisfied) in the background 
template selection for the
invariant-mass distribution of EM clusters passing (failing) the electron reconstruction procedure.

\begin{figure}[h!]
\centering

\includegraphics[width=0.49\textwidth]{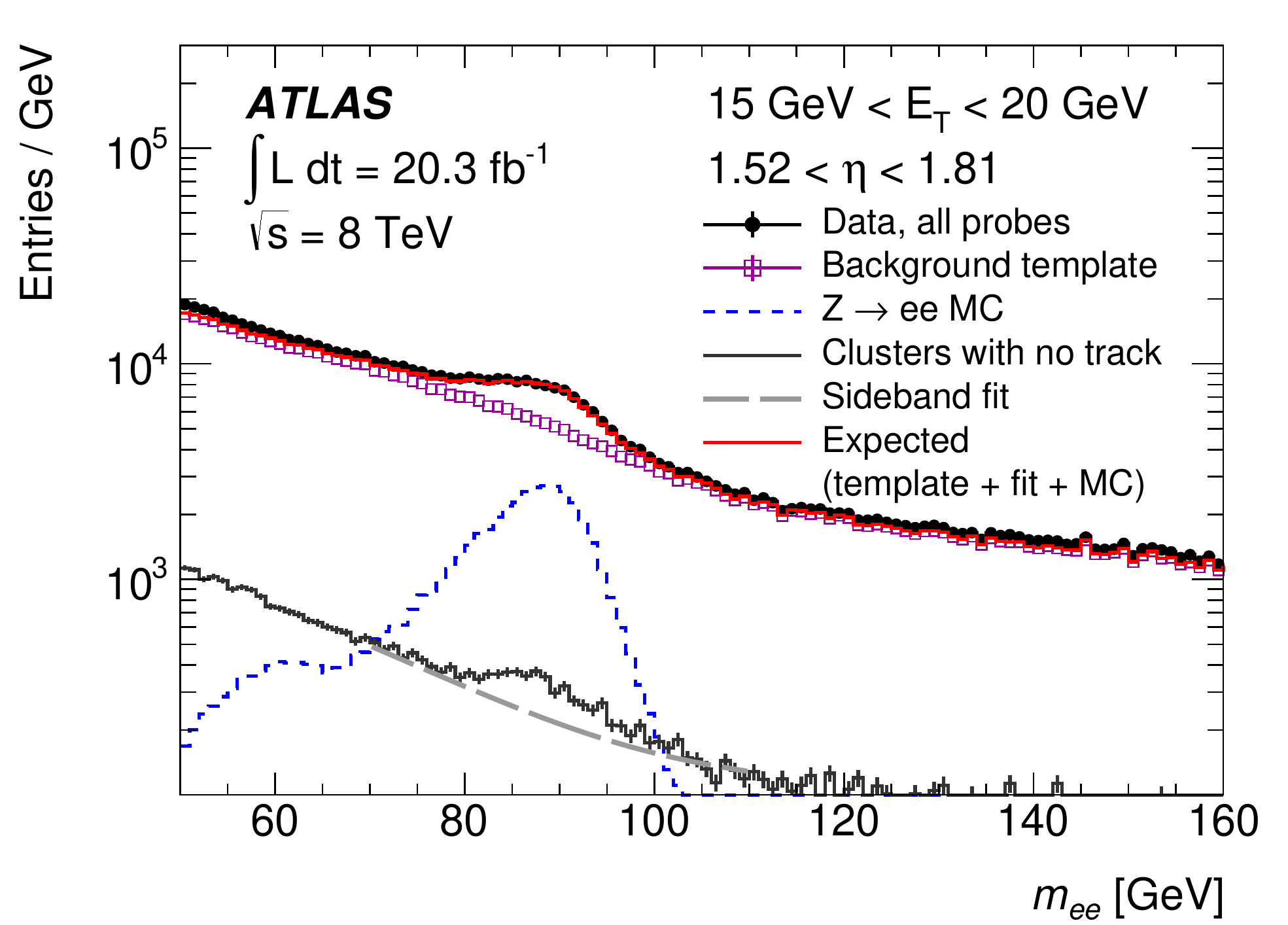}
\includegraphics[width=0.49\textwidth]{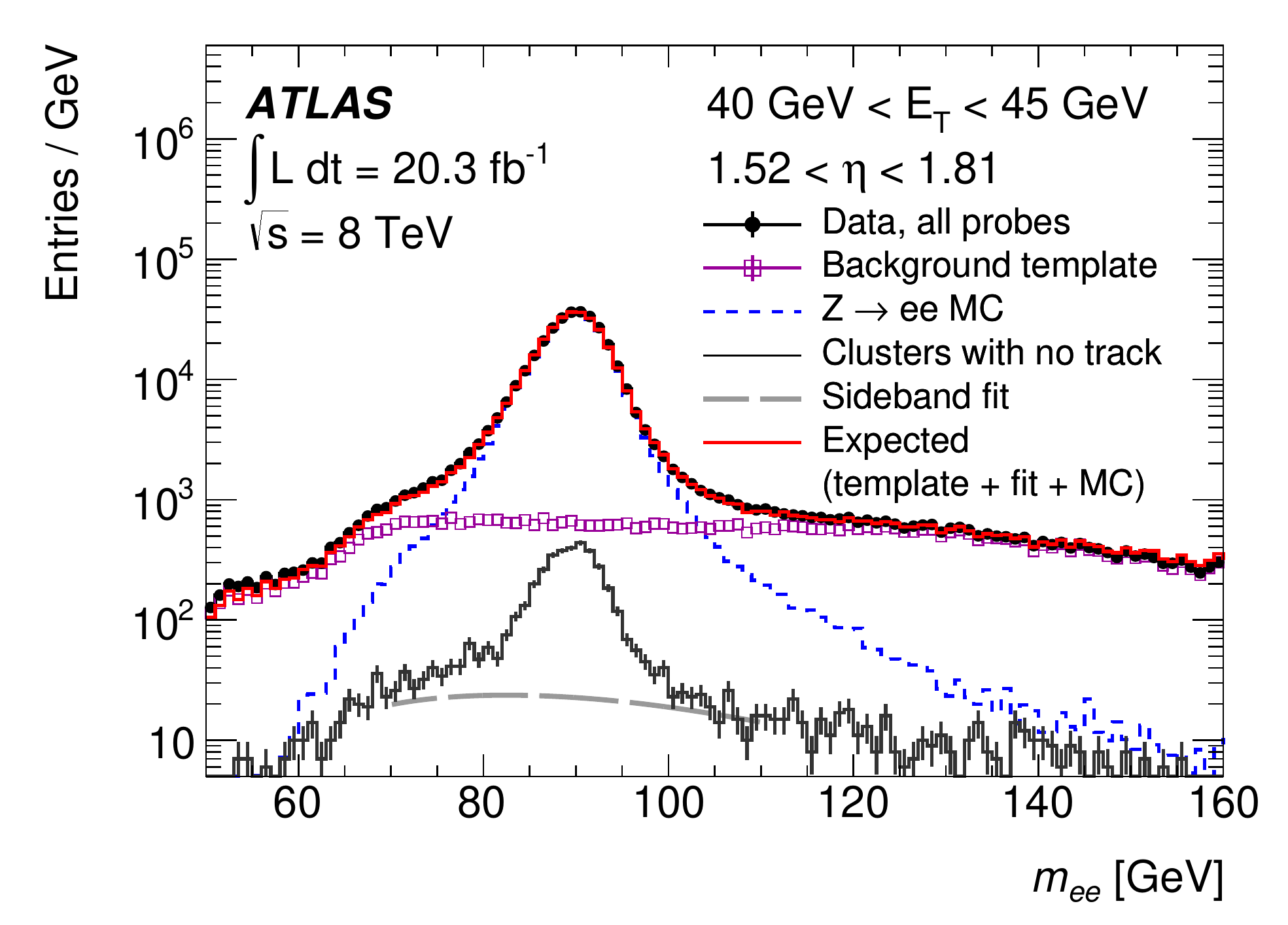}
\includegraphics[width=0.49\textwidth]{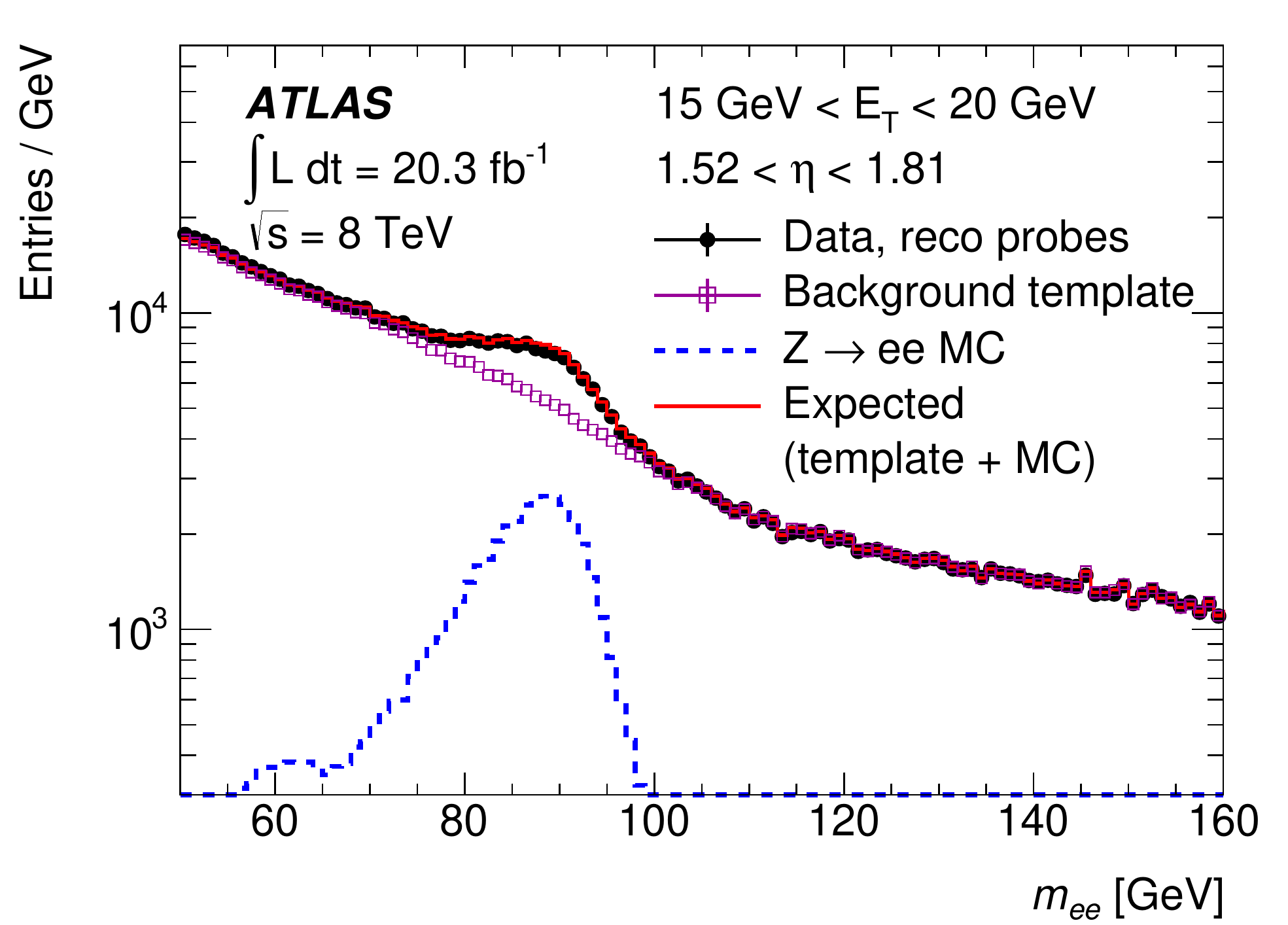}
\includegraphics[width=0.49\textwidth]{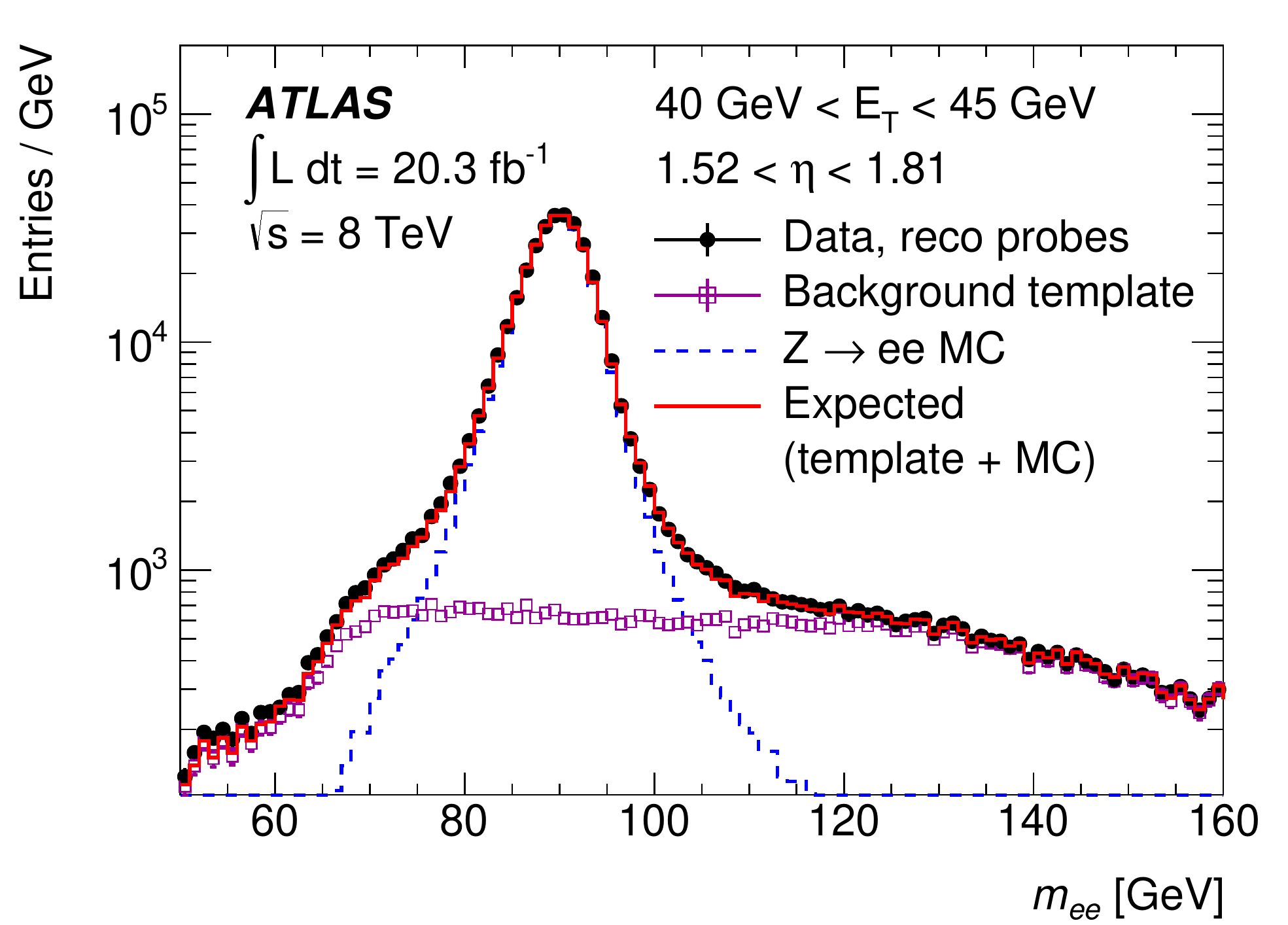}

\caption{
Invariant-mass distributions of the \tnp\ pairs for probe EM clusters with 1.52~$<$~\eta~$<$~1.81 and $15$~\GeV$ < \et < 20$~\GeV\ (left)  or $40$~\GeV$ <
\et < 45$~\GeV\ (right), before (top) and after (bottom) applying the reconstruction criteria. The data (black dots with error
bars) at the \textit{all probes} level is composed of two components: clusters with no matching track (dark grey histogram with error bars) and clusters with a matching track. The background is evaluated separately for these two components. A third-order polynomial (grey dashed line spanning the region from 70~\GeV\ to 110~\GeV) depicts the estimated photon background from a fit performed in the sideband regions as explained in Section \ref{sec:Reco_Bkg_notrk} and shown in Figure \ref{fig:main_ZeeRecoPhotonFit}. A background template normalized in this case to the high-mass tail (magenta markers) is used to estimate the background with a matching track.  
This background template is obtained by requiring some of the identification criteria not to be satisfied. 
Additionally, probes must pass or fail the track quality selection requirements depending on whether the background to the electrons passing or failing the reconstruction requirements is determined (see Section \ref{sec:Reco_Bkg_withtrk}). The shown magenta distribution is the sum of both components. For illustration only, the signal prediction of the MC simulation (blue dashed line) is also displayed. The sum of the normalized background template and the signal
prediction of the MC simulation (red line, shown for comparison but not used in the measurement) agrees well with the data points.
}
\label{fig:main_ZeeReco_InvMassDistribution}
\end{figure}

Figure~\ref{fig:main_ZeeReco_InvMassDistribution} 
shows the invariant-mass distributions of the \tnp\ pairs for probe EM clusters 
(composed of clusters with or without a track match at the probe level)
for two selected bins both at the probe level and before and after applying the reconstruction criteria to the probe electron candidate.
The estimates of the two background components are also depicted. 
As demonstrated by the figure, the measured data agree well with the prediction, and the background subtraction procedure performs well.

The systematic uncertainty is estimated as for the identification efficiency. In addition to the variations listed in Section~\ref{sec:Zmass_bkgsub}, the sidebands for the polynomial fit used for the estimation of the background to electrons without an associated track are varied among these choices: $[70,80 \GeV]$ and $[100,110 \GeV]$, $[60,80 \GeV]$ and $[100,120 \GeV]$, $[50,80 \GeV]$ and $[100,130 \GeV]$, $[55,70 \GeV]$ and $[110,125 \GeV]$. 

\subsection{Results}

\begin{figure}
\centering
\includegraphics[width=0.49\textwidth]{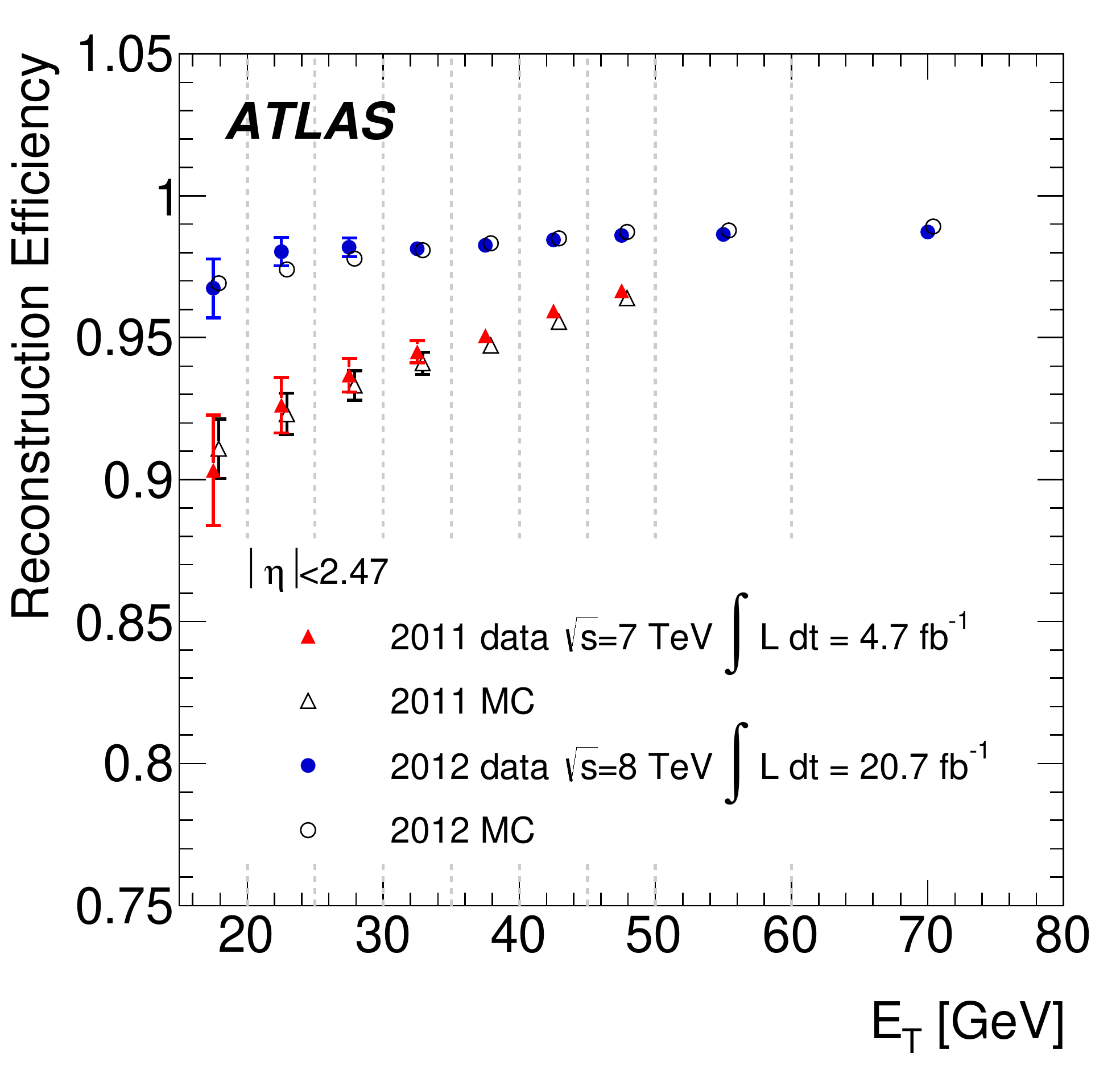}
\includegraphics[width=0.49\textwidth]{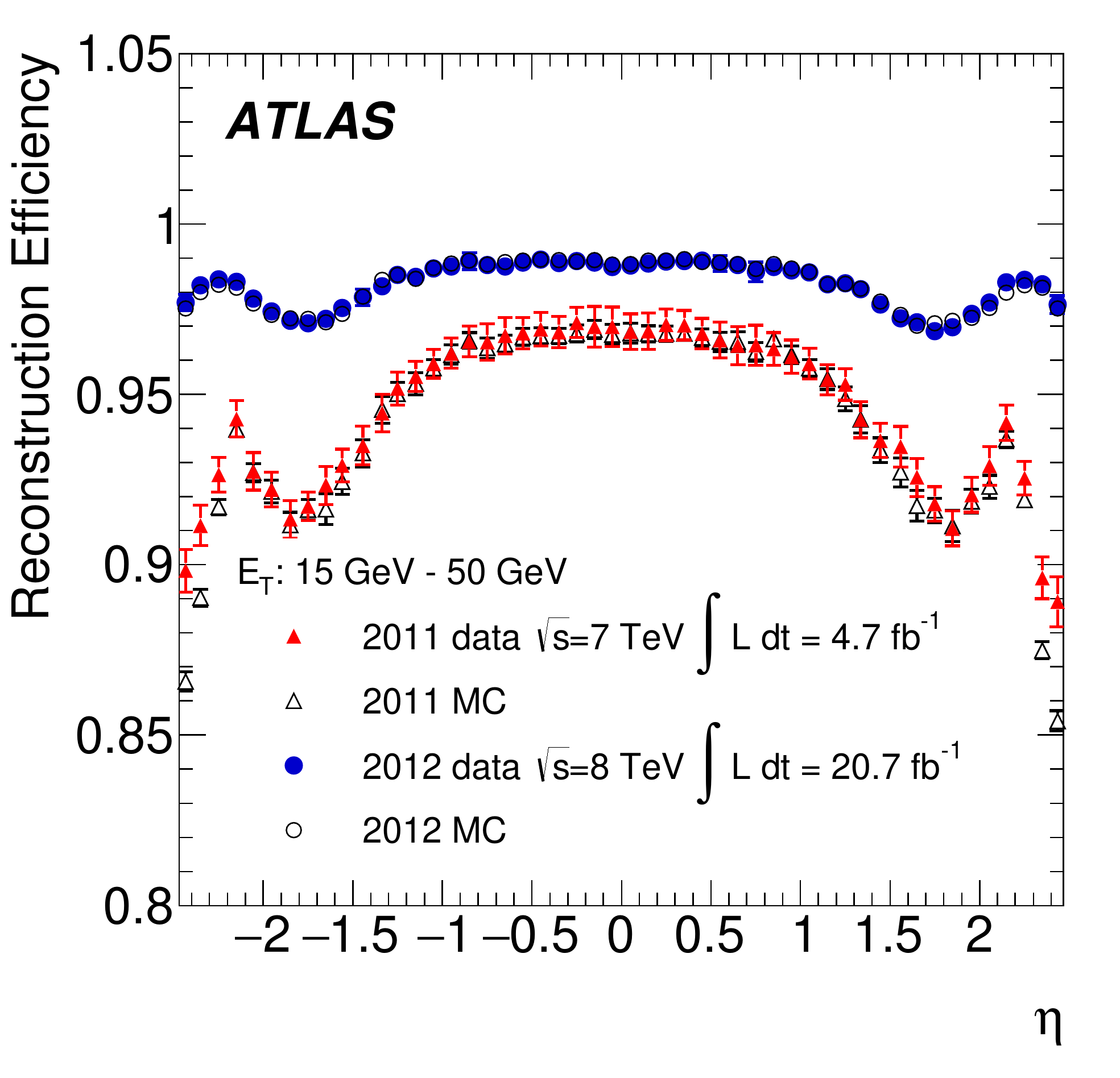}
\caption{  Measured reconstruction efficiencies as a function of \et\  integrated over the full pseudorapidity range (left) 
and as a function of $\eta$ for 15~\GeV~$<$~\et~$<$~50~\GeV\ (right) for the 2011 (triangles) and the 2012 (circles) data sets.
For illustration purposes a finer $\eta$ binning is used.  The dashed lines in the left plot indicate the bins in which the efficiencies are calculated.}
\label{fig:main_ZeeReco_Reco20112012}
\end{figure}

The reconstruction efficiency, like the identification efficiency,  is measured differentially in (\et, $\eta$) bins. The
efficiency to reconstruct an electron associated with a track of good quality varies from 95\% to 99\% between the \endcap\ and
\barrel\ regions for low-\et\ electrons (\et~$<$~20~\GeV). For very high \et\ electrons (\et~$>$~80~\GeV) the efficiency is $\sim$99\%
over the whole $\eta$ range. The results are shown in Figure~\ref{fig:main_ZeeReco_Reco20112012}, projected in \et\ and \eta. 
 The measured efficiency agrees well
with the prediction of the MC simulation. The data-to-MC correction factors are at most $1$--$2$\% different from unity and in most of the measurements they are within only a few permille of one. 
The total uncertainty is $<$~0.5\% for electrons with \et\ between 25~\GeV\ and 80~\GeV. It is larger at lower
\et, varying between 0.5\% and 2.0\%. The statistical and systematic uncertainties are of the same order. 
Good data--MC agreement observed for $\et > 15$~\GeV\ gives confidence in the description of the MC simulation of
the detector response, which is relied on for electrons with $\et < 15$~\GeV.
In this low-\et\ region, the data-to-MC correction factor is assumed to be 1.0 
with an uncertainty of 2\% in the barrel and 5\% in the endcap region.

\begin{figure}[!ht]
\centering
\includegraphics[width=0.49\textwidth]{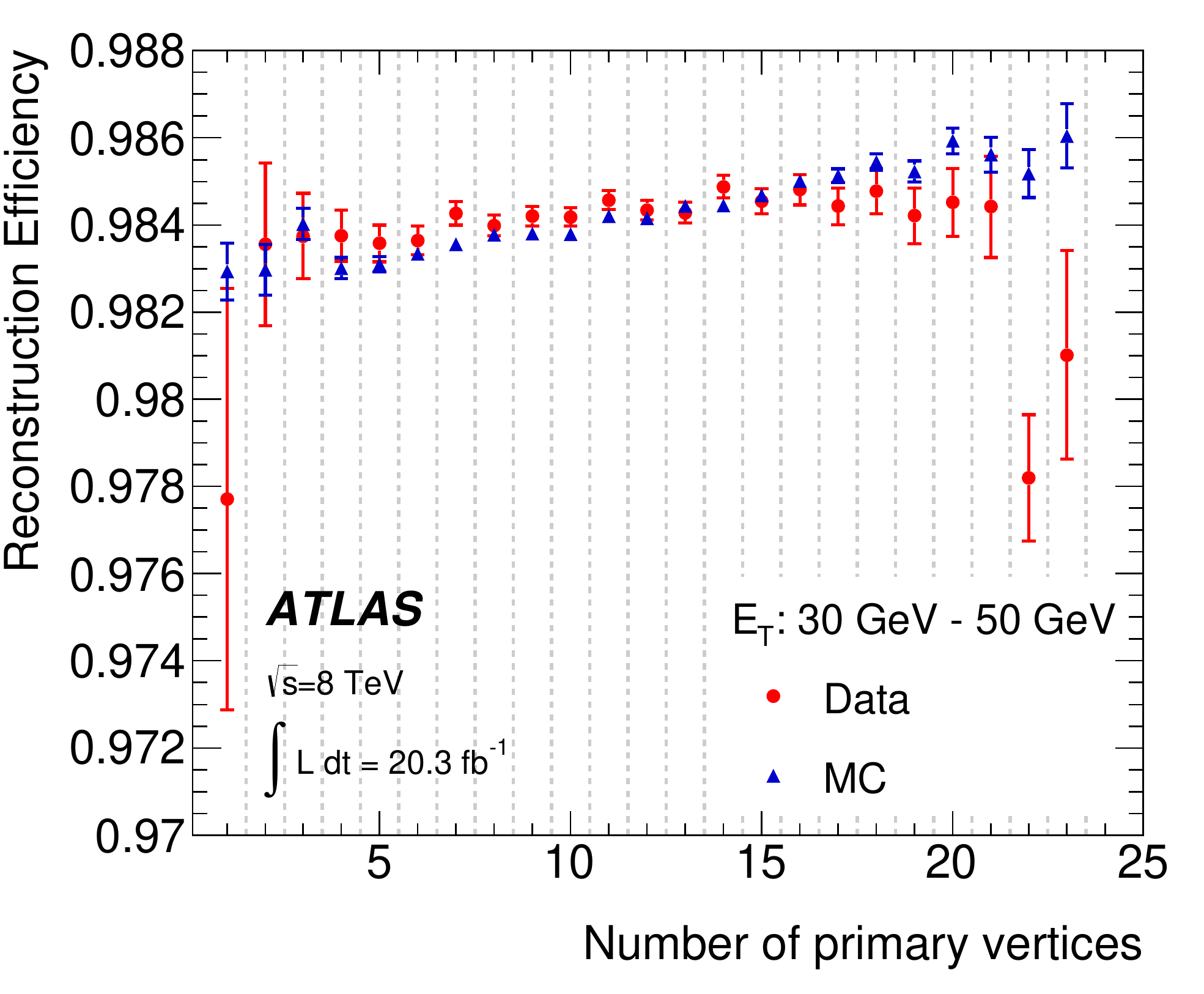}
\caption{Measured reconstruction efficiency (red circles) 
as a function of the number of reconstructed primary vertices
for 30~\GeV~$<$~\et~$<$~50~\GeV\ and integrated over $\eta$, compared to the prediction of the MC simulation (blue triangles). 
The uncertainties are 
statistical+systematic. The dashed lines indicate the bins in which the efficiencies are calculated.
}
\label{fig:main_ZeeReco_eff_nvx}
\end{figure}

As described in Section~\ref{sec:Reconstruction}, for the 2012 data, a new track reconstruction algorithm has been introduced in order to
improve the reconstruction of electrons that have undergone significant bremsstrahlung.
Figure~\ref{fig:main_ZeeReco_Reco20112012} also compares the reconstruction efficiencies measured in the 2011 and 2012 data. The new track fitting
algorithm improves the overall electron reconstruction efficiency by $\sim$5\%. Most of this improvement is in the low-\ET\ range, where the electron reconstruction efficiency increases by more than $\sim$7\%. This constitutes a significant gain for important measurements such as the determination of Higgs boson properties in the channel $H \rightarrow ZZ^* \rightarrow 4\ell$~\cite{Aad:2014eva}. 

The gain in efficiency from the new track reconstruction algorithm flattens the distribution of the reconstruction efficiency in $\eta$. For the 2011 data, a large drop in efficiency was observed for the endcap regions, where more
bremsstrahlung occurs due to a higher amount of material. For the 2012 data, this drop has become much smaller. Furthermore, the 2012
results are more precise than the final 2011 results, partly because of the increase in the size of the available data sample, 
but also due to improvements in the background subtraction method. 

The efficiencies are also measured as a function of the number of primary vertices in order to investigate the dependence of the
electron reconstruction on pile-up. Figure~\ref{fig:main_ZeeReco_eff_nvx} shows that for data, the reconstruction efficiency for electrons
with \et~$>$~30~\GeV\ does not change with the number of primary vertices.

\section{Combined reconstruction and identification efficiencies}
\label{sec:CombiRecoID}

\begin{figure}[!ht]
\centering
\includegraphics[width=0.49\textwidth]{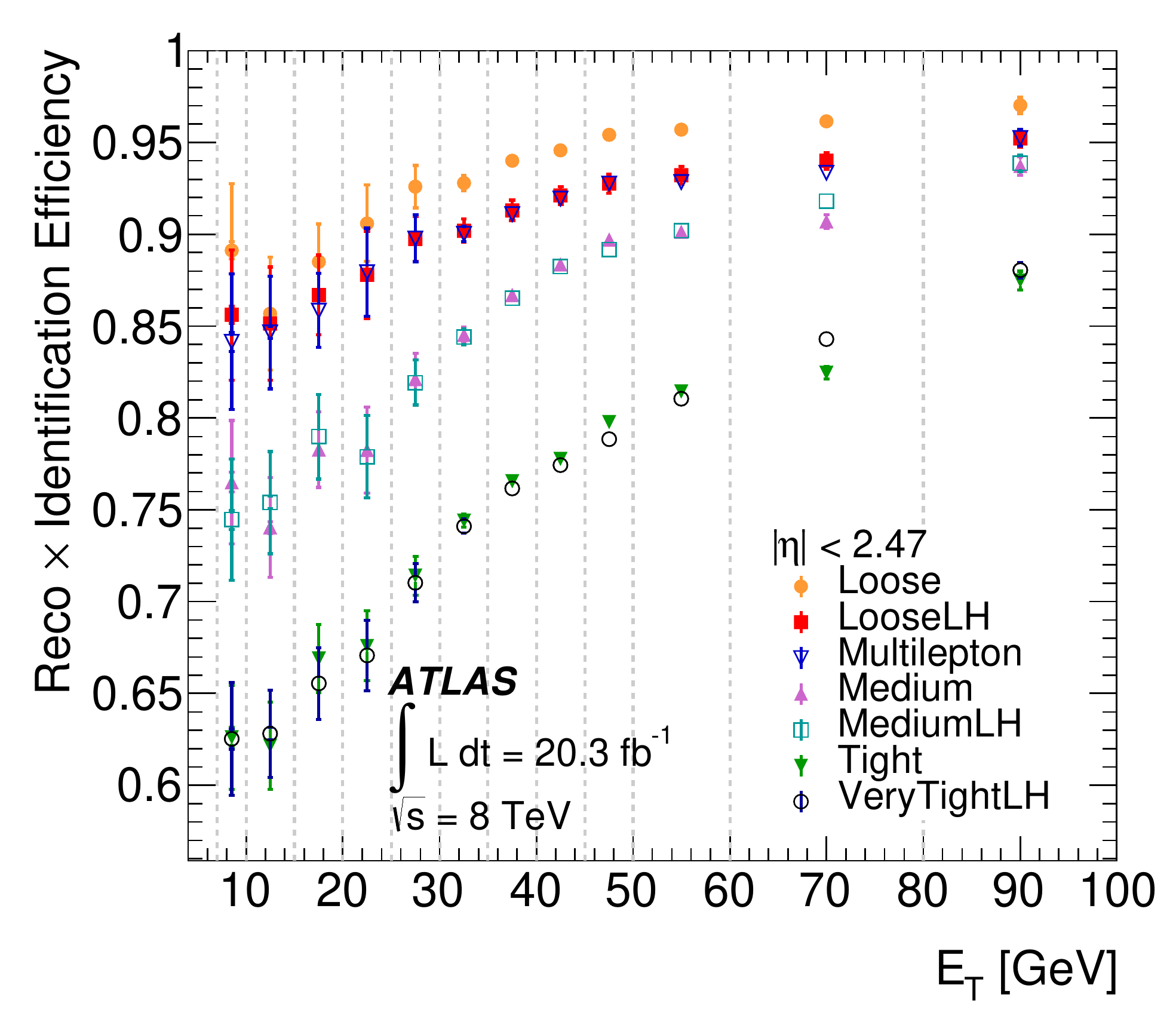}
\includegraphics[width=0.49\textwidth]{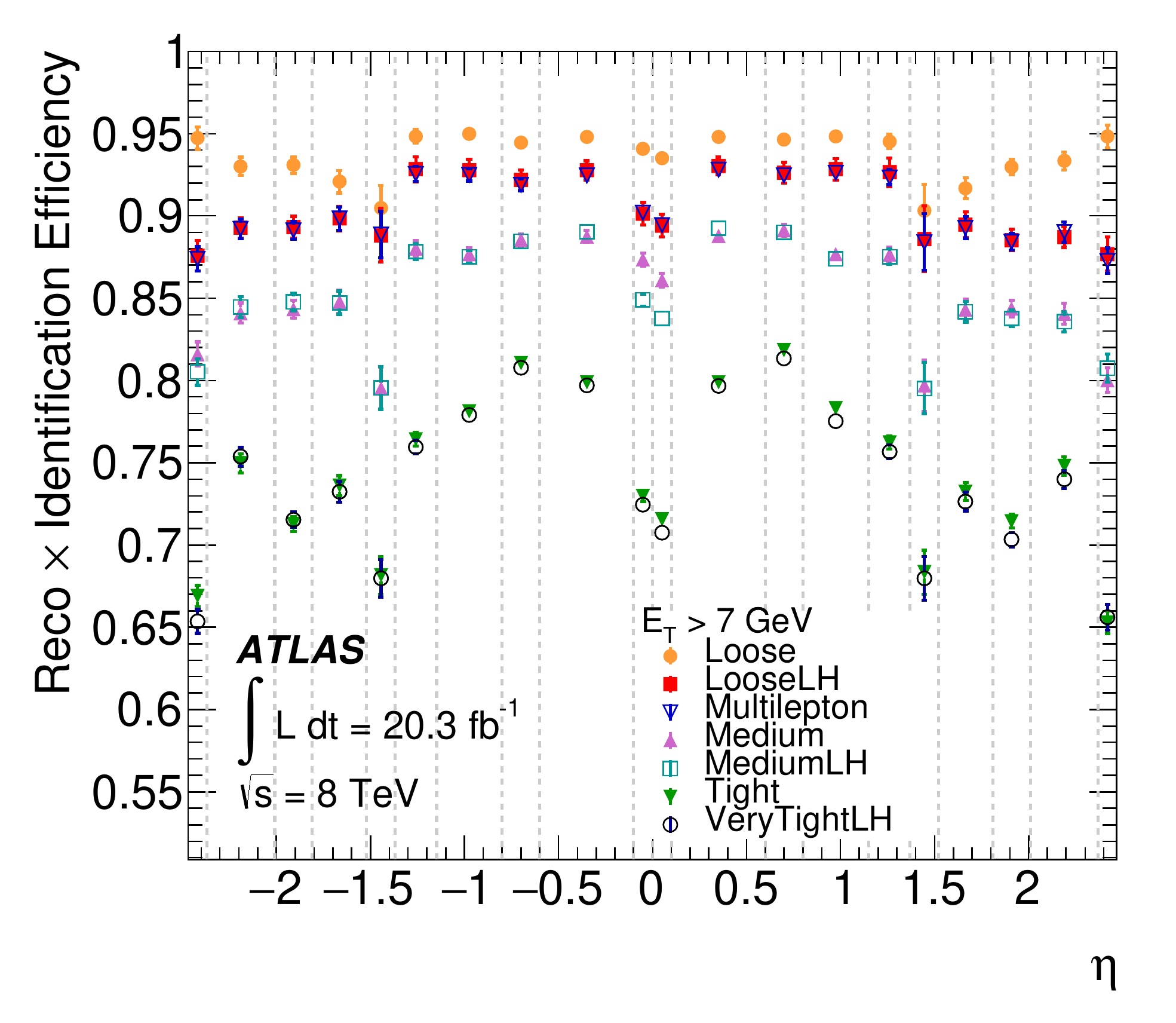}
\caption{
  Measured combined reconstruction and identification efficiency  for the various cut-based and LH selections 
as a function of \et\ (left) and $\eta$ (right) for electrons.
The data efficiency is derived from the measured data-to-MC efficiency ratios and the prediction of the MC simulation from \Zee\ decays. 
The uncertainties are statistical (inner error bars) and statistical+systematic (outer error bars). 
The last \et\ bin includes the overflow.
}
\label{fig:main_CombiRecoID_Data}
\end{figure}

\begin{figure}[!ht]
\centering
\includegraphics[width=0.49\textwidth]{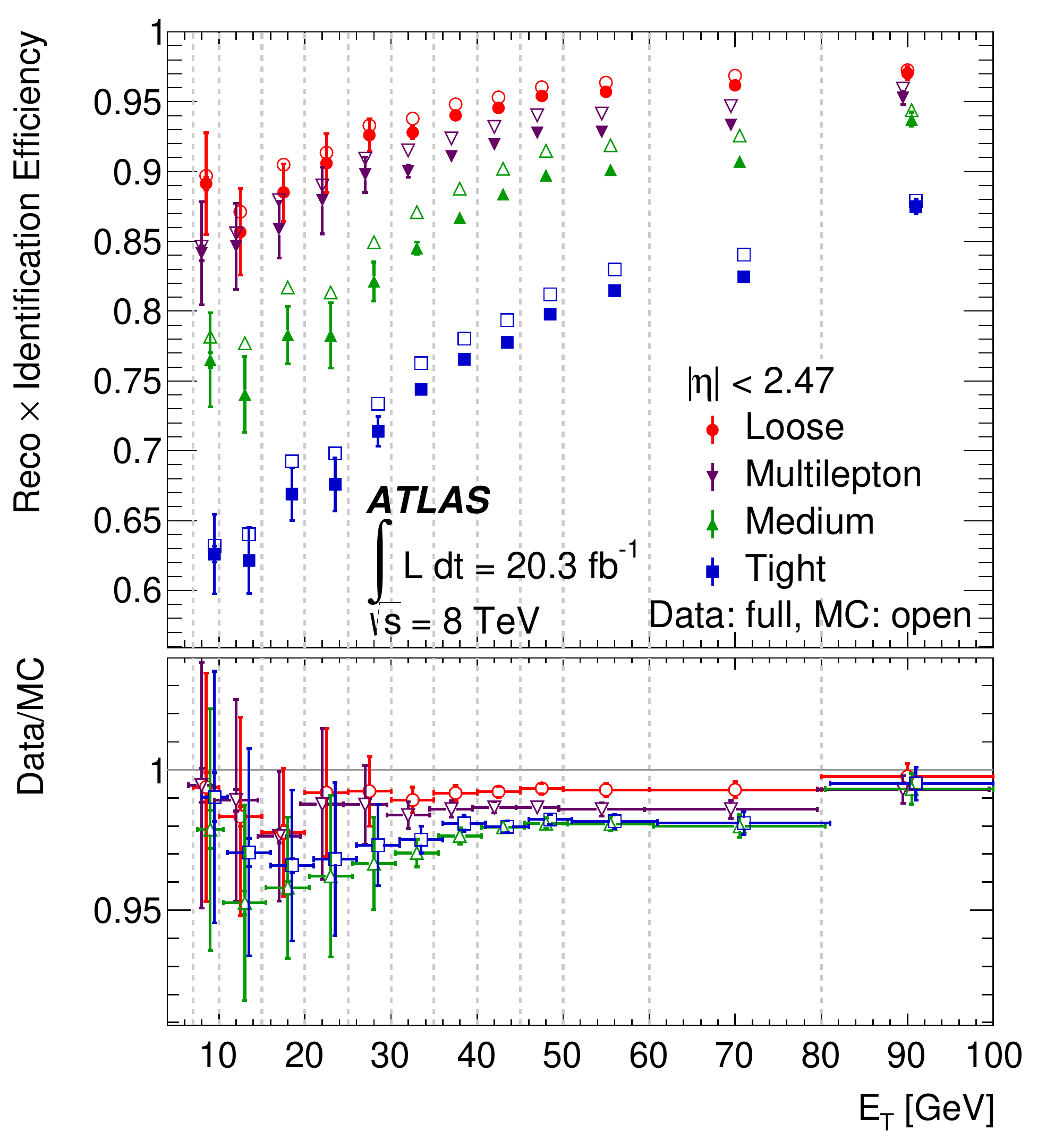}
\includegraphics[width=0.49\textwidth]{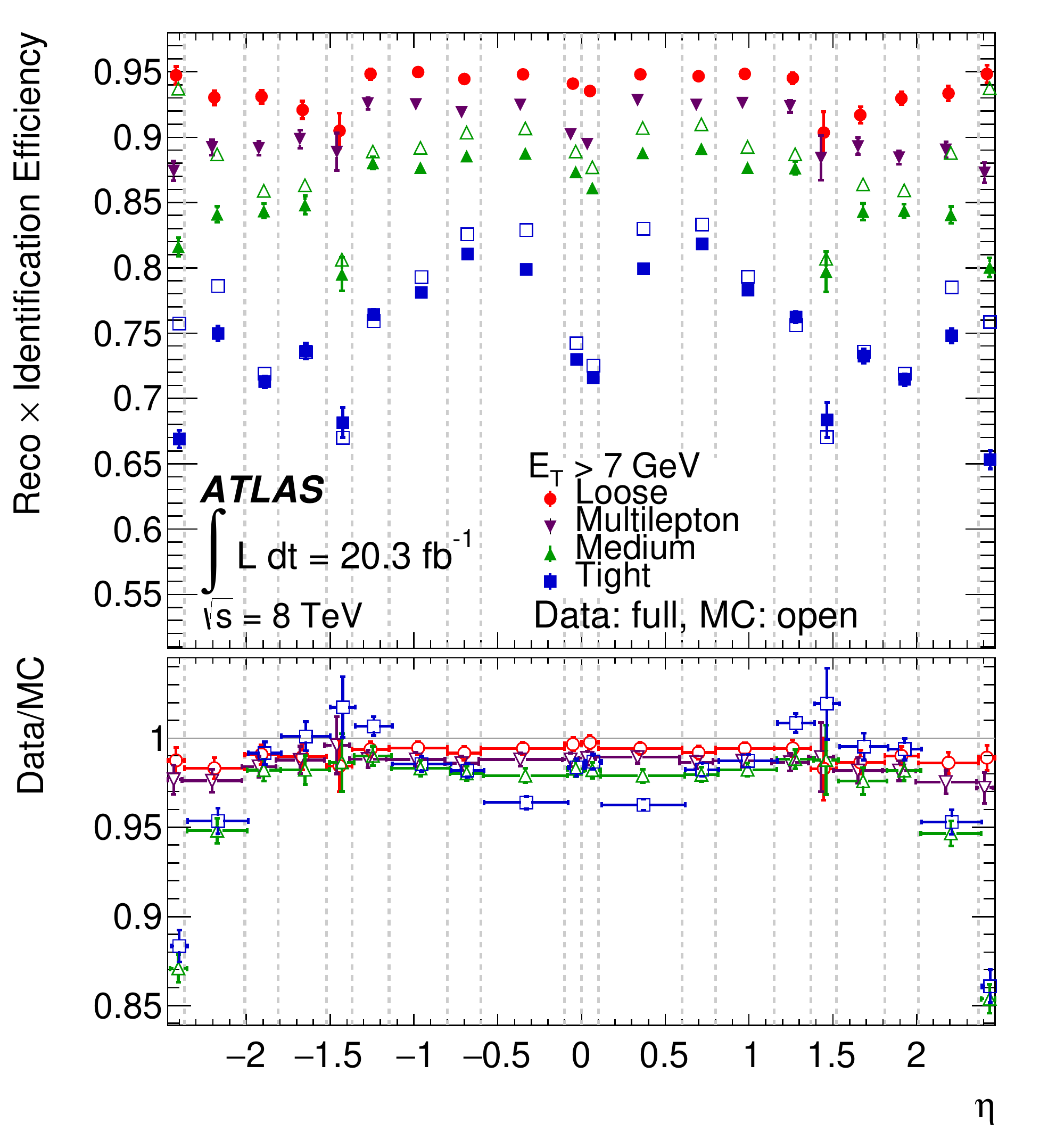}
\caption{
 Measured combined reconstruction and identification efficiency as a function of \et\ (left) and $\eta$ (right)
for the cut-based \loose, \multilepton, \medium\ and \tight\ selections, compared to expectation of the MC simulation for electrons from \Zee\
decay. The lower panel shows the data-to-MC efficiency ratios. 
The data efficiency is derived from the measured data-to-MC efficiency ratios and the prediction of the MC simulation for electrons from \Zee\
decays. The uncertainties are statistical (inner error bars) and statistical+systematic (outer error bars). The last \et\ bin includes the overflow.
}
\label{fig:main_CombiRecoID_DataMC_cuts}
\end{figure}

\begin{figure}[!ht]
\centering
\includegraphics[width=0.49\textwidth]{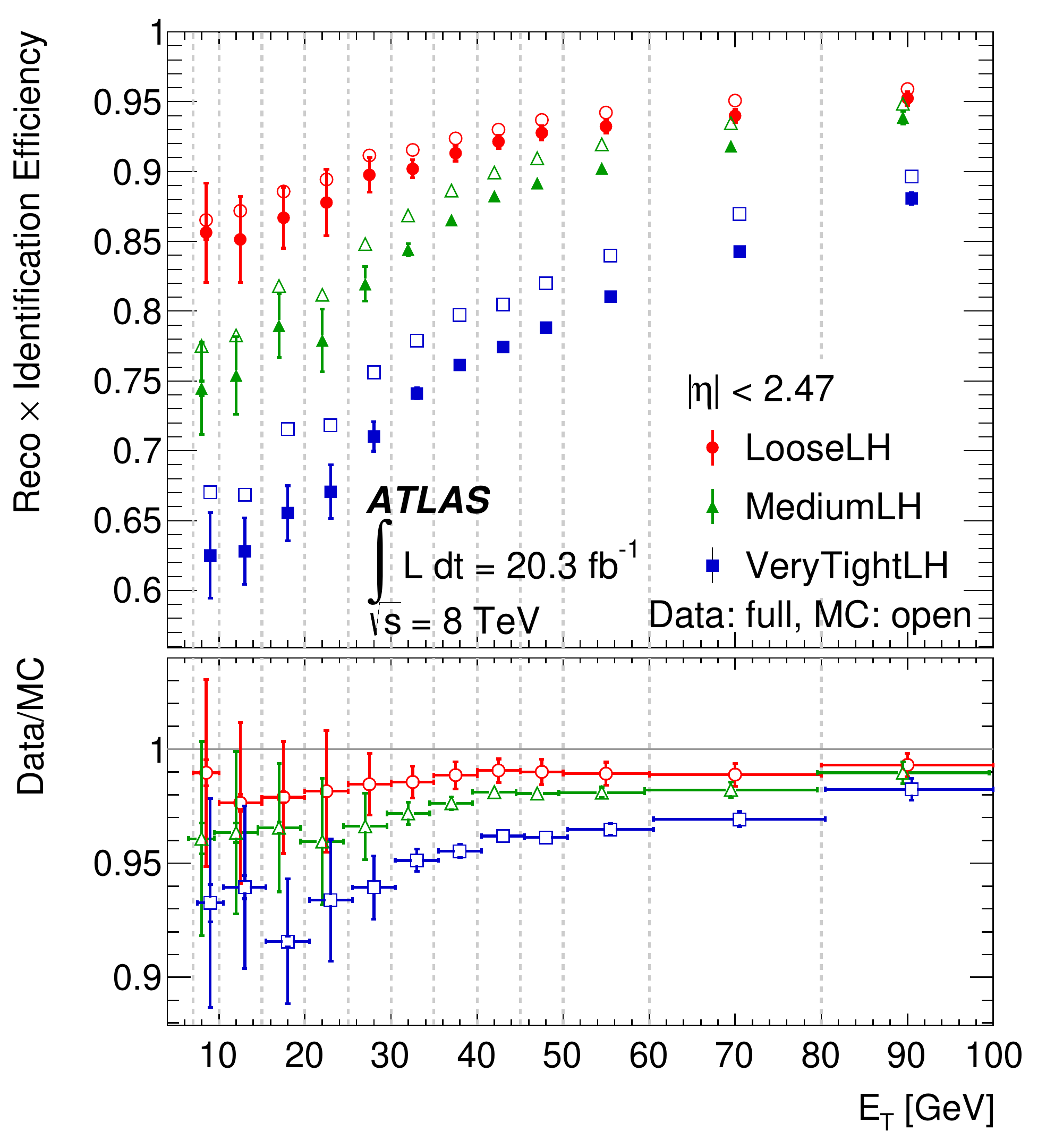}
\includegraphics[width=0.49\textwidth]{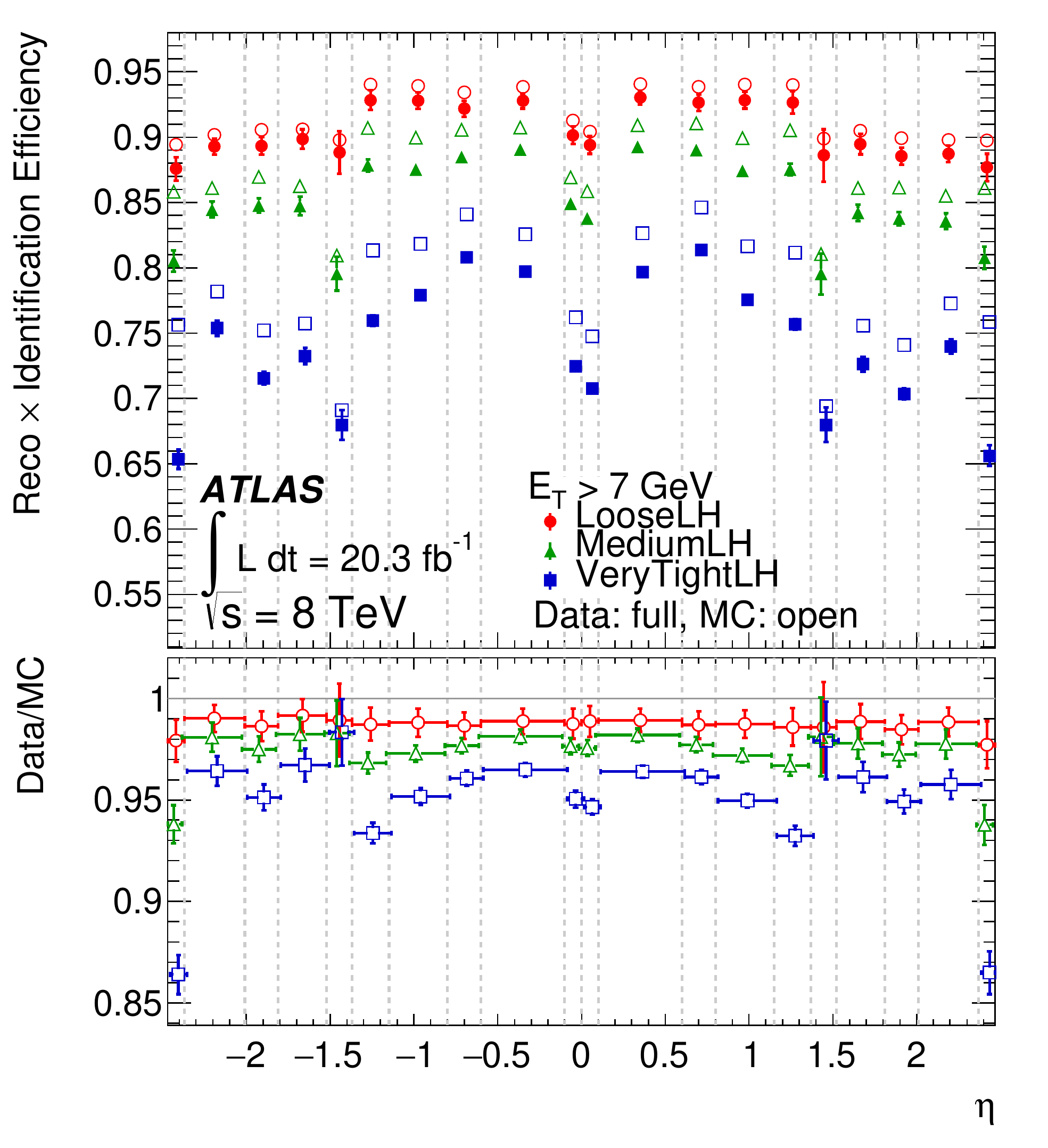}
\caption{
 Measured combined reconstruction and identification efficiency as a function of \et\ (left) and $\eta$ (right)
for the \looseLLH, \mediumLLH\ and \veryTightLLH\ selections, compared to predictions of the MC simulation for electrons from \Zee\
decay. 
The lower panel shows the data-to-MC efficiency ratios.
The data efficiency is derived from the measured data-to-MC efficiency ratios and the prediction of the MC simulation for electrons from \Zee\
decays.
The uncertainties are statistical (inner error bars) and statistical+systematic (outer error bars). The last \et\ bin includes the overflow.
}
\label{fig:main_CombiRecoID_DataMC_LH}
\end{figure}

Figure~\ref{fig:main_CombiRecoID_Data} shows the combined efficiencies to reconstruct and identify electrons with respect to reconstructed energy clusters in the EM calorimeter for all identification selections. The efficiencies are shown as a function of \et\ and $\eta$. As described in Section~\ref{sec:main_id_results}, the measured data-to-MC correction factors are applied
to a simulated \Zee\ sample. The resulting efficiencies correspond to the measured data efficiencies and can be compared to the
efficiencies of simulated electrons in \Zee\ events as done in Figure~\ref{fig:main_CombiRecoID_DataMC_cuts} and ~\ref{fig:main_CombiRecoID_DataMC_LH}. For electrons with $\et < 15$~\GeV, the reconstruction efficiency cannot be measured and is taken instead from the MC simulation. 

The combined efficiency to reconstruct and identify an electron from \Zee\ with \et\ around 25~\GeV\ is about 92\% for the \loose\ cut-based identification and around 68\% for the \tight\ cut-based identification as well as the \veryTightLLH\ selection. It is lower (higher) at lower (higher) \et, with a sharper turn-on as well as a greater $\eta$ dependence for the tighter selections. 
Since the reconstruction efficiency is constant, the shapes are mainly determined by the variation of the identification efficiency (see Sections~\ref{sec:MeasureReco} and~\ref{sec:MeasureID}).

\FloatBarrier
\section{Summary}\label{sec:Conclusions}

Using the full 2012 data set, 20.3~\ifb\ of 8~\TeV\ $pp$ collisions produced by the LHC, the reconstruction, identification, and charge misidentification efficiencies of central electrons in the ATLAS detector are 
determined using a \tnp\ method. Reconstruction and charge misidentification efficiencies are measured for electrons from \Zee\ decays.  The identification
efficiency measurements from \Jpsi\ and \Z\ decays are combined using data-to-MC efficiency ratios, improving the precision of
the results. 

In 2012, a new track reconstruction algorithm and improved track--cluster matching were introduced to recover efficiency losses due to electrons undergoing brems\-strahlung. As a result, the overall electron reconstruction efficiency
is increased by roughly 5\% with respect to the 2011 efficiency. Averaged over $\eta$, it is about 97\% for electrons with $\et=15$~\GeV\ and reaches about 99\%
at $\et = 50$~\GeV.  For electrons with \et~$>$~15~\GeV, the reconstruction efficiency varies from 99\% at low $|\eta|$ to 95\% at high $|\eta|$.

The uncertainty on the reconstruction efficiency is below 0.5\% for \et~$>$~25~\GeV, and between $0.5$--$2$\%
at lower transverse energy.  Below 15~\GeV, the reconstruction efficiency is not measured due to the overwhelming background contamination of the sample.

The electron identification was improved by loosening the selection criteria for the shower shapes in the EM
calorimeter that are most affected by the increased instantaneous luminosities provided by the LHC in 2012. 
To compensate for the
loss in rejection power, new selection criteria were introduced and requirements on variables less sensitive to pile-up were tightened.
Additionally, new identification selections were developed: the cut-based \multilepton\ selection, optimized for low-energy
electrons, as well as an identification based on the likelihood (LH) approach. Using the LH identification selections, 
the background rejection is significantly improved while maintaining the same signal efficiency as that of the cut-based selections.
The identification efficiency has a strong dependence on \et\ and, for the tighter criteria, on $\eta$. Calculated with respect to
reconstructed electrons satisfying quality criteria for their tracks, it averages between 96\% (cut-based \loose) and 78\%
(\VeryTightLLH) for electrons with \et~$>$~15~\GeV. 
The measured pile-up dependence is below 4\% for $1$--$30$ reconstructed primary collision vertices per bunch crossing for all sets of selection criteria. 
Some differences between the behaviour in data and MC simulation are observed, but understood.
The total uncertainties in the identification efficiency measurements are 5--6\% (1--2\%) for electrons below (above) \ET~=~25~\GeV. 

Charge misidentification of electrons in the probed \ET\ range is mostly caused by the emission of brems\-strahlung. The charge misidentification depends strongly on the applied selection criteria as well as on the $\eta$ of the electron. For representative selections the probability is at sub-percent level for $|\eta| < 1$ and can be as high as 10\% for $|\eta| \sim$ 2.5.

The measured data-to-MC efficiency ratios are applied as correction factors in analyses, such as the measurement of the properties of the Higgs boson, and their uncertainties are propagated accordingly. The scale factors are close to unity with deviations larger than a couple of percent from unity occurring only for low-\et\ or high-$|\eta|$ regions.

\section*{Acknowledgements}

We thank CERN for the very successful operation of the LHC, as well as the
support staff from our institutions without whom ATLAS could not be
operated efficiently.

We acknowledge the support of ANPCyT, Argentina; YerPhI, Armenia; ARC, Australia; BMWFW and FWF, Austria; ANAS, Azerbaijan; SSTC, Belarus; CNPq and FAPESP, Brazil; NSERC, NRC and CFI, Canada; CERN; CONICYT, Chile; CAS, MOST and NSFC, China; COLCIENCIAS, Colombia; MSMT CR, MPO CR and VSC CR, Czech Republic; DNRF and DNSRC, Denmark; IN2P3-CNRS, CEA-DSM/IRFU, France; GNSF, Georgia; BMBF, HGF, and MPG, Germany; GSRT, Greece; RGC, Hong Kong SAR, China; ISF, I-CORE and Benoziyo Center, Israel; INFN, Italy; MEXT and JSPS, Japan; CNRST, Morocco; FOM and NWO, Netherlands; RCN, Norway; MNiSW and NCN, Poland; FCT, Portugal; MNE/IFA, Romania; MES of Russia and NRC KI, Russian Federation; JINR; MESTD, Serbia; MSSR, Slovakia; ARRS and MIZ\v{S}, Slovenia; DST/NRF, South Africa; MINECO, Spain; SRC and Wallenberg Foundation, Sweden; SERI, SNSF and Cantons of Bern and Geneva, Switzerland; MOST, Taiwan; TAEK, Turkey; STFC, United Kingdom; DOE and NSF, United States of America. In addition, individual groups and members have received support from BCKDF, the Canada Council, CANARIE, CRC, Compute Canada, FQRNT, and the Ontario Innovation Trust, Canada; EPLANET, ERC, ERDF, FP7, Horizon 2020 and Marie Sk{\l}odowska-Curie Actions, European Union; Investissements d'Avenir Labex and Idex, ANR, R{\'e}gion Auvergne and Fondation Partager le Savoir, France; DFG and AvH Foundation, Germany; Herakleitos, Thales and Aristeia programmes co-financed by EU-ESF and the Greek NSRF; BSF, GIF and Minerva, Israel; BRF, Norway; CERCA Programme Generalitat de Catalunya, Generalitat Valenciana, Spain; the Royal Society and Leverhulme Trust, United Kingdom.

The crucial computing support from all WLCG partners is acknowledged gratefully, in particular from CERN, the ATLAS Tier-1 facilities at TRIUMF (Canada), NDGF (Denmark, Norway, Sweden), CC-IN2P3 (France), KIT/GridKA (Germany), INFN-CNAF (Italy), NL-T1 (Netherlands), PIC (Spain), ASGC (Taiwan), RAL (UK) and BNL (USA), the Tier-2 facilities worldwide and large non-WLCG resource providers. Major contributors of computing resources are listed in Ref.~\cite{ATL-GEN-PUB-2016-002}.

\clearpage
\printbibliography
\clearpage
\newpage 
\begin{flushleft}
{\Large The ATLAS Collaboration}

\bigskip

M.~Aaboud$^\textrm{\scriptsize 137d}$,
G.~Aad$^\textrm{\scriptsize 88}$,
B.~Abbott$^\textrm{\scriptsize 115}$,
J.~Abdallah$^\textrm{\scriptsize 8}$,
O.~Abdinov$^\textrm{\scriptsize 12}$,
B.~Abeloos$^\textrm{\scriptsize 119}$,
O.S.~AbouZeid$^\textrm{\scriptsize 139}$,
N.L.~Abraham$^\textrm{\scriptsize 151}$,
H.~Abramowicz$^\textrm{\scriptsize 155}$,
H.~Abreu$^\textrm{\scriptsize 154}$
% ...
\bigskip
\\
% ...
$^{8}$ Department of Physics, The University of Texas at Arlington, Arlington TX, United States of America\\
% ...
$^{12}$ Institute of Physics, Azerbaijan Academy of Sciences, Baku, Azerbaijan\\
% ...
$^{88}$ CPPM, Aix-Marseille Universit{\'e} and CNRS/IN2P3, Marseille, France\\
% ...
$^{115}$ Homer L. Dodge Department of Physics and Astronomy, University of Oklahoma, Norman OK, United States of America\\
% ...
$^{119}$ LAL, Univ. Paris-Sud, CNRS/IN2P3, Universit{\'e} Paris-Saclay, Orsay, France\\
% ...
$^{137}$ $^{(a)}$ Facult{\'e} des Sciences Ain Chock, R{\'e}seau Universitaire de Physique des Hautes Energies - Universit{\'e} Hassan II, Casablanca; $^{(b)}$ Centre National de l'Energie des Sciences Techniques Nucleaires, Rabat; $^{(c)}$ Facult{\'e} des Sciences Semlalia, Universit{\'e} Cadi Ayyad, LPHEA-Marrakech; $^{(d)}$ Facult{\'e} des Sciences, Universit{\'e} Mohamed Premier and LPTPM, Oujda; $^{(e)}$ Facult{\'e} des sciences, Universit{\'e} Mohammed V, Rabat, Morocco\\
% ...
$^{139}$ Santa Cruz Institute for Particle Physics, University of California Santa Cruz, Santa Cruz CA, United States of America\\
% ...
$^{151}$ Department of Physics and Astronomy, University of Sussex, Brighton, United Kingdom\\
% ...
$^{154}$ Department of Physics, Technion: Israel Institute of Technology, Haifa, Israel\\
$^{155}$ Raymond and Beverly Sackler School of Physics and Astronomy, Tel Aviv University, Tel Aviv, Israel\\
% ...
$^{a}$ Also at Department of Physics, King's College London, London, United Kingdom\\
$^{b}$ Also at Institute of Physics, Azerbaijan Academy of Sciences, Baku, Azerbaijan\\
$^{c}$ Also at Novosibirsk State University, Novosibirsk, Russia\\
$^{d}$ Also at TRIUMF, Vancouver BC, Canada\\
$^{e}$ Also at Department of Physics {\&} Astronomy, University of Louisville, Louisville, KY, United States of America\\
% ...
\end{flushleft}




\end{document}